\documentclass[aps,pra,superscriptaddress,twocolumn,10pt,longbibliography]{revtex4-1}

\usepackage{amsmath}
\usepackage{graphicx}
\usepackage{bm}
\usepackage{color}

\date{\today}

\newcommand{\sgn}{\mathrm{sgn}}
\newcommand{\di}{{i\mkern1mu}}

\definecolor{darkblue}{rgb}{0.1,0.2,0.6}
\definecolor{darkred}{rgb}{0.8,0.1,0.2}
\usepackage[colorlinks,citecolor=darkblue,linkcolor=darkred,urlcolor=darkblue]{hyperref}
\usepackage[all]{hypcap}

\usepackage{etoolbox}
\usepackage{mathtools}
\usepackage{amsmath}
\usepackage{amsfonts}
\usepackage{amssymb}
\usepackage{bm}
\usepackage{bbold}

\begin{document}

\title{Generalized quantum measurements with matrix product states:\\
Entanglement phase transition and clusterization}
\author{Elmer~V.~H.~Doggen}
\email[Corresponding author: ]{elmer.doggen@kit.edu}
\affiliation{\mbox{Institute for Quantum Materials and Technologies, Karlsruhe Institute of Technology, 76021 Karlsruhe, Germany}}
\affiliation{\mbox{Institut f\"ur Theorie der Kondensierten Materie, Karlsruhe Institute of Technology, 76128 Karlsruhe, Germany}}
\author{Yuval Gefen}
\affiliation{Department of Condensed Matter Physics, Weizmann Institute of Science, 7610001 Rehovot, Israel}
\author{Igor V.~Gornyi}
\affiliation{\mbox{Institute for Quantum Materials and Technologies, Karlsruhe Institute of Technology, 76021 Karlsruhe, Germany}}
\affiliation{\mbox{Institut f\"ur Theorie der Kondensierten Materie, Karlsruhe Institute of Technology, 76128 Karlsruhe, Germany}}
\affiliation{Ioffe Institute, 194021 St.~Petersburg, Russia}
\author{Alexander D.~Mirlin}
\affiliation{\mbox{Institute for Quantum Materials and Technologies, Karlsruhe Institute of Technology, 76021 Karlsruhe, Germany}}
\affiliation{\mbox{Institut f\"ur Theorie der Kondensierten Materie, Karlsruhe Institute of Technology, 76128 Karlsruhe, Germany}}
\affiliation{Petersburg Nuclear Physics Institute, 188300 St.~Petersburg, Russia}
\affiliation{Landau Institute for Theoretical Physics, 119334 Moscow, Russia}
\author{Dmitry G.~Polyakov}
\affiliation{\mbox{Institute for Quantum Materials and Technologies, Karlsruhe Institute of Technology, 76021 Karlsruhe, Germany}}

\begin{abstract}
 We propose a method, based on matrix product states, for studying the time evolution of many-body quantum lattice systems under continuous and site-resolved measurement. Both the frequency and the strength of generalized measurements can be varied within our scheme, thus allowing us to explore the corresponding two-dimensional phase diagram. The method is applied to one-dimensional chains of nearest-neighbor interacting hard-core bosons. A transition from an entangling to a disentangling (area-law) phase is found. However, by resolving time-dependent density correlations in the monitored system, we find important differences between different regions at the phase boundary. In particular, we observe a peculiar phenomenon of measurement-induced particle clusterization that takes place only for frequent moderately strong measurements, but not for strong infrequent measurements.
\end{abstract}

\maketitle

\section{Introduction} \label{Sec:Intro}

Open quantum systems \cite{Breuer2002,Rotter2015a} form the bridge between the world of unitary, deterministic evolution of closed quantum systems \cite{Polkovnikov2011a}, and the familiar experience of our macroscopic world. Recently, open quantum systems have received renewed interest in the context of quantum information processing and quantum circuits. The coupling to the environment can lead to decoherence in arrays of qubits, which limits the fidelity of quantum operations. A sufficiently high fidelity is essential for the performance of programmable quantum devices, in particular for ``quantum supremacy,'' which was reported to have been achieved recently~\cite{Arute2019a}.

Quantum measurements, via their back-action on the measured system, can mimic the effect of an environment. In a sense, the environment also ``measures'' the system, but without ``recording'' the extracted information. To that effect, coupling to the environment is a special type of ``blind measurement'' \cite{Roy2020}. Designing specific measurement protocols can be considered as engineering of the environment to which the quantum system is coupled. Improving our understanding of quantum measurement processes is therefore of immediate practical importance. At the same time, controlling the decoherence induced by the coupling to the environment may also help to advance our fundamental understanding of quantum measurements  \cite{Schlosshauer2005a, Wiseman2009}.

This important role of measurements in the quantum-information context, as well as their relation to decoherence and entanglement spreading \cite{Calabrese2004, Horodecki2009, Laflorencie2016a}, has led to a flurry of activity on the subject, especially with regard to entanglement transitions using a quantum circuit description \cite{Li2018a, Chan2019a, Skinner2019a, Li2019a, Szyniszewski2019a, Szyniszewski2020a, Bao2020a, Gullans2020a,  Jian2020a, Jian2020b, Choi2020a, Fan2020a, Chen2020a, Sang2020a, Zabalo2020a, Ippoliti2021a, Lavasani2021a, Ippoliti2021b}. The effect of local quantum measurements on entanglement has also been considered for systems described by a lattice Hamiltonian, in particular, for many-body localized systems \cite{Lunt2020a}, the quantum Ising chain \cite{Lang2020a, Rossini2020a, Biella2021a, Turkeshi2021},  non-interacting spinless fermionic models \cite{Cao2019a, Alberton2020a, Buchhold2021}, Hubbard-type interacting chains with short-range \cite{Fuji2020a} and long-range \cite{Minato2021a} interactions, and ultracold gases \cite{Goto2020a}. The question as to whether the effect of measurements on quantum circuits is qualitatively the same for real many-body systems is a subject of ongoing studies, as we detail below.

A key feature that has emerged in the pioneering studies of monitored systems is the interplay between the entangling effect of time evolution and the disentangling effect of the measurement, leading to an entanglement phase transition \cite{Li2018a,  Chan2019a, Skinner2019a, Li2019a}. Various types of measurements have been considered: local projective measurements, local weak measurements that only slightly perturb the system (for their recent applications in other contexts, see, e.g., Refs. \cite{Roy2020, Snizhko2020a, Snizhko2020b, Kumar2020b, Gebhart2020a, Xu2020a, Ivanov2020a, Manousakis2020a, Munoz2020a, Monroe2021a, Wang2021a}), non-local measurements of several sites of the system, and global measurements that act on the many-body system as a whole. The main diagnostic tool for the measurement-induced entanglement transition is the behavior of the entanglement entropy averaged over the measurement runs, but other indicators, such as mutual information or entanglement negativity, have also been used to explore the phenomenon. However, manifestations of the entanglement phase transition in more conventional density correlations are not yet sufficiently explored.

The entanglement transition was discussed for measurement-only dynamics, where non-local measurements produce both entangling (by non-locality) and disentangling (by projection) trends \cite{Ippoliti2021a, VanRegemortel2021a}. The measurement-induced transition was argued to be related to the ``purification transition'' \cite{Gullans2020a}, which can be employed for quantum-state preparation, control, and manipulation by means of quantum measurements. In addition, the properties of the entanglement transition have been linked to the theory of error corrections in quantum information processing \cite{Choi2020a, Fan2020a, Gullans2020b, Ippoliti2021a, Ippoliti2021b, Sang2020b, Li2021a}. In the presence of additional symmetries and constraints, a more sophisticated phase diagram may emerge, where the entanglement transition is accompanied by other types of phase transitions, see, e.g., Refs.~\cite{Sang2020a,Bao2021a}.

A key open question concerns the degree of universality between the various types of measurement protocols, applied to various different types of systems. In particular, it has been argued \cite{Bao2020a, Szyniszewski2020a} that the effect of continuous (weak, or ``generalized'') quantum measurements on quantum circuits is, by and large, analogous to the effect of rare projective measurements. A generalized phase diagram of hybrid quantum circuits in the plane of frequency vs. strength of measurements was analyzed in Refs.~\cite{Szyniszewski2019a, Szyniszewski2020a}, where a transition between entangling (for weak or infrequent measurements) and disentangling (strong or frequent) phases was established. At the same time, indications of a possible essential difference between the transitions at strong and weak measurements were reported in Ref.~\cite{Szyniszewski2019a}. Thus, it remains a challenging task to explore the universality of the entanglement transition for various measurement setups.

Another important---and still open---question concerns the properties of different phases around the transition. On the entangling side of the transition, volume-law scaling of the entanglement entropy was found in various hybrid unitary circuits, whereas in continuously monitored fermionic and spin systems both volume-law and logarithmic-law \cite{Fuji2020a, Alberton2020a, Buchhold2021, Turkeshi2021, Jian2021a} types of entangling scaling were reported. Logarithmic scaling, familiar from models described by conformal field theory \cite{Calabrese2004, Calabrese2009, Laflorencie2016a}, corresponds to the emergence of a critical entangled steady state in the thermodynamic limit. Volume-law behavior in this picture corresponds to the properties of small systems \cite{Alberton2020a, Buchhold2021}. A critical phase characterized by the conformal scaling of observables was also found in free models subject to non-unitary evolution governed by a non-Hermitian Hamiltonian \cite{Chen2020a,Gopalakrishnan2020a}. Concurrently, however, entangled phases were argued \cite{Cao2019a} to be unstable to arbitrarily weak measurements in the case of continuously monitored non-interacting fermionic chains that are measured at all sites.

One difficulty in determining the degree of universality in this large array of different systems, however, is that numerical studies thus far have mostly focused on fine-tuned models and not generic many-body systems. This means that, for instance, the role of interactions between particles, integrability of the model, as well as the interplay between many-body and the measurement-induced effects are largely unclear. Another common drawback of numerical studies of the measurement-induced transitions in correlated many-body systems is a limited accessibility of large system sizes, which is especially crucial for exploring the predicted change of behavior \cite{Alberton2020a} with increasing system size in the entangling phase.

A particularly promising direction, therefore, is to use the versatile approach of matrix product states (MPS) \cite{Schollwock2011a, Paeckel2019a} to simulate the dynamics, as applied recently to quantum measurements \cite{Tang2020a, Goto2020a, VanRegemortel2021a}. Specifically, MPS are a class of variational Ans{\"a}tze that approximate the exponential complexity of generic many-body states through a polynomial number of parameters, restricted to low- to moderately-entangled states.

In the present paper, we propose an MPS-based approach that describes quantum measurements in a continuous way. The dynamics of the monitored system is represented by the combination of unitary (governed by a many-body Hamiltonian) and stochastic non-unitary (effective non-Hermitian Hamiltonian) evolution. The latter models a local coupling to external degrees of freedom, thus bridging the concepts of environment-induced decoherence and quantum measurements. The advantage of the protocol we will formulate here is that it can be applied to any problem that permits an MPS description and hence includes interacting models. Importantly, the MPS approach can be controllably used for interacting many-body models with system sizes considerably larger than those accessible to exact methods, which are restricted to $\approx 20$ lattice sites.

The model we consider here is an interacting many-body system of hard-core bosons on a lattice, which constitutes a Luttinger liquid in the low-energy, continuum limit. We study the measurement-induced dynamics starting from the (moderately entangled) ground state of the system. Depending on the strength and frequency (probability) of the measurement process, the dynamics is either entangling or disentangling, in agreement with recent predictions based on the aforementioned quantum circuit descriptions. However, we also uncover a distinct feature of the system under study: in a certain parameter range near the transition between the entangling and disentangling phases, a clustering of particles occurs such that extended regions of particles and holes emerge over time. This signifies that, while the overall phase diagram breaks down into two distinct phases according to the entanglement scaling, the properties of the phases and the transition between them, quantified in other observables, depend on the measurement implementation.

The paper is organized as follows. In Sec.~\ref{Sec:Model}, we specify the system Hamiltonian and introduce the local measurements through a stochastic non-unitary evolution. We also compare our implementation of measurements with the existing approaches. In Sec.~\ref{Sec:Numerics}, we introduce the observables that we numerically calculate using the MPS and describe the simulation results. The obtained results for the entanglement entropy and density correlations are discussed in terms of phase diagrams in Sec.~\ref{Sec:Phase}. Finally, we summarize our findings in Sec.~\ref{Sec:Discuss}.

\section{Model and method \label{Sec:Model}}

\subsection{System}

We consider a hard-core boson model on a lattice of length $L$ (sites $x = 1, 2, \ldots, L$) characterized by the following Hamiltonian:
\begin{equation}
\mathcal{H}_0 = \sum_{x=1}^{L-1} \Big[ -\frac{J}{2}\Big(\hat{b}_x^\dagger \hat{b}_{x+1} + \mathrm{H.c.} \Big) + \Delta\, \hat{n}_x \hat{n}_{x+1} \Big]. 
\label{eq:ham}
\end{equation}
Here $\hat{b}_x^\dagger$, $\hat{b}_x$ are the bosonic creation and annihilation operators and $\hat{n}_x \equiv \hat{b}_x^\dagger \hat{b}_x$ on the $\{0,1\}$-manifold of local occupation. We set $J=1$ and $\hbar=1$. For $|\Delta| \leq 1$, the ground state is a Luttinger liquid (otherwise, it is ferro- or antiferromagnetic). Below, we focus on this range of interaction, choosing (for the most part of the paper, except for a brief discussion of other values of $\Delta$ in Sec.~\ref{s4b2}) attractive interaction with $\Delta = -0.5$. The model \eqref{eq:ham} is known as the $t$-$V$ model in the spinless fermion language and the XXZ Heisenberg chain in the spin-$1/2$ language. Throughout the paper, we consider the case of half-filling; namely, we take $L$ to be even and fix the number of particles to be $L/2$.

\subsection{Measurement}
\label{s2b}

The Hamiltonian \eqref{eq:ham} defines the deterministic unitary dynamics of the quantum state. At the same time, measurements induce a non-unitary stochastic dynamics (quantum trajectory). In line with this, we model the measurement-induced dynamics of a pure state by a non-Hermitian stochastic Hamiltonian, with the stochasticity controlled by random outcomes.

We break the time axis into measurement intervals of the same duration $T$, i.e., the $j$-th measurement occurs between the times $t_j$ and $t_{j+1}=t_j+T$. The measurement in the $j$-th interval is implemented as a two-component process: (i) a quantum quench by adding at time $t_j$ and removing at time $t_{j+1}$ a random purely imaginary (anti-Hermitian) on-site potential $\mathcal{H}_\mathrm{meas}^{j}$ to the Hamiltonian $\mathcal{H}_0$ and (ii) a continuous---between $t_j$ and $t_{j+1}$---restoration of the wave function norm through a global renormalization of the MPS. Specifically, $\mathcal{H}_\mathrm{meas}^{j}$ is of the form
\begin{equation}
\mathcal{H}_\mathrm{meas}^{j}
= \di M \sum_x p_x^{j}\, \sgn \Big( n_x - m_x^{j} \Big) \hat{n}_x,
\label{eq:measHam}
\end{equation}
where $n_x \equiv \langle \hat{n}_x \rangle$ is the expectation value of the on-site density, $p_x^j=\{ 0, 1 \}$ is a binary random variable with values 0 and 1, and $m_x^j\in (0, 1)$ is a random variable uniformly distributed between 0 and 1. The constant $M$ is real and positive. 

The quantity $p_x^j$ indicates whether the measurement is performed at the site $x$ within the $j$-th time interval. Its stochastic distribution is characterized by the probability $P_x^j$ that the site $x$ is measured at the $j$-th time step (which corresponds to $p_x^j=1$). In this paper, we take $P_x^j$ to be time independent and the same over all lattice sites, with $P_x^j\equiv P$. The value of $P$ thus parameterizes the measurement rate, with $P=1$ describing the limit of frequent measurements at each time step.

The probabilistic character of quantum-measurement outcomes is enforced by the randomness of the quantity $m_x^j$. The stochastic behavior of the factor $\sgn (n_x - m_x^{j})$, which depends on $n_x$, is then akin to the Born rule. In the limiting cases of $n_x=1$ and 0 at a given time, the sign function is equal to 1 in the former case and $-1$ in the latter, irrespective of $m_x^j$, which corresponds to deterministically measuring the presence ($n_x=1$) or absence ($n_x=0$) of a particle. For intermediate values of $n_x$, the measurement outcome is probabilistic, depending on both $n_x$ and the random choice of $m_x^j$.

The quantity $M$ in Eq.~(\ref{eq:measHam}) characterizes the measurement strength. The limit of $MT\gg 1$ describes a projective measurement of the particle density, whereas the opposite limit corresponds to a weak measurement. Indeed, our implementation of the measurement can be formalized in terms of a sequence of imaginary-time evolution events governed by $\mathcal{H}_\mathrm{meas}^{j}$, superimposed onto the real-time evolution induced by $\mathcal{H}_0$. The two types of dynamics compete with each other as far as the density distribution in space is concerned. The unitary dynamics drives the system towards a spatially homogeneous state with a local occupancy of 1/2. By contrast, a fast, corresponding to large $M$, imaginary-time dynamics at a given site drives the local occupancy to either fully occupied or fully empty, thus embodying the notion of a projective measurement.

Consecutive measurements describe a stochastic process, with $\mathcal{H}_\mathrm{meas}^{j+1}$ at step $j+1$ characterized by a randomly chosen set of $p_x^{j+1}$ and $m_x^{j+1}$, which is independent of $p_x^j$ and $m_x^j$, and the running value of the expectation value $n_x$. As already mentioned above, the latter evolves in time under the action of both the Hermitian and anti-Hermitian parts of the Hamiltonian. Crucially, we complement the nonunitary dynamics of $n_x$ by a constraint that maintains wave-function normalization at each point in time (note that, without the constraint, the wave function norm may both decrease and increase depending on the sign of ${\rm Im}\,\mathcal{H}_\mathrm{meas}^{j}$). The measurement in our formalism is thus the combined effect of the generalized (unitary + nonunitary) Hamiltonian dynamics complemented by restoration of the norm of the state. A detailed discussion of the physical justification of the measurement Hamiltonian \eqref{eq:measHam}, as well as a comparison with a more conventional implementation of measurements is provided in Appendix \ref{sec:appendix}.

By construction, our formalism describes the evolution of the wave function conditioned on the measurement record [a particular sequence of ``readouts'' -- random numbers $\text{sgn}(n_x-m_x^j)=\pm 1$], and as such is suitable for the calculation of quantities like the entanglement entropy, which are not describable in terms of the Lindblad equation for the density matrix averaged over quantum trajectories. The fully ``Hamiltonian" theoretical framework complements that of quantum circuits. Our model provides a conceptually particularly effective (and efficient) formalism to study the effect of measurement in the plane of two variables: the measurement rate ($P$) and the measurement strength ($M$), both encoded in Eq.~(\ref{eq:measHam}).

It is worth noting that the measurement procedure induces correlations throughout the system, which travel with a velocity that is bounded from above. This is similar to the Lieb-Robinson bound; however, the maximum velocity in the presence of strong measurements is limited by $M$ rather than by $J$. We thus require the discretization time step of the numerical integrator $\delta t$ to satisfy $\delta t \ll \min\{M^{-1}, T\}$. This also clarifies the meaning of the term ``continuous" when we refer above to the continuous restoration of the wave function norm---the restoration is discretized with the time step $\delta t$, which is the smallest time scale in the system.

Before we proceed in Sec.~\ref{s2c} to describe the measurement protocol from the numerical perspective in more detail, let us comment on our choice of the initial state, which we take to be the ground state. As a starting point, we compute, using the density matrix renormalization group (DMRG), the ground state of the Hamiltonian \eqref{eq:ham}. Time evolution is then implemented using the time-dependent variational principle (TDVP) \cite{Haegeman2016a}, where we use the same hybrid approach as in Ref.~\cite{Doggen2020a}, combining the one-site and two-site implementation of the TDVP. These methods are of the MPS class of algorithms, which are restricted to a variational subspace of the full Hilbert space, targeting low-entanglement states.

As a matter of principle, the MPS framework outlined above allows for a simulation of the crossover from weak to strong measurements for arbitrary interacting lattice Hamiltonians and arbitrary initial states. We opt to start the dynamics not from an unentangled product state, as is conventional, but from the ground state of the Hamiltonian \eqref{eq:ham}. This has the following key advantage. The ground state is only moderately entangled, with a characteristic logarithmic scaling of entanglement entropy \cite{Laflorencie2016a}. In the absence of measurements, the system remains in the ground state under unitary time evolution. Thus, any increase or reduction in entanglement is  solely due to the measurement (and its interplay with $\mathcal{H}_0$). This should be contrasted with the case of an initial high-energy product state that rapidly entangles, reaching a volume-law entangled phase under the dynamics of $\mathcal{H}_0$ in the absence of any measurement. From a technical perspective, our choice of the initial condition speeds up the simulation of the dynamics, since a relatively modest size of the variational manifold suffices. However, the MPS ansatz is restricted to moderate entanglement. This ansatz captures the weakly entangled ground state, with only logarithmic growth of the entanglement with system size, very well, even with a modest bond dimension. As mentioned, absent any measurement, we can then of course capture the dynamics quasi-exactly up to late times. Likewise, any disentangling dynamics can be captured accurately. Conversely, if the measurement protocol leads to increased entanglement over time, then the numerical simulations are only quantitatively accurate up to a limited time. In practice, for our choice of parameters this means that the numerical results shown below are quantitatively accurate up to values of the bipartite von Neumann entropy $S \lesssim 3$. It should be emphasized that in this regime the consequence of the truncation of the entanglement spectrum, for the problem at hand (see Fig.~\ref{fig:entropyscaling}), is a reduction of the entropy. This implies we can correctly capture the regime where entangling dynamics occurs, as increasing the bond dimension will only increase the entropy.

\subsection{Protocol}
\label{s2c}

We simulate the dynamics of the monitored chain governed by $\mathcal{H}_0 + \mathcal{H}_\mathrm{meas}$, using various choices of measurement strength $M$, probability of measuring each site $P$, and system size $L$, in a time window $t \in [0, 50]$. We start from the ground state of $\mathcal{H}_0$ with $\Delta = -1/2$. We choose the measurement interval $T = 1$ and the time step $\delta t = 0.005$ for the TDVP integrator. At $t = 50$, we switch off the measurement Hamiltonian \eqref{eq:measHam} and continue evolving according to the Hamiltonian \eqref{eq:ham} in a small time interval $t \in [50, 60]$. The size of the variational manifold is controlled by a numerical parameter called the bond dimension $\chi$ \cite{Schollwock2011a}. For finding the ground state we use $\chi = 500$, and for the dynamics we mainly use $\chi = 64$ (but for some values of $L$, $P$ and $M$ we benchmark our results using larger bond dimensions, see below). For each set of parameters, we repeat the procedure at least $R = 40$ times.

At each of the measurement steps $t = j= 0, 1, \ldots$, we generate new measurement outcomes through the evolution of $n_x(t)$ and by taking new random numbers $m_x^{(j)}$. Hence, it is possible, depending on the dynamics and random chance, for one particular site to be measured as having a particle present, a hole present, or not being measured, and these outcomes can change during the time evolution. The dynamics under $\mathcal{H}_0$ alone tends to ``delocalize'' the system (a random product state will evolve towards a volume-law entangled state that is homogeneous in density), while the dynamics governed by $\mathcal{H}_\mathrm{meas}$ rather tends to localize particles at the measured sites. We therefore expect a competition between both mechanisms that can potentially lead to a transition between delocalized (entangling) and localized (disentangling) types of behavior.

\subsection{Comparison to existing approaches}

Our protocol is aimed at mimicking the coupling to a measurement apparatus. This can be described by the quantum trajectory approach  with matrix product states (cf.~Refs.~\cite{Tang2020a, Goto2020a}). 
As an essential ingredient of this approach, one frequently uses the Lindblad master equation (for more details, see the review \cite{Daley2014a}):
\begin{equation}
    \dot{\bar{\rho}} = - \frac{i}{\hbar} [\mathcal{H}, \bar{\rho}] - \frac{1}{2} \sum_m \gamma_m [L_m^\dagger L_m \bar{\rho} + \bar{\rho} L_m^\dagger L_m -2 L_m \bar{\rho} L_m^\dagger], \label{eq:lindblad}
\end{equation} 
where $\bar{\rho}$ is the averaged-over-trajectories density matrix, $\mathcal{H}$ is the Hamiltonian of the system, and the operators $L^\dagger_m$ and $L_m$ are the jump operators with a corresponding coupling strength $\gamma_m$ in a Markovian approximation of the coupling to the environment. It should be emphasized that the Lindblad equation describes the evoluton of the  density matrix averaged over measurement outcomes. Such an averaged density matrix is not sufficient for evaluation of the entanglement entropy (or other characteristics of the entanglement), since the entropy is a non-linear function of the density matrix. To study measurement-induced evolution of the entanglement (and, in particular, entanglement phase transitions), one should follow individual quantum trajectories (preserving purity of the state), then calculate the entanglement, and only then perform the averaging. The Lindblad equation is not suitable for these purposes: systems characterized by the same $\bar{\rho}$ may have totally different entanglement.

The emergence of an effective non-Hermitian Hamiltonian to describe the measurement process in conjunction with quantum jumps is an inherent part of the quantum trajectory approach \cite{Dalibard1992a, Daley2014a}, with recent examples being Refs.~\cite{Cao2019a,Fuji2020a, Goto2020a,Alberton2020a,Biella2021a,Turkeshi2021} for various choices of the effective Hamiltonians and jump operators. 
In particular, the quantum trajectory approach was applied for numerical simulations of monitored non-interacting systems in Refs.~\cite{Cao2019a, Alberton2020a}, where unraveling is employed to obtain a stochastic Schr\"odinger equation from the Lindblad equation.
This procedure makes use of the formal equivalence of the term
$L_m^\dagger L_m \bar{\rho} + \bar{\rho} L_m^\dagger L_m$
in the Lindblad equation (\ref{eq:lindblad}) to an anticommutator of the density matrix and an effective anti-Hermitian Hamiltonian $\propto i L_m^\dagger L_m$. 
For interacting systems, this approach was used in, e.g.,  Ref.~\cite{Fuji2020a}, where exact diagonalization was employed, which is restricted to relatively small systems of $L \approx 20$. A continuous approach applied to random quantum circuits was also recently proposed in Ref.~\cite{Szyniszewski2020a} where it was implemented for system sizes $L \le 20$.

The MPS implementation of the quantum-trajectory approach can be viewed as a continuum (in the time domain) analog of discrete quantum circuit models that have been studied recently (see Sec.~\ref{Sec:Intro}). Similarly, we are capable of following individual quantum trajectories without averaging over the measurement outcomes encoded in the sequence of random variables $m_x^{(j)}$. A key difference between the approach based on quantum jumps is that we now replace the instantaneous jumps by continuous ones. In our protocol the ``measurement'' and the time evolution therefore occur simultaneously. Hence, there is a direct interplay between them in the continuum time domain, in contrast to the approaches of Refs.~\cite{Tang2020a, Goto2020a}.

In contrast to random hybrid circuits, the unitary part of the evolution in our scheme is governed by the physical Hamiltonian of the system, and, hence, the ``unitaries'' applied to the site at consecutive time steps are not random but rather are determined by the same fixed (time-independent) Hamiltonian $\mathcal{H}_0$. In addition, at each time step, we maintain half-filling, which provides a global constraint on the hybrid evolution of the system. The advantage of our approach is that it permits the investigation of generic many-body systems on a lattice for an arbitrary strength of measurements. Note that our approach differs from that based on the unravelling of the Lindblad equation in that the non-Hermiticity in our case is not emergent but is introduced at the fundamental level of the original stochastic Hamiltonian (modeling wave-function collapse). In this regard, our approach bears a certain similarity to that of Ref.~\cite{Chen2020a}, where stochastic non-unitary dynamics of free fermions was addressed by introducing a random anti-Hermitian part of the Hamiltonian. However, in contrast to that work, random fluctuations between different realizations of the stochastic Hamiltonian emerge in our formalism because of the application of the Born-like rule for density measurements, and hence the update rules depend on the system state.

A detailed analysis of similarities and differences between our method and previous approaches is presented in Appendix \ref{sec:appendix}. In particular, we explain there the origin of non-Hermiticity of the effective measurement Hamiltonians. For comparison with other approaches, we derive there a Lindblad-type master equation for $\bar{\rho}$ corresponding to our protocol. This equation is non-linear, at variance with the conventional Lindblad equation \eqref{eq:lindblad} and in similarity with a master equation obtained in Ref.~\cite{Patel2017a}, where the wave-function collapse for a single strong measurement  was modelled in terms of multiple weak measurements. As we point out in Appendix \ref{sec:appendix}, the non-linearity is in fact a general property of the master equation for $\bar{\rho}$ in the case of protocols with a post-selection. Furthermore, the non-linearity is generic also for master-equations describing the evolution of averaged moments of the density matrix in monitored systems \cite{Buchhold2021}, which are needed to characterize the entanglement.

\section{Numerical results \label{Sec:Numerics}}

In this section, we introduce the physical observables and present our numerical results used to probe the entanglement transition and dynamics of a monitored correlated system.

\subsection{Signatures of measurement-induced transition}

\subsubsection{Entanglement entropy}

A measure that is useful as a diagnostic tool is the von Neumann entanglement entropy  $S$ \cite{Laflorencie2016a}. An ergodic, thermalizing system is characterized by the volume-law growth of entanglement, whereas a localized system shows area-law growth of entanglement. In the case of a one-dimensional system, the volume law corresponds to a scaling $\propto L$, and the area law corresponds to just a constant. The von Neumann entropy of entanglement for a bipartition into subsystems $A$ and $B$ is given by:
\begin{equation}
    S = -\mathrm{Tr}(\rho_A \ln \rho_A), \quad \rho_A \equiv \mathrm{Tr}_B |\Psi \rangle \langle \Psi |,
    \label{S}
\end{equation}
where $\rho_A$ is the reduced density matrix of subsystem $A$.

The initial state we consider here---the ground state of a Luttinger liquid---has the feature that $S$ is neither volume-law-entangled nor area-law-entangled, with $S$ showing an intermediate behavior: a logarithmic growth with system size. This allows us to distinguish the entangling phase from the disentangling one by comparing the entropy after a sufficiently long time to the initial entropy. (Of course, strictly speaking, this requires the $L\to \infty$ limit, whereas in practice we are limited to large but finite $L$. The large-$L$ requirement becomes particularly stringent close to the transition.)  Below we calculate the time evolution of the entanglement entropy for individual quantum trajectories, as well as the entropy averaged over $R$ runs.

Throughout this work, we measure the entropy $S$ in units of $\ln 2$, which corresponds to a replacement $ \ln \to \log_2$ in Eq.~\eqref{S}.

\subsubsection{Particle density and clustering} \label{sec:cluster}

Another useful---and experimentally accessible---measure  is the particle density. In the limit of $M \gg 1$, ``chokepoints'' of high or low density (particles and holes) are generated that serve as blockades to correlations. It turns out to be instructive to consider the \emph{cluster size}, which is a commonly used diagnostic tool in percolation transitions. Here we define a cluster as a set of consecutive sites with a density at most $0.2$ from the extreme values ($0$ and $1$). We then compute the maximum cluster length for each realization.

\subsection{Simulations} \label{Sec:Simulations}

\subsubsection{Strong frequent measurement: $M \gg 1$, $P = 1$}
\label{sec:strong_frequent}

\begin{figure}
    \centering
    \includegraphics[width=\columnwidth]{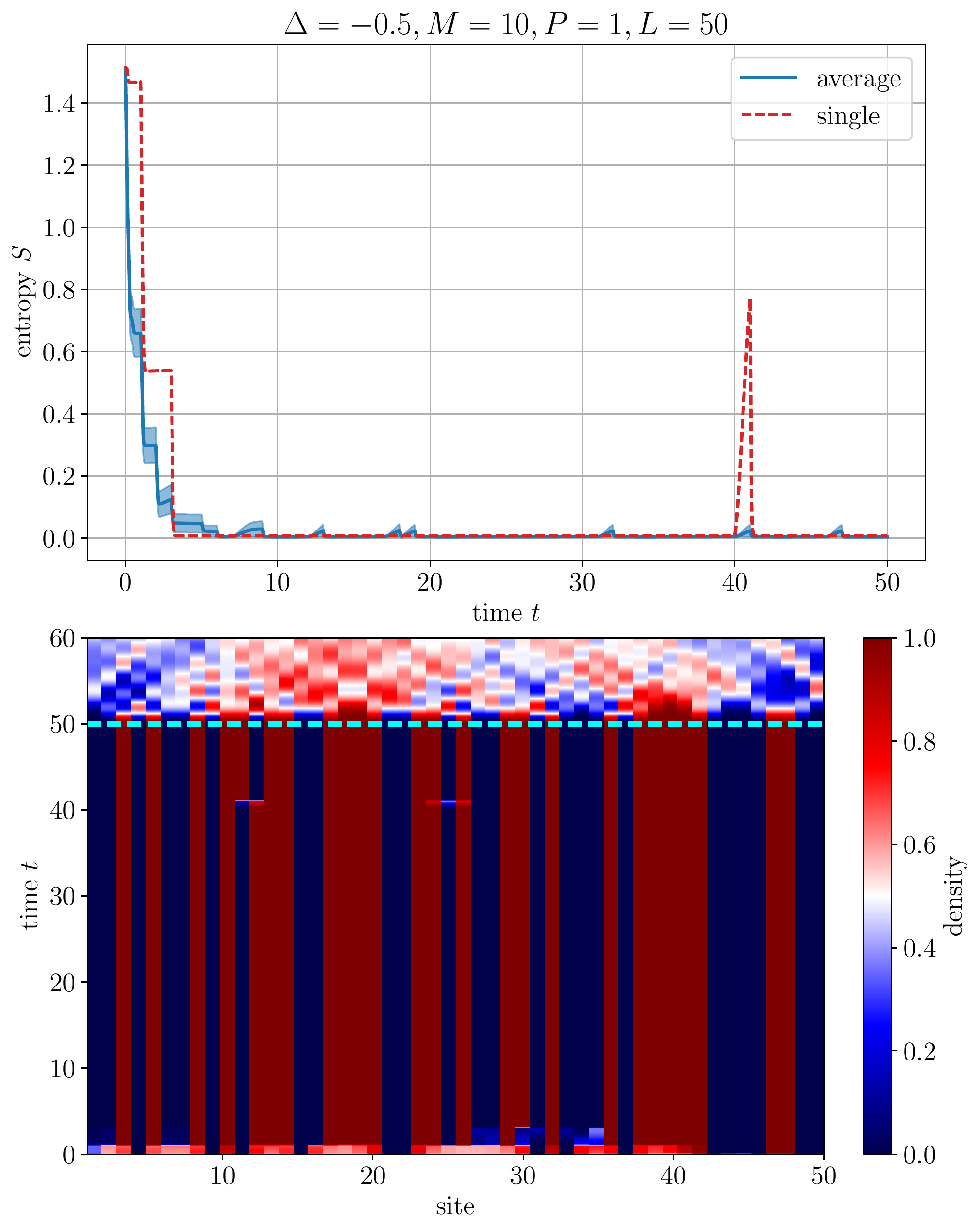}
    \caption{Time evolution of the entropy $S$ (top panel), showing the average over $R=40$ realizations for $L = 50$, in the limit of strong frequent ($M=10,P=1$) measurement. The dashed red line corresponds to the entropy for an arbitrarily chosen single realization, with the corresponding density evolution in the bottom panel. The shaded area indicates a $1\sigma$ error, estimated through the standard deviation of $S(t)$ across the different realizations. The dashed line in the bottom panel indicates the time at which the measurement Hamiltonian \eqref{eq:measHam} is switched off.}
    \label{fig:M10P1}
\end{figure}

First, we consider the case where measurement is strong, $M = 10$, and each site is always measured, $P=1$. The results for the entanglement entropy and particle density are shown in Fig.~\ref{fig:M10P1}. The entropy approaches zero on a time scale of order unity and remains very close to zero for the whole duration of the measurement run, with occasional spikes in the entropy representing rare fluctuations (``glitches''). The small, time-independent value of entropy clearly indicates that the system is characterized by the area-law scaling of entanglement, corresponding to the disentangling phase.

In terms of the density, a random configuration of particles and holes is chosen at the very beginning of each run, $t\sim 1$, and we observe the quantum Zeno effect \cite{Li2018a}. The density profile in the $x$-$t$ plane forms stripes of occupied and unoccupied states. Whether a given site is occupied or not is essentially determined by the random variable $m^{(j=1)}_x$ at the first step (so that there are no correlations between different sites). The pattern established at $t \sim 1$ remains almost unchanged at later times.

However, the particles are not exactly frozen: since $M$ is finite, the dynamics driven by $\mathcal{H}_0$ slightly perturbs the product state, leading to a nonzero probability of a flip in the sign of $\mathcal{H}_\mathrm{meas}$, as can be seen to occur in the Fig.~\ref{fig:M10P1}. Nonetheless, long-range correlations are strongly suppressed and we observe an area-law state with essentially zero entanglement between distant parts of the system. A substantial entanglement appears only between neighbouring sites and only during the rare processes of particle hopping. When such flips occur close to the center of the system, where the bipartition is taken, they affect the entropy. This leads to rare peaks in the individual traces of $S(t)$ that lead to a finite value of entropy in the averages.  Only in the limit $M \rightarrow \infty$, which corresponds to a projective measurement, does the quantum Zeno state become fully robust  (i.e., strictly time-independent).

After switching off the measurement Hamiltonian, the density quickly settles to a homogeneous state (see the density plot above the dashed line in the lower panel of Fig. \ref{fig:M10P1}). The striped density pattern (appearing as an exact product state) at time $t=50$ induces light-cone structures in the density evolution for $t>50$. Since the energy of the measurement-stabilized striped phase is high, the entropy for $t>50$ (not shown in Fig. \ref{fig:M10P1}) grows towards a volume-law value that is much higher than the initial one (that corresponded to a weakly entangled ground state of the Luttinger \mbox{liquid}, $S \propto \ln L$).

\subsubsection{Strong infrequent measurement: $M \gg 1$, $P \ll 1$}
\label{sec:strong_infrequent}

\begin{figure}
    \centering
    \includegraphics[width=\columnwidth]{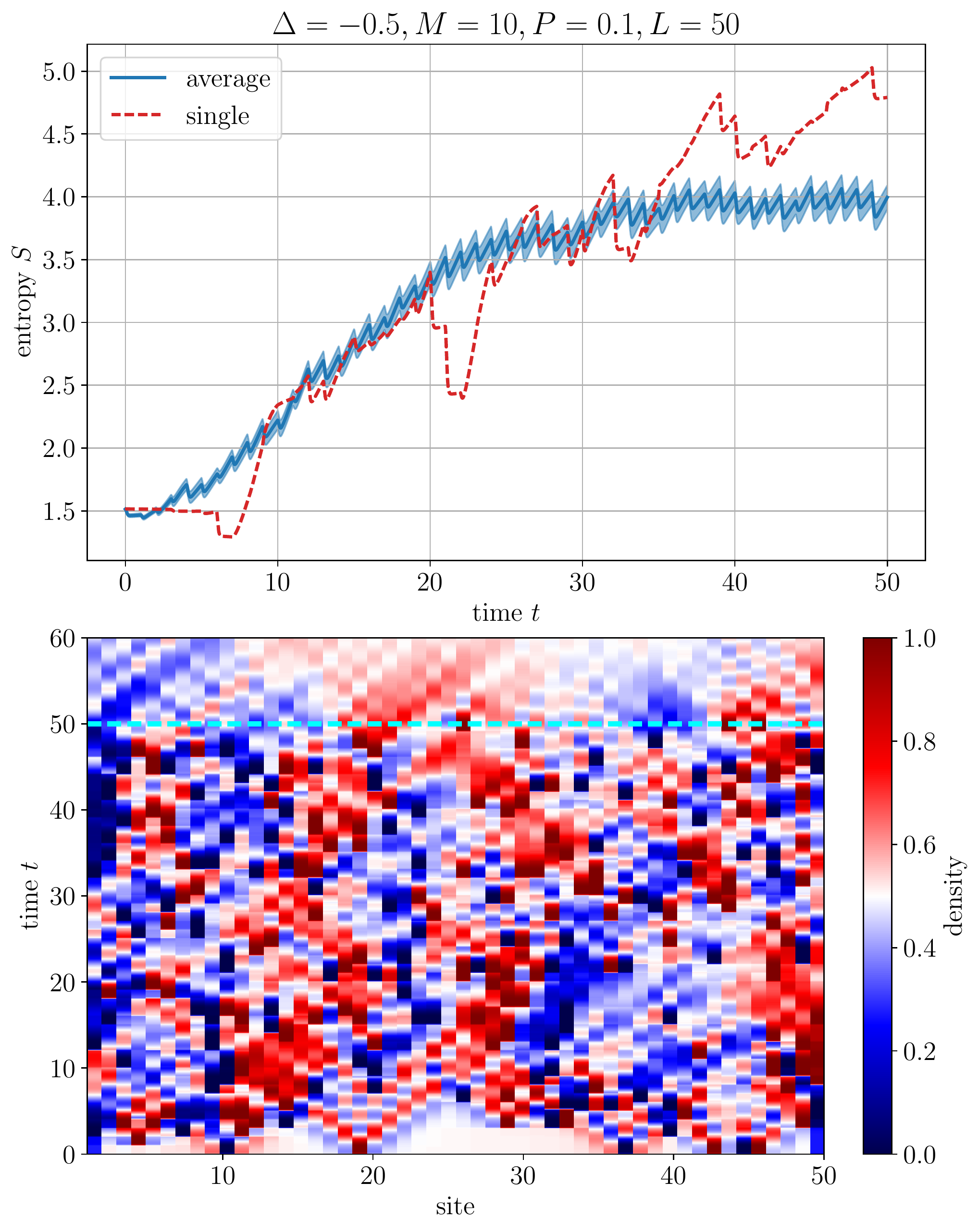}
    \caption{Same as in Fig.~\ref{fig:M10P1}, but for a strong infrequent ($M=10,P=0.1$) measurement.
    The initial density profile (local half-filling represented by the white color) develops into a random red-blue pattern on the time scale $t\sim 5$. Bond dimension cutoff effects become important for $t \gtrsim 20$ due to the fast growth of entanglement, see also Sec.~\ref{s2b} and Fig.~\ref{fig:entropyscaling}.}
    \label{fig:M10P01}
\end{figure}

Next, we consider the case where measurement is strong, as in the previous case, but the probability of measurement is much lower than unity. The result is shown in Fig.~\ref{fig:M10P01} for $M=10$ and $P=0.1$. In this case, the rare measurements do create locally polarized sites but they are not sufficient to suppress entanglement across the system. On the contrary, the entanglement, while noisy, rapidly grows, reaching values substantially larger than the initial one. The system is thus in the entangling phase. The entanglement growth stems from quenching the system by the measurements that distort the initial homogeneous ground state by introducing rare polarized regions. These regions then develop in time according to the unitary dynamics governed by the many-body Hamiltonian \eqref{eq:ham}.

As a result of this dynamics, the density becomes strongly inhomogeneous at later times; the inhomogeneity is further enhanced by subsequent rare strong measurements. Note that through the maintaining of global half-filling, local strong measurements affect neighboring sites: this can be clearly seen, e.g., at $t\sim 1$ around $x=10$ and $x=37$ in the lower panel of Fig.~\ref{fig:M10P01}, where the projection on the globally-half-filled state induces an excess (reddish) density at the neighboring sites. We thus see that, under the global constraint in a realistic system, a local strong measurement can induce correlations, similar to non-local measurements \cite{Roy2020, Ippoliti2021a}.

The zigzag fluctuation pattern in the average entropy on the scale of a single time step corresponds to the decrease in the entropy caused by the strong measurement in the vicinity of the bipartition cut. Eventually, the average entropy saturates, as occasional decreases induced by the measurement and the entangling dynamics of the unitary Hamiltonian balance out, see Fig.~\ref{fig:M10P01}.

\subsubsection{Weak frequent measurement: $M \ll 1$, $P=1$}
\label{sec:weak_frequent}

\begin{figure}
    \centering
    \includegraphics[width=\columnwidth]{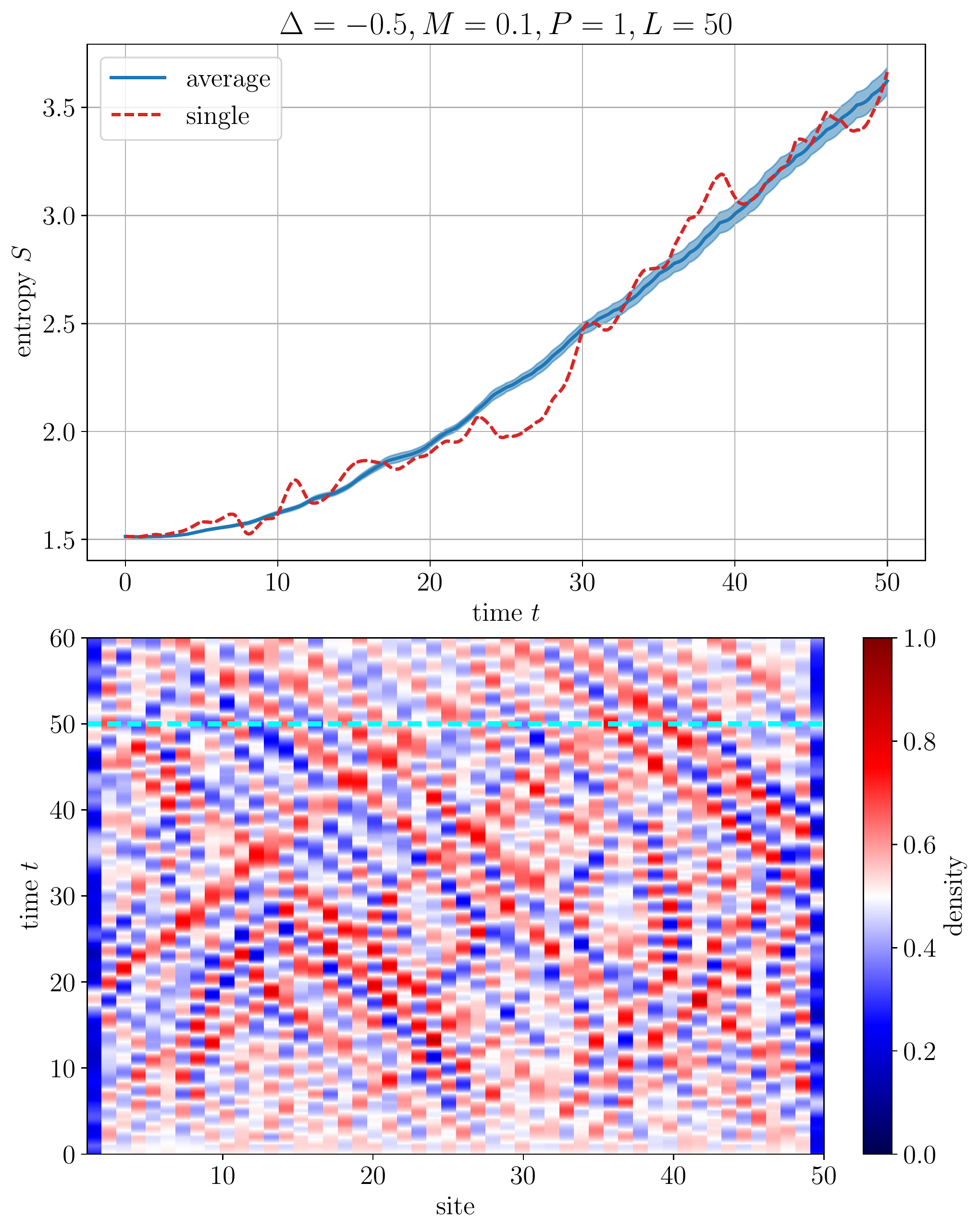}
    \caption{Same as in Fig.~\ref{fig:M10P1}, but for a weak frequent ($M=0.1,P=1$) measurement. Measurement-induced perturbations of the density propagate ballistically through the system and are reflected at the boundaries.}
    \label{fig:M01P1}
\end{figure}

We now consider the case where measurement is frequent but weak, see Fig.~\ref{fig:M01P1}, where $P = 1$ and $M = 0.1$. In this case, the characteristic timescale for projecting onto a particle or hole state under the dynamics of $\mathcal{H}_\mathrm{meas}$ is substantially larger than the timescale $T=1$ associated with the duration of the measurement. This leads to a situation where the initial state, which has homogeneous density (except close to the edges of the system, in view of open boundary conditions) is only weakly perturbed by the measurement. As a result, we observe particle and hole fluctuations induced by the measurement, which traverse the system ballistically.

As seen in the upper panel of Fig.~\ref{fig:M01P1}, the entanglement entropy rapidly grows with time and becomes considerably larger than its initial value, which is a signature of the entangling phase (see Sec.~\ref{Sec:Entscaling}). The physics of the entanglement growth can be understood as follows. The process of weak measurement continuously heats the system---the imaginary-time propagation does not conserve energy---and the system trends toward a highly entangled state at high energy, in qualitative similarity to the case of strong infrequent measurements depicted in Fig.~\ref{fig:M10P01}. In the latter case, however, the heating (quenching) process is strongly inhomogeneous in space, whereas it is fairly uniform for a frequent, weak measurement. Consequently, the entropy curve shows a much smoother behavior.

\subsubsection{Weak infrequent measurement: $M \ll 1$, $P \ll 1$} \label{sec:weakinfreq}
\label{sec:weak_infrequent}

\begin{figure}
    \centering
    \includegraphics[width=\columnwidth]{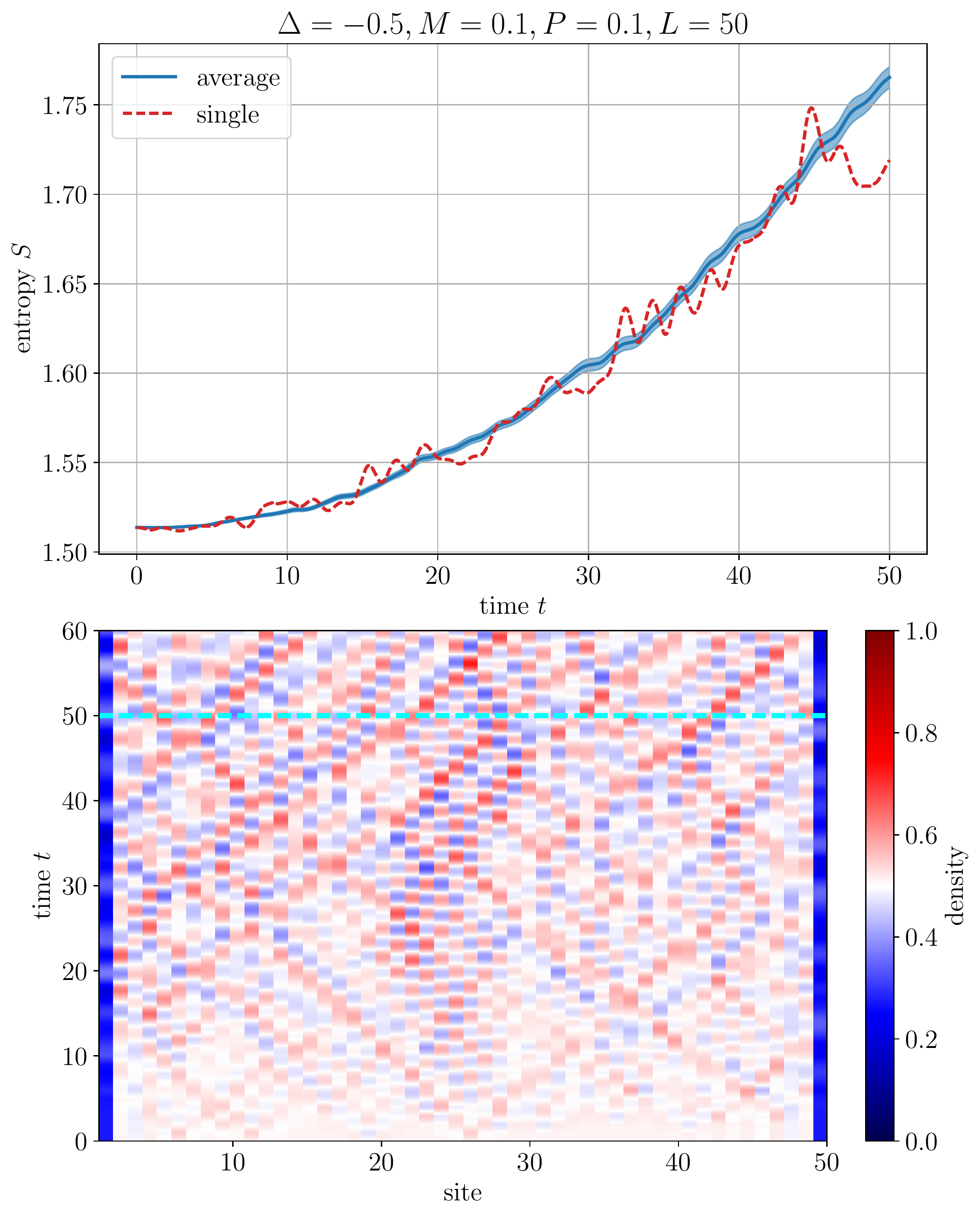}
    \caption{Same as in Fig.~\ref{fig:M10P1}, but for a weak infrequent ($M=0.1,P=0.1$) measurement.}
    \label{fig:M01P01}
\end{figure}

Finally, we consider the case where the measurement is both weak, $M=0.1$, and infrequent, $P =0.1$, see Fig.~\ref{fig:M01P01}. This case is rather similar to the preceding one ($M \ll 1$ and $P=1$, Sec.~\ref{sec:weak_frequent}), except that the less frequent measurement naturally leads to less heating, so that the growth of the entropy is slower than in the case of frequent measurement. Nevertheless, the entanglement exhibits a clear growing trend, and the system is expected to eventually reach qualitatively the same, highly entangled high-energy state. Thus, at arbitrarily small values of $M$ and $P$ the entanglement will eventually grow to large values as the initial state at zero temperature is gradually heated by the (effective) coupling to the environment. Clearly, the time that is required for the entanglement to reach the saturation value becomes progressively longer when $M$ and $P$ are reduced.

The density pattern (lower panel of Fig.~\ref{fig:M01P01}) is distorted already by a weak infrequent measurement, forming a structure of overlapping light-cone rays. Interestingly, the contrast appears to increase with time, which can be regarded as a result of the interplay between the unitary and measurement-induced dynamics: the measurements have a tendency to magnify density fluctuations.

\section{Phase diagram \label{Sec:Phase}}

\subsection{Entanglement entropy}

We are now in a position to carry out a characterization of the dynamics of our monitored system in the parameter plane spanned by the measurement strength $M$ and the measurement probability $P$. In section \ref{Sec:Simulations}, we considered four limiting regimes corresponding to ``corners'' of this phase diagram, for the choices of the Hamiltonian parameters considered in this work. In the limit of large $M$ and $P$, entanglement in the system is destroyed and an area-law phase appears. In the other three regimes, we found that entanglement \emph{grows} with respect to the initial zero-temperature state because of the addition of energy to the system by measurements. Hence, we expect to find a transition between the two phases (disentangling and entangling)   in the $M$--$P$ plane. For hybrid quantum circuits, the phase diagram of this type was analyzed in Ref. \cite{Szyniszewski2019a},
where the transition line connected two points, one located on the axis of weak constant measurement ($P=1$) and the other corresponding to the limit of rare projective measurements ($M=\infty$) .

\subsubsection{Continuous measurements of all sites: constant $P=1$}

\begin{figure}
    \centering
    \includegraphics[width=\columnwidth]{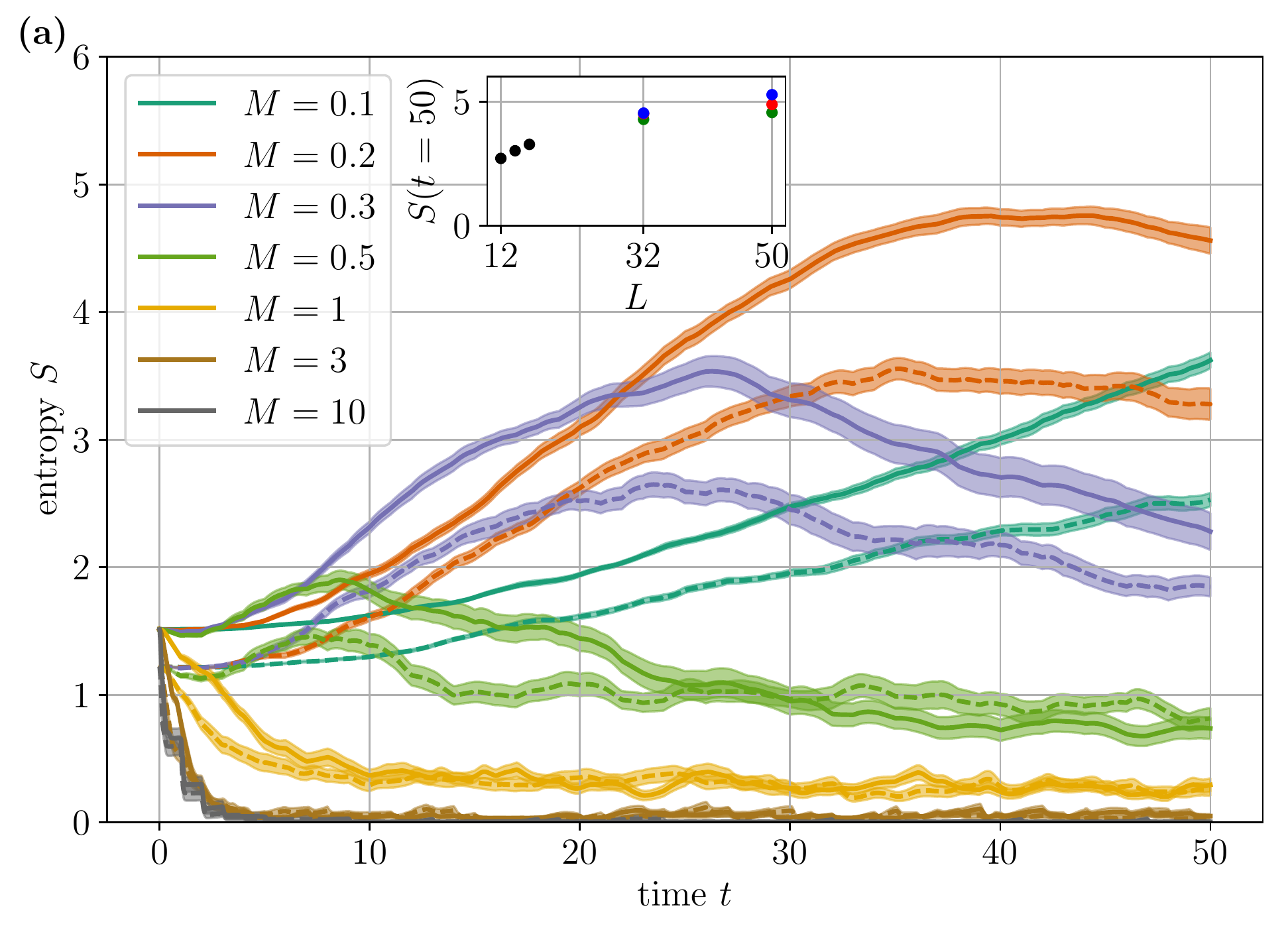}
    \includegraphics[width=\columnwidth]{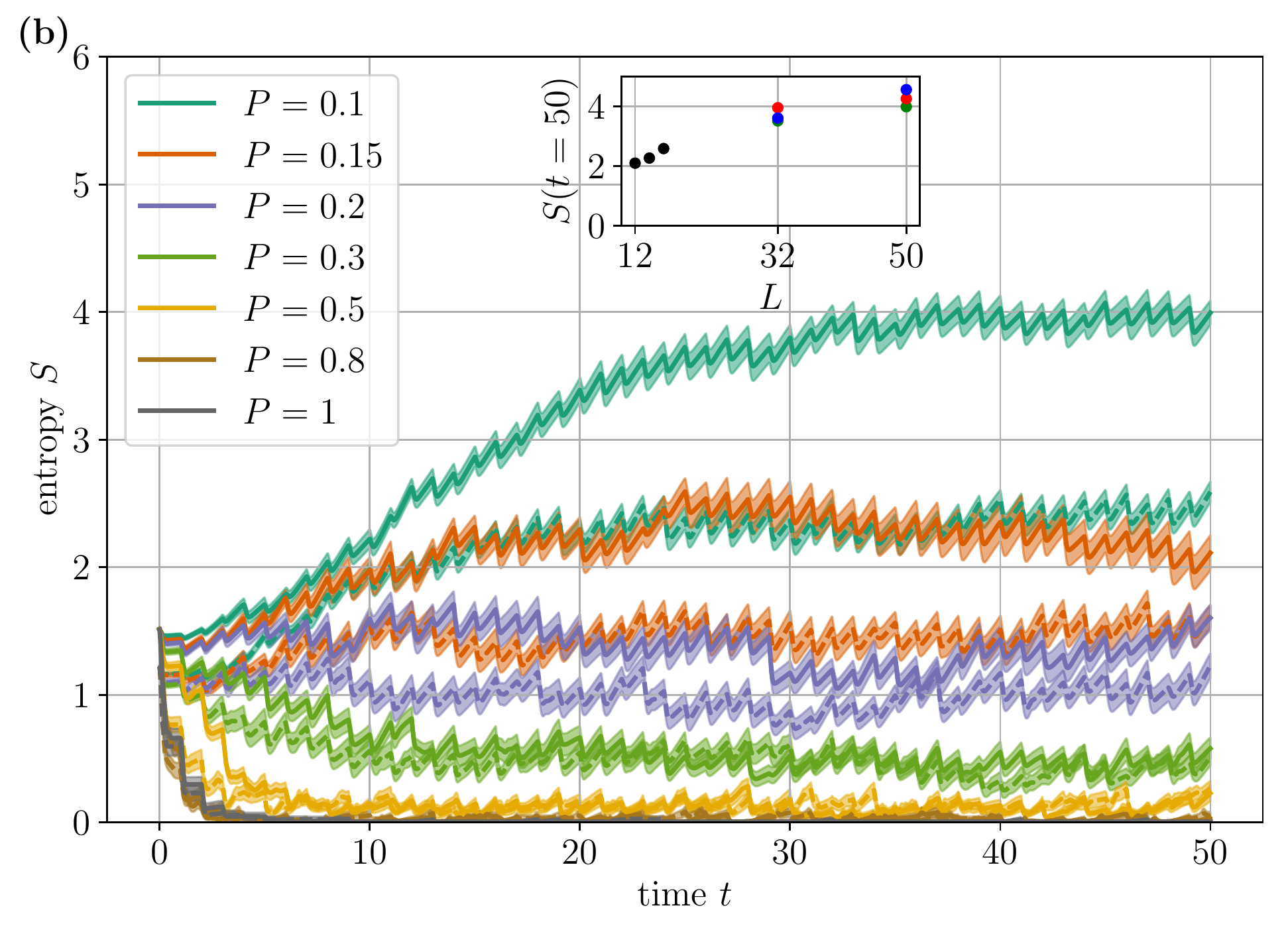}
    \caption{\textbf{a)} Average entropy $S(t)$ as a function of time for various measurement strengths $M$ and fixed $P = 1$. The solid (dashed) lines show $L = 50$ ($L = 16$). Inset: $S(t=50)$ as a function of system size for $M = 0.2$, where exact results are shown for $L = 12, 14, 16$ and the colors indicate bond dimension for $L = 32, 50$, with $\chi = 64$ (green), $\chi = 96$ (red), $\chi = 128$ (blue). \textbf{b)} As in panel \textbf{a)}, but for fixed $M = 10$ and varying $P$. The inset shows results for $P = 0.1$. Bond dimension cutoff effects are relevant for the $L = 50$ curves reaching $S \gtrsim 3$.}
    \label{fig:entropytime}
\end{figure}

In Fig.~\ref{fig:entropytime}a, we show the average entropy $S(t)$ for $P = 1$ and various choices of $M$ from very weak ($M=0.1$) to strong, nearly projective ($M=10$) measurements. To probe the dependence of the entropy on the system size $L$, we compare in this plot the data for $L = 50$ and $L=16$. For $L=16$ we choose $\chi = 256$, so that the dynamics is simulated exactly (the exact simulation with MPS requires $\chi = 2^{L/2}$).

For large $M$, we see that the entropy decreases over time as was discussed in Sec.~\ref{sec:strong_frequent}, which is consistent with the disentangling (area-law) phase. This is also confirmed by the fact that the long-time saturation value of $S$ is independent on $L$ (within the uncertainty that results from fluctuations of the average in our finite ensemble). Conversely, for small $M$, we see an increase of $S$ with respect to its initial value as was discussed in Sec.\ref{sec:weak_frequent}. Furthermore, the entropy $S$ increases with increasing system size $L$. These are hallmarks of the entangling phase. Our data indicate that the system is in the disentangling (area-law) phase for $M=10$, 3, 1, and 0.5, and in the entangling phase for $M=0.1$, 0.2, and 0.3.   We thus estimate the position of the transition point on the $P=1$ axis as $M_c \approx 0.4$.

Interestingly, for a certain range of values of $M$ (see the curves for $M = 0.2, 0.3, 0.5$ in Fig.~\ref{fig:entropytime}a), the entropy as a function of time exhibits a maximum before saturation. A similar effect was observed in Ref.~\cite{Goto2020a} where, however, a very different initial condition was chosen (a high-energy, strongly inhomogeneous state, as opposed to the low-energy, homogeneous one considered here).

\subsubsection{Strong measurements with varying measurement frequency: constant $M=10$}

In Fig.~\ref{fig:entropytime}b,  the time dependence of the average entropy $S(t)$ is shown for strong measurements. Specifically, in this figure we fixed $M = 10$, while the measurement frequency $P$ was varied from $P=0.1$ to $P=1$. Qualitatively, the evolution is quite similar to that shown in Fig.~\ref{fig:entropytime}a. For small $P$, we find entangling behavior (Sec.~\ref{sec:strong_infrequent}), while for large $P$ it is disentangling (Sec.~\ref{sec:strong_frequent}). As in Fig.~\ref{fig:entropytime}a, we have two complementary criteria for identification of the entangling phase: (i) the long-time value of the entropy is higher than its initial value; (ii) the entropy increases with system size ($L=50$ vs $L=16$). Both criteria yield consistent results, allowing us to identify the points $P=1$, 0.8, 0.5, and 0.3 as belonging to the disentangling phase, and the points $P=0.1$, 0.15, and 0.2 as belonging to the entangling phase, with the critical value being close to the latter point, $P_c \approx 0.2$--0.25.

A comment on the criterion (i) is in order here. The entanglement entropy at $t=0$ is that of the ground state of a Luttinger liquid, $S(t=0; L) \approx (1/6)\ln L$ \cite{Laflorencie2016a}. In the disentangling phase, $S$ tends to an $L$-independent value at large $t$, i.e. $S(L) \equiv S(t\to \infty; L) < S(t=0; L)$, at least for large $L$. At the same time, one cannot a priori exclude a possibility that there is a part of the disentangling phase where $S(L)$ grows with $L$ more slowly than $(1/6)\ln L$ (e.g., logarithmically with a smaller coefficient). Our data do not support such a possibility (although of course we cannot rule it out completely on the basis of finite-$L$ and finite-$t$ simulations). Indeed, the fact that the criteria (i) and (ii) agree implies that, whenever $S(L)$ increases with $L$, it is larger than $S(t=0;L)$, i.e. the increase of $S(L)$ with $L$ in the disentangling phase is faster than for the initial entanglement.  In fact,  as we argue in Sec.~\ref{Sec:Entscaling}, the observed increase of $S(L)$ in the disentangling phase is of volume-law character, at least within the range of $L$ and $t$ accessible to our simulations.

\subsubsection{Overall phase diagram}

\begin{figure}
    \centering
    \includegraphics[width=\columnwidth]{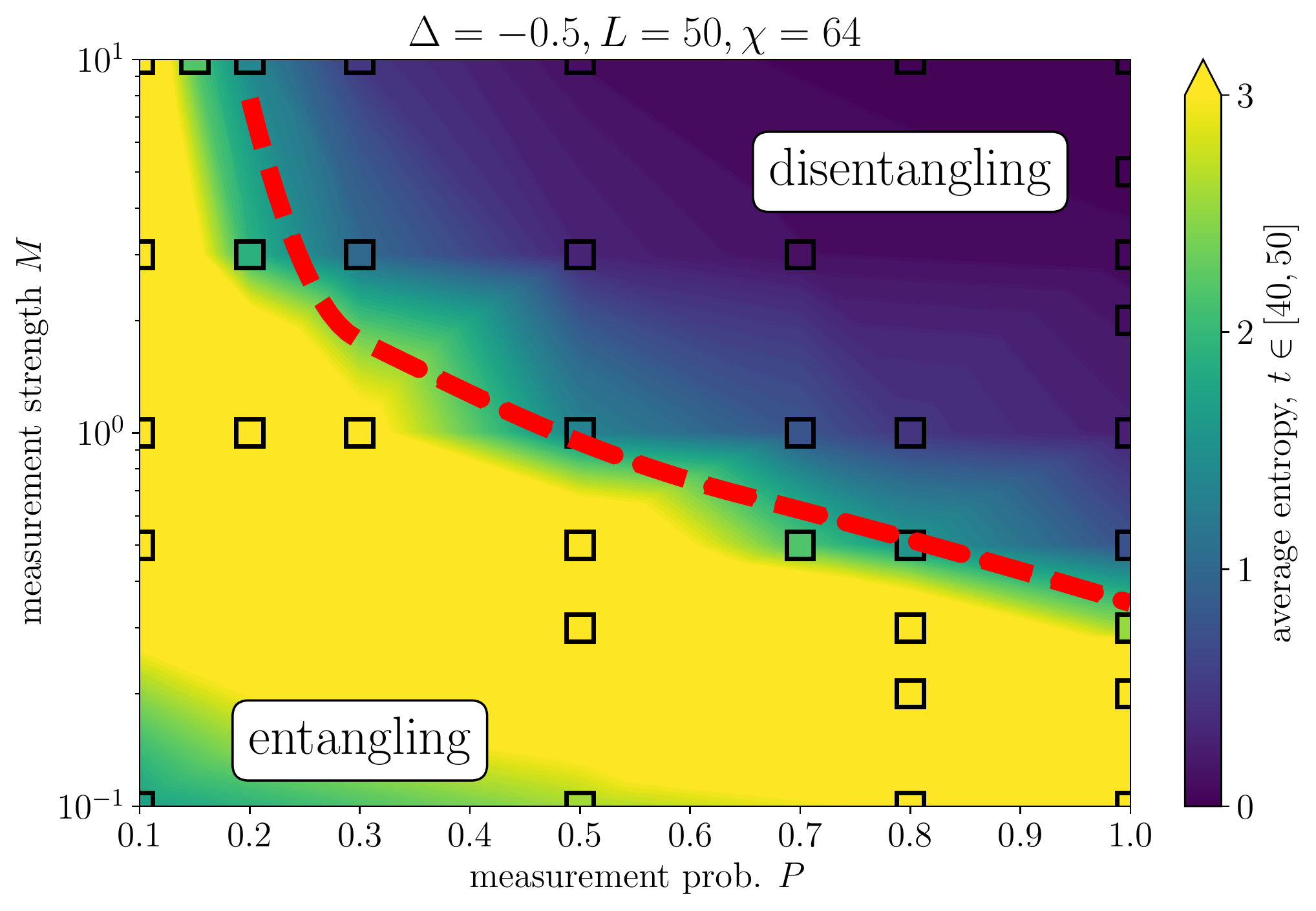}
    \caption{Phase diagram, showing the averaged entropy $S$ for $t \in [40, 50]$ for $R=40$ realizations. The black squares indicate data points and the background color depicts interpolated values between the data points. The estimated phase boundary, corresponding to an approximate contour of constant entropy equal to the initial value, is shown as dashed red line.}
    \label{fig:phasediag}
\end{figure}

In Fig.~\ref{fig:phasediag} we summarize the results for the entropy $S$, averaged over the time interval $[40,50]$ and over $R=40$ realizations, in the  whole $P$--$M$ parameter plane. In analogy with the above estimates of critical points on the $P=1$ and $M=10$ lines, we have estimated the transition line in the $P$--$M$ plane, which is also shown in the figure.

It is worth commenting on the bottom left corner of the phase diagram (rare weak measurements), where the entropy is above its initial value but smaller than in the most of the entangling phase. The reason for this was discussed in Sec.~\ref{sec:weakinfreq}: the entropy grows, but it takes a time much longer than the duration of the protocol to saturate (corresponding to a strongly entangled state).

In the bright yellow region in Fig.~\ref{fig:phasediag} we have indicated the region where the entropy reaches values $S \geq 3$. As discussed in Sec.~\ref{s2b}, the results for these data points should be interpreted in a qualitative sense; strong growth of the entanglement is observed, and the MPS ansatz does not capture the exact dynamics over the latter part of the time window simulated. We hence extrapolate this growth to longer times. The precise value of the entanglement deep in the entangling regime is not relevant for the position of the phase boundary.

\subsubsection{Entropy scaling in the entangling phase} 
\label{Sec:Entscaling}

An important question is the dependence of the large-$t$ saturation value of the entanglement entropy $S(L)$ on the system size $L$ in the entangling phase. Most of works on entanglement transitions in quantum circuits indicate a volume-law scaling, $S(L) \propto L$, with a prefactor $L$ depending on the measurement strength.

In the insets of Fig.~\ref{fig:entropytime} we show  the behavior of the entropy at $t = 50$ as a function of system size, for two particular choices of parameters in the entangling phase ($M = 0.2$, $P = 1$ and $M = 10$, $P = 0.1$), which correspond to the largest values of $S$ in  the upper and lower panels of Fig.~\ref{fig:entropytime}, respectively. For both these points in the parameter plane, saturation of the entropy as a function of time is reasonably reached by our largest time $t=50$. In order to better quantify the rate of the entanglement growth, we have calculated the values of $S(t=50)$ for these choices of $P$ and $M$ also for larger values of the bond dimension, $\chi=96$ and 128. For $L=32$ the obtained values of the entropy for $\chi = 64$ and $\chi = 128$ are close in value (within statistical error bars that are related to a finite size of realizations and are of the order of the symbol size), which is a signature of saturation of $S(t)$ with $\chi$. At the same time, for $L=50$ (where the entropy is larger), the obtained value of $S$ for $\chi=128$ is substantially above than that for $\chi=64$. This drift indicates that, for these points in the $P$--$M$ plane, the actual values of $S(t=50)$ are still somewhat above the $\chi=128$ results (shown by blue color in the insets); presumably by an amount of the order of a distance between the $\chi=128$ and $\chi=64$ points. (To find more accurately the saturated value, one would need a calculation with $\chi \approx 256$, which is in principle possible but requires very substantial computational time.)  Keeping this in mind, we see that the values of the entanglement entropy at $L=50$ are broadly consistent with volume-law trends based on the data for $L=16$ and $L=32$. We cannot exclude, however, a different type of scaling might emerge for larger system sizes or time scales.

It is worth noticing that the slopes of the $S(L)$ dependencies that can be estimated in this way are somewhat smaller than those that would be found based only on the data for small systems (accessible to exact diagonalization). This indicates that finite-size effects are sizeable, so that supplementing exact-diagonalization numerical studies by approximate approaches (such as the MPS method used in this work) that can be applied to larger systems is crucial.

\begin{figure}
    \centering
    \includegraphics[width=\columnwidth]{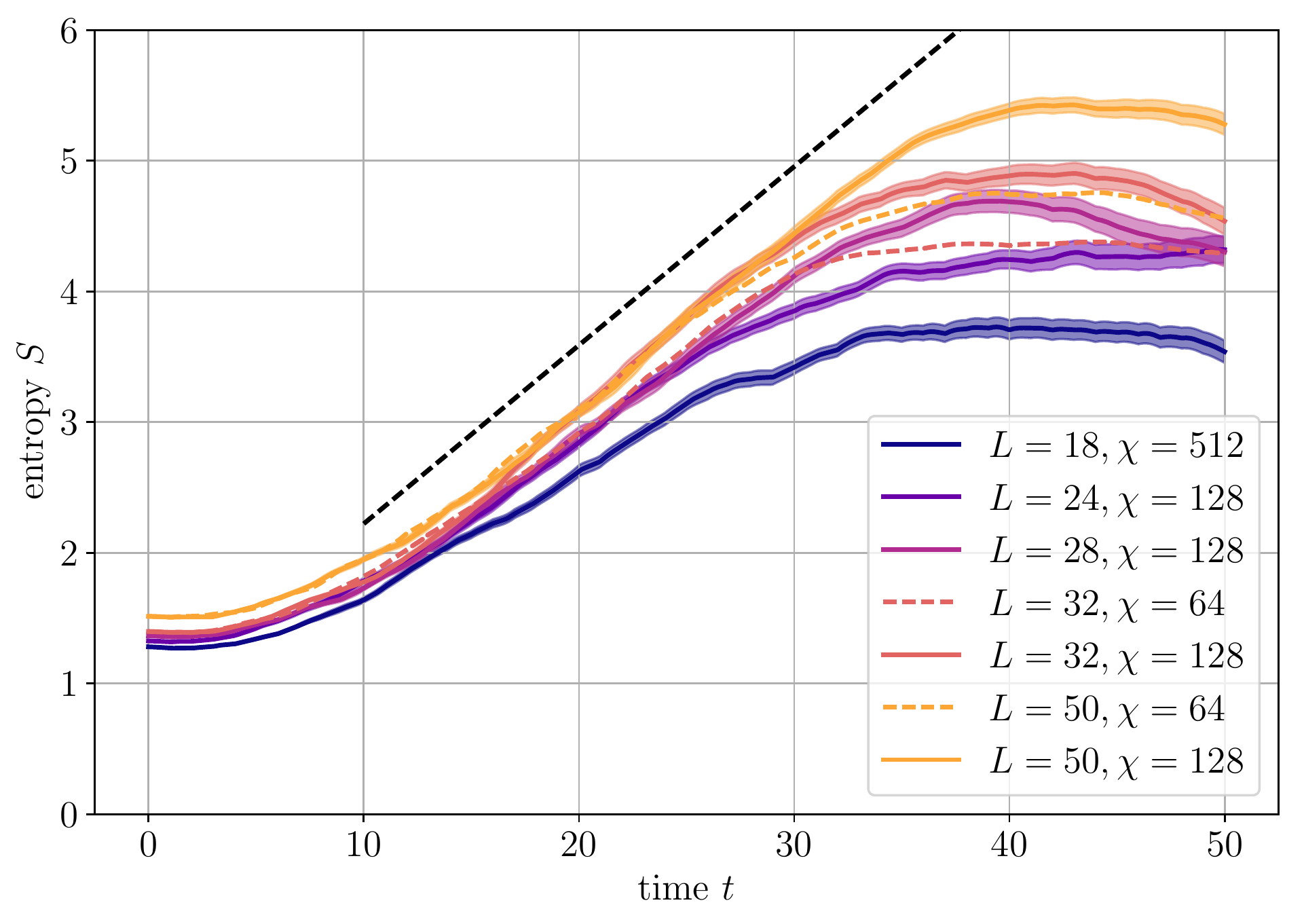}
    \caption{Time dependence of the entanglement entropy $S(t)$ at at $P=1$, $M=0.2$ for system sizes from $L=18$ to $L=50$.  For the two largest system sizes, the data with bond dimensions $\chi=64$ and 128 are shown, where $\chi = 64$ is indicated by a dashed line. The straight black dashed line is a guide to the eye obtained through a linear fit of the $L = 50, \chi = 128$ data in the linear regime $t \in [15, 25]$, yielding a slope $\approx 0.14$. Truncation effects due to the finite bond dimension start to become apparent at $t \approx 25$ for $L = \{32, 50\}$.}
    \label{fig:entropyscaling}
\end{figure}

The volume-law behavior of the entanglement entropy $S$ in the entangling phase is also supported by the time dependence $S(t)$. In Fig.~\ref{fig:entropyscaling} we show the dependence $S(t)$ at $P=1$, $M=0.2$ for several system sizes (and for two values of the bond dimension $\chi$ for the largest sizes $L=32$ and 50). We see that, for lengths $L \ge 24$, a linear increase $S(t) \propto t$ is found, with an $L$-independent slope. This is expected in the volume-law phase \cite{Skinner2019a}: the entanglement increases as $S(t) \simeq s v_S t$ until it saturates at a time $t \simeq L / 2 v_S$ at a value $S(L) \simeq sL/2$, where $v_S$ is the entanglement propagation velocity.  When the bond dimension is insufficient, it leads to a cutoff of the linear increase of $S(t)$ and to the saturation before system-size effects kick in. The slope in Fig.~\ref{fig:entropyscaling} yields $sv_s \simeq 0.14$ for this point in the phase diagram. At the same time, the $L$-dependence of $S(L)$ in the inset of  Fig.~\ref{fig:entropytime}a yields an estimate $s = 2(dS/dL) \simeq 0.15$\:--\:0.2 for the same values of the parameters $P$ and $M$. From these data, we estimate the entanglement propagation velocity, $v_S \equiv (sv_S)/s \simeq 0.7$\:--\:1.0. This velocity is close (or possibly identical) to the velocity of ballistic propagation of density fluctuations as observed, e.g., in Fig.~\ref{fig:M01P1}, which is consistent with predictions from Refs.~\cite{Calabrese2004, DeChiara2006a}.

It was proposed in Refs.~\cite{Alberton2020a,Buchhold2021} that the volume-law scaling of $S(L)$  is only of transient character and crosses over to a $\ln L$ behavior for large $L$ at the transition. The system sizes that we can access are not sufficient to rigorously test the validity of this conjecture.

\subsection{Clusterization}
\label{sec:clusterization}

\subsubsection{Clusterization at $\Delta = -0.5$}
\begin{figure}
    \centering
    \includegraphics[width=\columnwidth]{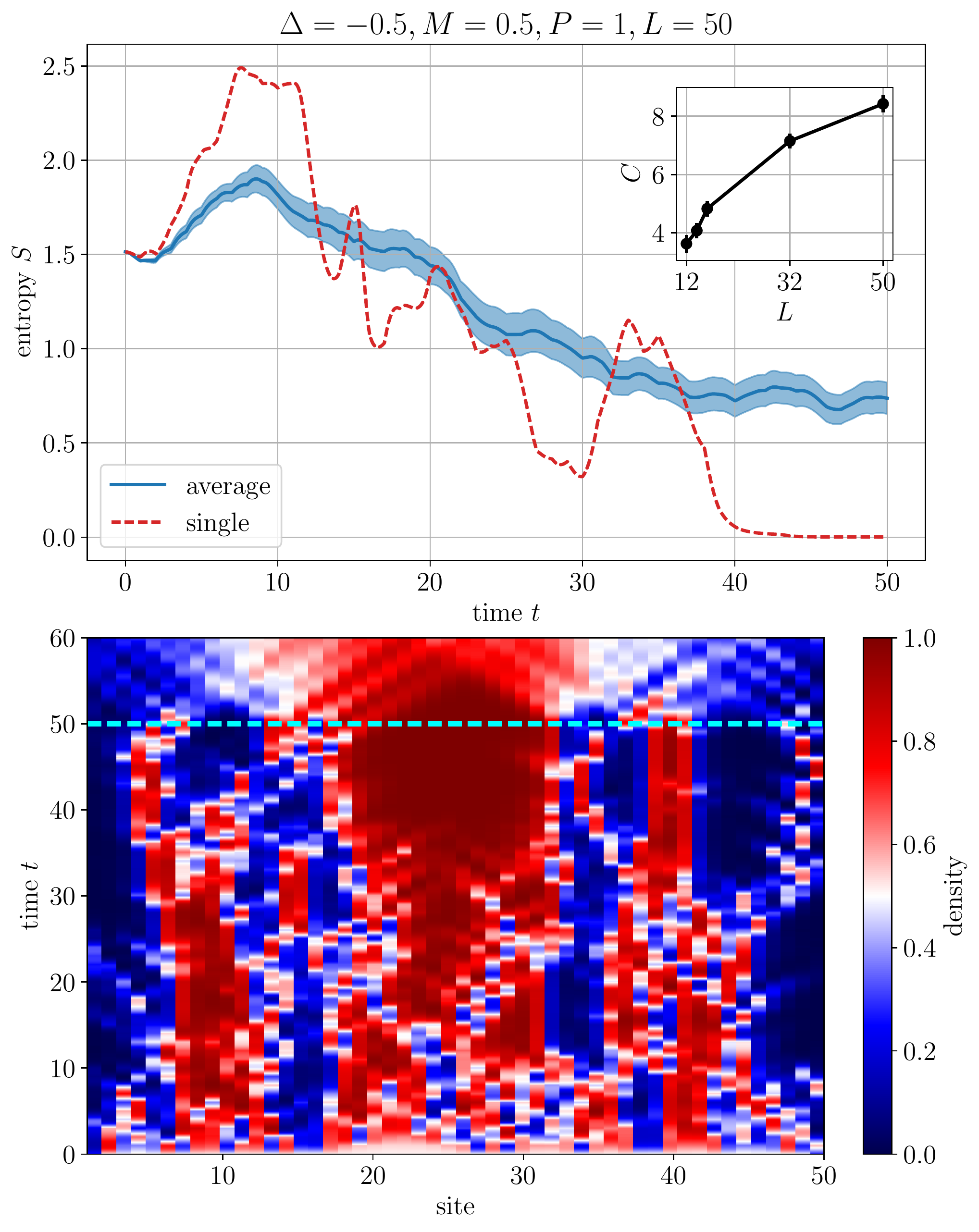}
    \caption{As in Fig.~\ref{fig:M10P1}, but for $M=0.5$, $P=1$. In the bottom panel, the realization with the largest maximum cluster size is chosen. Inset: average maximum cluster length $C$ at $t = 50$ as a function of system size. Error bars indicate a $1\sigma$-interval.}
    \label{fig:M05P1}
\end{figure}

We discuss the clustering of particles (and holes), see Sec.~\ref{sec:cluster}. Interestingly, we observe, close to the transition from the disentangling to the entangling phase, the emergence of large domains resulting from the interplay of $\mathcal{H}_0$ and the measurement protocol. As an illustration, we show in Fig.~\ref{fig:M05P1} the dynamics for a single realization with $M = 0.5$ and $P = 1$. As discussed above, this point is on the area-law side of the transition, close to the phase boundary. For the single realization plotted in the lower panel---the one with the largest maximum cluster length---we observe formation of large polarized domains at time $t \approx 35$. As the dotted line in the upper panel shows, the entanglement entropy for this realization practically vanishes at $t \ge 40$. Hence, the large polarized regions effectively block transport, analogous to what is observed in models with local constraints \cite{Doggen2021a}.

\begin{figure}
    \centering
    \includegraphics[width=\columnwidth]{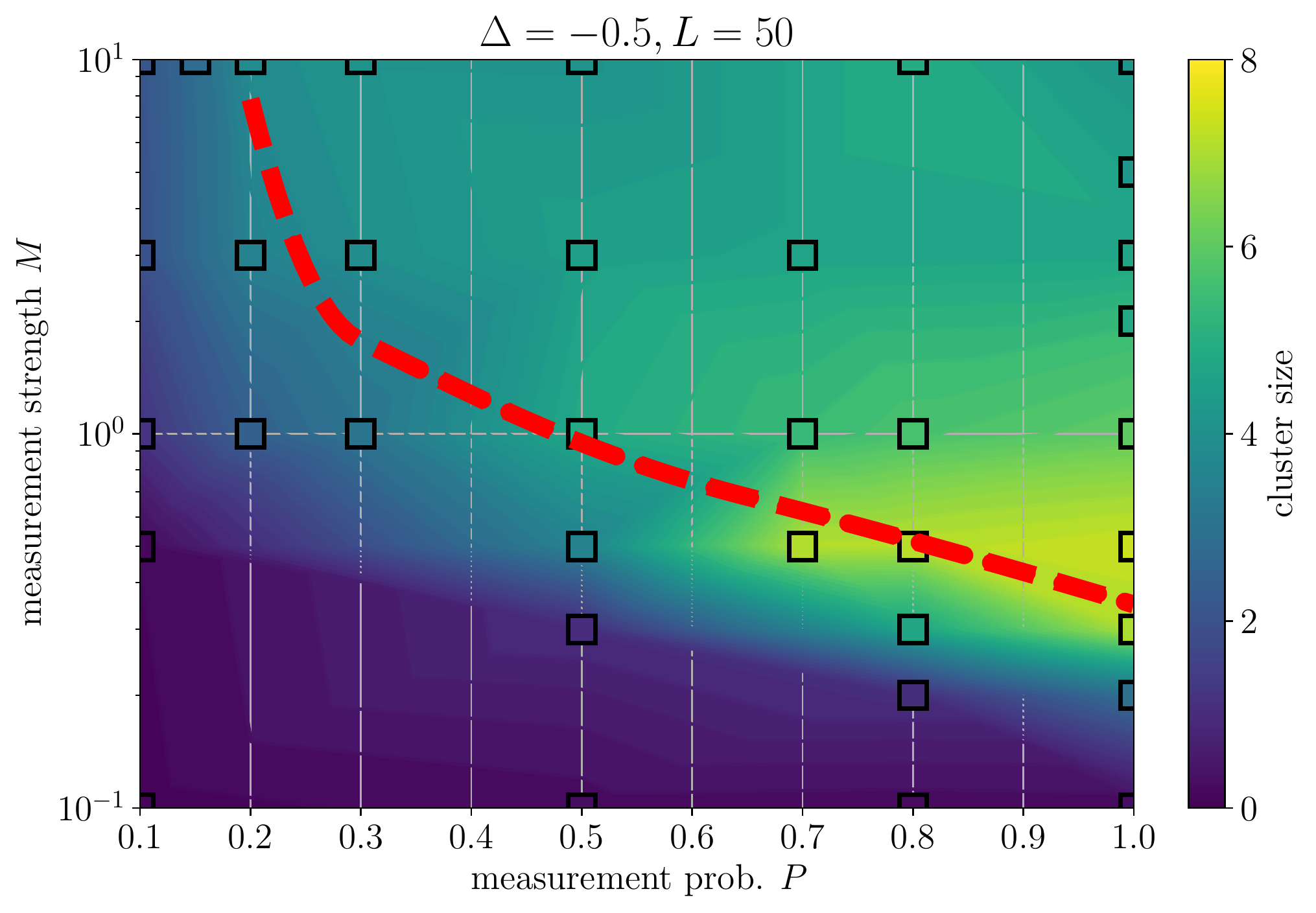}
    \caption{Average maximum cluster length as a function of measurement probability $P$ and measurement strength $M$, evaluated at $t = 50$. Note that the region with high cluster size is located around a section of the phase boundary between entangling and disentangling phases (Fig.~\ref{fig:phasediag}), which is also shown in this plot.}
    \label{fig:phasediag_cluster}
\end{figure}

To characterize the clustering quantitatively, we consider for every realization the maximum cluster length and then average it over all realizations. The resulting averaged maximum cluster length is denoted $C(t)$. In the inset of Fig.~\ref{fig:M05P1}, we show the system-size dependence of $C(t=50)$ for the same parameters $M = 0.5$,  $P = 1$. A clear increase of $C$ with system size $L$ is observed. The data suggests sublinear growth $C \sim L^y$, with $y \approx 0.5$.
The results for $C(t=50)$ for various values of $P$ and $M$ are shown in Fig.~\ref{fig:phasediag_cluster}. They exhibit a peak---which reveals the clusterization phenomenon---on the right side of the diagram, around the phase boundary between the entangling and disentangling phases. It should be emphasized that this clusterization is observed only near the portion of the phase boundary that corresponds to frequent measurements ($P$ close to unity).  No such peak is observed in the opposite corner of the phase diagram. This indicates that at least some important aspects of the entanglement transition are not fully universal and, in particular, differ qualitatively between the regimes of weak and strong measurements. 

Clusterization can be viewed as a result of an enhancement of the attractive interaction by coupling of the system to the environment via the measurements, cf.~Ref.~\cite{Buchhold2021}). As shown in the following section, the clusterization effect exists (in a weaker form) also in a non-interacting system ($\Delta=0$), where the measurements create an effective attractive interaction. One can also note a certain similarity between the clustering and the dynamical quantum Zeno transition \cite{Kumar2020,Biella2021a}. Indeed, the rapid emergence of clusterization in the lower panel of Fig.~\ref{fig:M05P1} at $t\approx 35$ suggests a kind of dynamical phase transition induced by constantly monitoring the system at moderately weak measurement strength.

Qualitatively, one can understand the clusterization phenomenon as follows. Upon measuring a certain spatial region, an extended fluctuation may emerge, where the density is, e.g., increased compared to the background (the overall half-filling condition is then maintained at the expense of distant regions). The unitary dynamics leads to a spreading of this density seed to neighbouring sites, which are then more likely to have a higher density in the subsequent time evolution. For not too weak measurements, the ``contrast'' of the density fluctuation is enhanced by the measurement back-action. Over time, this leads to clusters of particles (or holes) with similar density. This can be dubbed a ``dynamical quantum Zeno effect.''

It is no surprise that this phenomenon requires the measurement and unitary parts of the Hamiltonian to be of roughly equal importance, as is the case near the crossover from the disentangling to entangling regime. After all, for $M \gg 1$ the unitary dynamics plays almost no role and sites are effectively locked at a fixed filling (a static quantum Zeno effect), with only occasional ``jumps.'' This results in freezing the striped structure of the density with a random distribution of the stripe sizes determined by the outcome of the first measurements, cf. Fig.~\ref{fig:M10P1}, without any preference to increased clusters.
The measurement also needs to be sufficiently frequent, otherwise the density correlations will be lost during the unitary time evolution. In the limit $M \ll 1$ the measurement can be considered only a weak perturbation that acts as a local heating process. Similarly, if $P \ll 1$ it is unlikely that neighboring sites are measured. Since the spreading of the density fluctuations to neighbouring sites is driven by the hopping term $J$, we expect only a modest effect of interactions on clusterization, which is confirmed in the following section.

\subsubsection{Role of interactions}
\label{s4b2}

\begin{figure}
    \centering
    \includegraphics[width=\columnwidth]{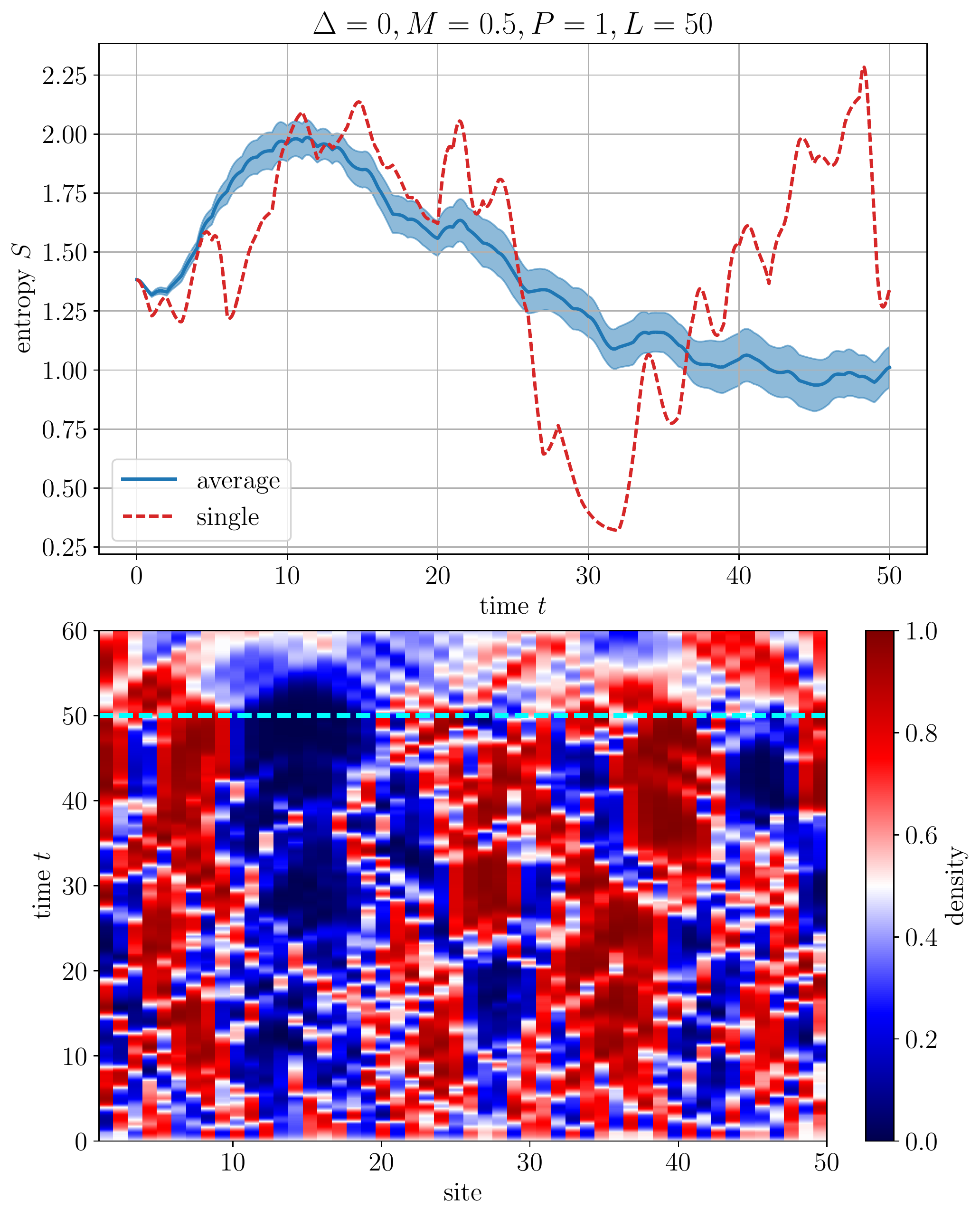}
    \caption{As in Fig.~\ref{fig:M10P1}, but for $M=0.5$, $P=1$ and $\Delta = 0$, such that the system is non-interacting.}
    \label{fig:non-interacting}
\end{figure}

We proceed by discussing the role of interaction in the clusterization phenomenon. In the vicinity of the entanglement-transition phase boundary (in the regime of frequent, intermediate-strength measurements), the measurement effectively enhances the ferromagnetic interaction, which shows up in the clusterization, as discussed above. A natural question to ask is what the role of interaction is in this phenomenon. If the measurement can effectively enhance the interaction, then perhaps it can also create an effective interaction in an otherwise non-interacting system? To test this hypothesis, we have performed the analysis for a non-interacting system ($\Delta = 0$) that has otherwise the same parameters as in Fig.~\ref{fig:M05P1}, i.e., $M = 0.5$, $P = 1$ (which are close to the point where the maximum degree of clusterization for the interacting system is observed). The results shown in Fig.~\ref{fig:non-interacting} illustrate an enhanced clusterization that is indeed observed also for a non-interacting system. At the same time, the average maximum cluster size over $R = 40$ realizations, $C = 7.15 \pm 0.28$, turns out to be lower than for the interacting system (see inset of Fig.~\ref{fig:M05P1}), as expected.

\begin{figure}
    \centering
    \includegraphics[width=\columnwidth]{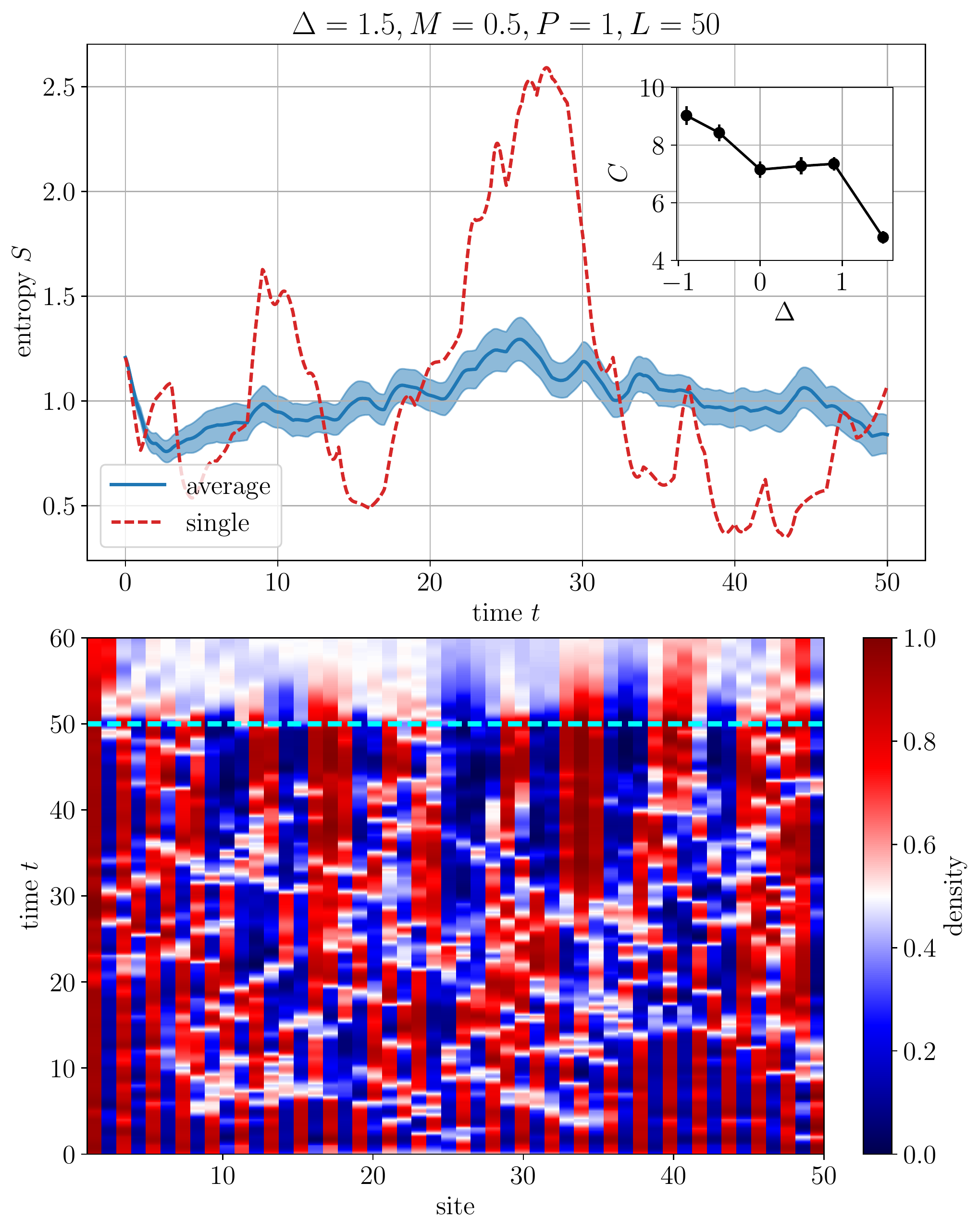}
    \caption{As in Fig.~\ref{fig:M10P1}, but for $M=0.5$, $P=1$ and $\Delta = 1.5$, such that the dynamics starts from an antiferromagnetic ground state. Inset: average maximum cluster length $C$ as a function of interaction strength $\Delta$ for $L = 50$, evaluated at $t = 50$. Error bars are $1\sigma$-intervals.}
    \label{fig:antiferro}
\end{figure}

For values of the interaction $|\Delta| > 1$, the ground state of the system is ferromagnetic (for $\Delta<-1$) or antiferromagnetic (for $\Delta > 1$). A typical simulation for the antiferromagnetic case is shown in Fig.~\ref{fig:antiferro} for $\Delta = 1.5$. One sees that the system still develops clusters, counteracting the antiferromagnetic correlations. The inset in Fig.~\ref{fig:antiferro} shows $C$ as a function of interaction strength for $L=50$, confirming that clusterization is enhanced by attractive interactions. Within the regime $|\Delta| < 1$, we find a modest but statistically significant growth of the cluster size for attractive interactions, while $C$ is constant within error bars in the repulsive regime $0 < \Delta < 1$. In the ferrogmagnetic regime (not shown in the figure), the initial state already has a cluster of $L/2$ particles, which is maintained during time evolution, yielding $C = 24.85 \pm 0.06$ using $\Delta = -1.5$.

\section{Summary and discussion \label{Sec:Discuss}}

In conclusion, we have proposed an MPS-based method for simulation of the dynamics of quantum many-body systems under continuous monitoring. The monitoring process is modelled as a site- and time-dependent non-Hermitian term in the Hamiltonian. The measurement protocol is controlled by two key parameters: the probability $P$ that a given site is measured at a given time interval (with $0 < P \le 1$) and the measurement strength $M$ (with $M \ll 1$ corresponding to weak measurement and $M\gg 1$ to strong, nearly projective measurement). In contrast to recent approaches, our protocol starts from the ground state of the original Hamiltonian, so that the observed evolution of the initial state  is entirely due to the effect of the measurements (including, of course, the interplay with the unitary dynamics).

We have applied the method to a 1D interacting many-body system, the ground state of which is a Luttinger liquid with a moderate entanglement ($S \propto \ln L$). The local measurement induces two competing processes: sufficiently strong measurements tend to disentangle the system through quasi-projections. At the same time, the measurement also leads to effective local heating of the initial zero-temperature ground state, which can lead to stronger entanglement. If the measurement leads to an area-law (disentangling) phase, it reduces the entanglement entropy with respect to the initial state (in the limit of large $L$). On the other hand, if the measurement drives the system to the volume-law (entangling) phase, it enhances the entanglement entropy compared to the initial state. Note that the competing effects of measurement reported in Ref.~\cite{Ippoliti2021a} involve nonlocal measurements, as opposed to our protocol.

Exploring systems with a length up to $L=50$, we have determined the phase diagram of the entanglement transition in the $P$--$M$ plane  (Fig.~\ref{fig:phasediag}). For sufficiently strong and at the same time sufficiently frequent measurement   (as in Fig.~\ref{fig:M10P1}), we find a disentangling phase: the entanglement entropy gets suppressed by measurement  down to an $L$-independent value (area law). On the other hand, if the measurement is sufficiently weak and/or sufficiently rare (as in Figs.~\ref{fig:M10P01}, \ref{fig:M01P1}, and \ref{fig:M01P01}), the system is in the entangling phase. Our results for the entanglement entropy in this phase are consistent, for the system sizes studied, with a volume law ($S \propto L$), though we cannot strictly exclude sublinear or logarithmic scaling in the thermodynamic limit.
The obtained phase diagram of the entanglement transition for the Hamiltonian system is qualitatively similar to earlier results for quantum circuits. Our findings thus indicate that such entangling-to-disentangling transitions occur generically in quantum many-body systems, though the position of the transition will depend on the microscopic parameters of the Hamiltonian.

Furthermore, we find that, close to the phase boundary in the range of frequent measurements ($P \approx 1$), the entanglement transition is accompanied by an increase of the size of clusters of particles and holes (Figs.~\ref{fig:M05P1} and \ref{fig:phasediag_cluster}).  We interpret this phenomenon as an enhancement of the attractive interaction by the measurements.   A similar phenomenon, although in a somewhat weaker form, is found also for a non-interacting system. The divergence of cluster size close to the entanglement transition may be a useful experimental probe, since particle densities are generally easier to measure than the entanglement entropy. Indeed, a setup similar to the one described in this work may be readily prepared in experiments on ultracold atoms or trapped ions, as the local particle density can be measured using quantum microscopy \cite{Bakr2009a}. It is an intriguing possibility that the divergence of cluster size we observe could be a signature of a transition to a dynamical quantum Zeno phase (cf. Ref.~\cite{Biella2021a}). At the same time, the precise connection between the clusterization effect and the measurement transition remains to be clarified.

Future work may focus on applying the method outlined here to various models of experimental relevance, such as the Hubbard model. An experimental implementation of the open quantum Ising chain on IBM's quantum hardware was realized very recently \cite{Kamakari2021a}.

\section*{Acknowledgments}

We thank M.~Buchhold, S.~Diehl, R.~Egger, E.~Medina Guerra, S.~Morales, P.~Kumar, A.~Romito, S.~Roy, K.~Snizhko, M.~Szyniszewski, O.~Yevtushenko, and A.~Zazunov for useful discussions. Numerical simulations were performed using the TeNPy library (version 0.6.1) \cite{tenpy}. We acknowledge financial support from the Deutsche Forschungsgemeinschaft (DFG): Project No. 277101999 -- TRR 183 (Project C01) and Grants No. EG 96/13-1 and No. GO 1405/6-1, as well as from the Israel Science Foundation.  Y.G. acknowledges support by the Helmholtz International Fellow Award. 

\vspace*{1cm}

\appendix

\section{Numerical details and convergence}

\begin{figure}[!htb]
    \centering
    \includegraphics[width=\columnwidth]{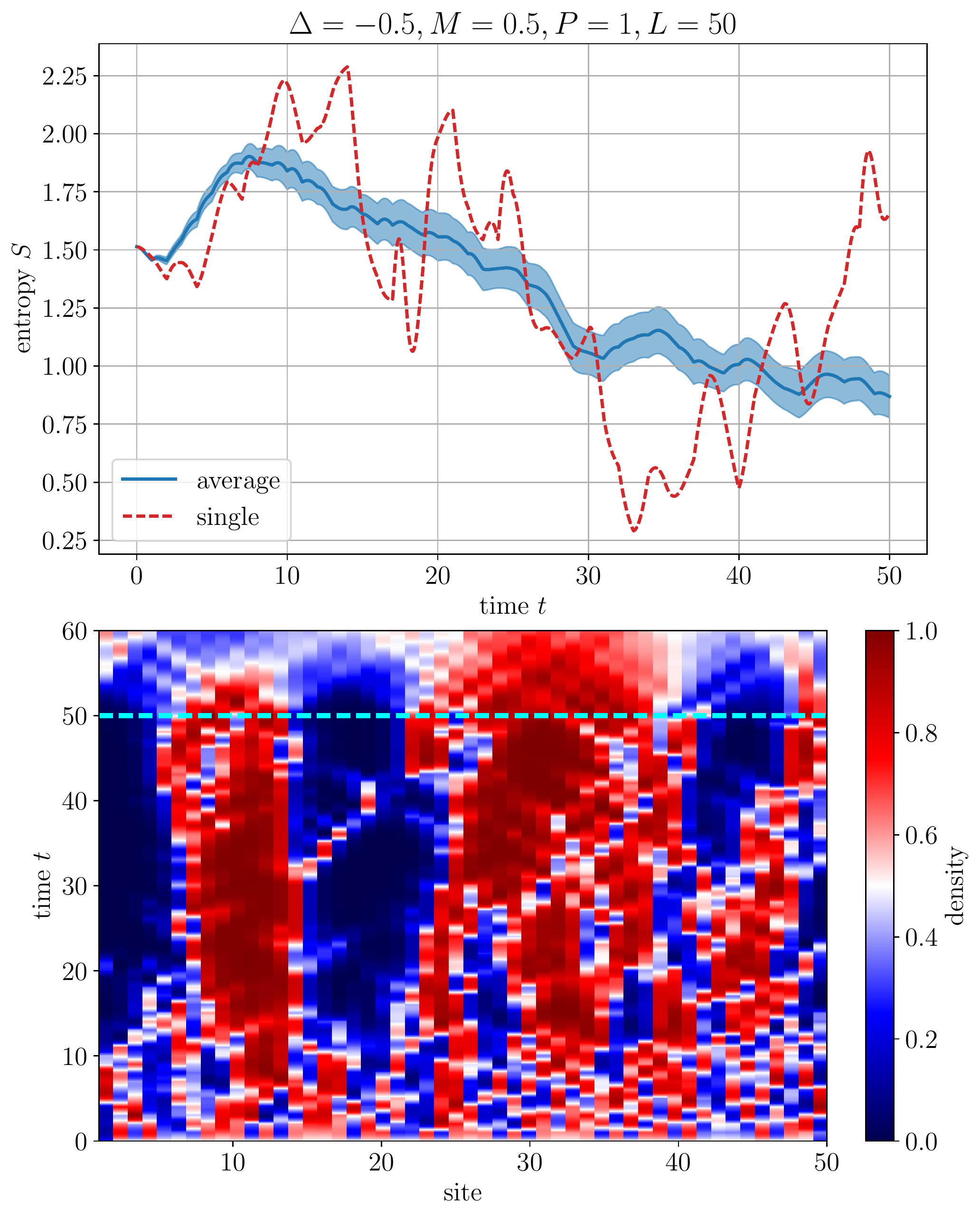}
    \caption{As in Fig.~\ref{fig:M05P1}, but for $\chi=128$. In the bottom panel, the realization with the largest cluster size is chosen.}
    \label{fig:benchmark}
\end{figure}

The numerical method outlined in the main text employs MPS \cite{Schollwock2011a} to simulate pure states. The key numerical convergence parameter is the bond dimension, which controls the size of the variational manifold. Deep in the entangling regime, we have shown that the entanglement grows rapidly, as expected, leading to eventual cutoff effects due to bond dimension truncation (see Fig.~\ref{fig:entropyscaling}). The results depicted in Fig.~\ref{fig:entropyscaling}, deep into the entangling regime, exhibit the fastest growth of entanglement within our parameter range; near the transition region and deep into the disentangling phase bond dimension cutoff effects are not relevant.

Let us consider the situation around the border between the observed disentangling and entangling regimes, in the region where we observe a peak in clusterization. We therefore take $L = 50$, $P = 1$ and $M = 0.5$. The results are shown in Fig.~\ref{fig:benchmark}. As estimated for the maximum cluster size at $t = 50$ we obtain $C = 8.43 \pm 0.29$ for $\chi = 64$ and $C = 8.18 \pm 0.28$ for $\chi = 128$ (with $1\sigma$ error bars), so that the results converged with bond dimension. Moreover, we can see that the curves for the average entropy match within error bars. Note that, since the measurement Hamiltonian \eqref{eq:measHam} explicitly depends on the many-body wave function itself and has a stochastic character, it is not possible to simulate exactly the same measurement outcomes. Thus, some small differences related to these random fluctuations remain even though convergence with the bond dimension is reached.

\section{Measurement protocol based on a non-Hermitian Hamiltonian} \label{sec:appendix}

In this Appendix, we compare the non-Hermitian measurement protocol (avoiding introduction of the detectors) with a more conventional measurement description. 
In particular, we explain the emergence of non-Hermiticity in the effective Hamiltonian.
Further, we derive the master equation for the evolution of the averaged density matrix within the non-Hermitian protocol, and discuss its similarity and difference with respect to the corresponding Lindblad equation for conventional weak measurements. 

In subsection \ref{sec1} we recap previous results for generalized measurements acting on a spin-$1/2$-chain, leading to a master equation of Lindblad type for the averaged density matrix $\bar{\rho}$ discussed in subsection \ref{sec2}.  Further, we discuss  in Sec.~\ref{sec2} the associated stochastic Schr\"odinger equation (which can be used to study the entanglement transition) and the emergence of non-Hermiticity.
We also point out in Sec.~\ref{sec2} that the master equation for $\bar{\rho}$ essentially depends on the possible post-selection protocol. 
We then proceed to compare these results to our approach. In Sec.~\ref{sec3} we discuss the stochastic time evolution of pure states under our non-Hermitian protocol. In Sec.~\ref{sec4} we derive the corresponding master equation for $\bar{\rho}$.

\subsection{Generalized measurements:\\ Spin-1/2 detector for a single site}\label{sec1}

\subsubsection{Basics of a standard measurement protocol}

Consider a single site with possible occupations $n=0$ or $n=1$ that will be the subject of measurement (this system can be equivalently represented as a single spin-$1/2$ with the $z$-projection being measured). A pure initial state of the system is a superposition of $n=0$ and $n=1$ states; note that, for a single site, we do not fix the filling (zero or one), otherwise there is nothing to measure. In the simplest measurement setting, the site is coupled to an ancillary system (or ``detector''), represented as a spin-$1/2$ \cite{Wiseman2009,Jacobs2006a,Roy2020,Turkeshi2021}. Projective measurements of the detector along a certain direction are made at discrete times; the detector state is then initialized anew in a certain state (assumed, for simplicity, to be the same pure state for all time intervals). 

We denote the basis states of the system ($s$) and the detector ($d$) as $|n\rangle$ and $|\sigma\rangle$, respectively, with the quantum numbers $n=0,1$ (occupation) and $\sigma=\pm$ (sign of the $z$-component of the detector's spin; sometimes we will use $\sigma=\pm 1$). The measurement protocol comprises the following steps:
\begin{itemize}
    \item[(i)] Initialization of 
the system and detector at time $t=0$ in a product state \begin{equation}
\rho(t=0)=\rho_s(t=0) \otimes |-\rangle\langle -|,
\end{equation}
where we choose, for definiteness, $\sigma=-$ for the initial detector state at each step. The initial state of the system will be assumed to be pure:
\begin{equation}
  \rho_s(t=0)=|\psi_s\rangle\langle \psi_s|,  
\end{equation}
where
\begin{equation}
|\psi_s\rangle=\alpha|1\rangle+\beta|0\rangle,
\qquad |\alpha|^2+|\beta|^2=1.
\label{psi_s}
    \end{equation}

    \item[(ii)] Unitary time evolution of the composite system governed by Hamiltonian 
    \begin{equation}
        H_\text{total}=H_s+H_d+H_{sd}
    \end{equation}
    up to time $T$. In what follows, we set the detector's own Hamiltonian $H_d$ to zero. Furthermore, for illustrative purposes, we will focus on the case when $H_s$ is also zero, so that the only nonzero part in $H_\text{total}$ is the one that coupes the system and the detector, $H_{sd}$. Note that, during this time, the system is, in general, in a mixed state (entangled with the detector) and, hence, is not described by a wave function.
 \item[(iii)] Instantaneous projective measurement of the detector with outcome $\sigma$, which results in 
\begin{equation}
    \rho(T)
    \quad \to \quad  \rho_s^{(\sigma)}(T)\otimes |\sigma\rangle\langle \sigma|.
\end{equation}  
The normalized system density matrix, conditioned
to the outcome $\sigma$, is given by
\begin{equation}\label{povm2}
   \rho_s^{(\sigma)}(T) = \frac{K_\sigma \rho_s(0) K_\sigma^\dagger}{{\rm tr}_s \left(K_\sigma  \rho_s(0) K_\sigma^\dagger\right)},
\end{equation}
where 
\begin{equation}\label{povm3}
  K_\sigma= \langle \sigma|U(T)|-\rangle,
\end{equation}
is the measurement (Kraus) operator acting on the states of the system,  $$\sum_{\sigma=\pm} K^\dagger_{\sigma}K_{\sigma}=1,$$ and $U(T)$ is the evolution operator at time $T$.
The probability $\mathcal{P}_\sigma$ to get the detector outcome $\sigma$ is given by
 \begin{equation}\label{povm4}
     \mathcal{P}_\sigma = {\rm tr}_s  \left(K_\sigma \rho_s(0) K_\sigma^\dagger \right), 
 \end{equation}
which is just the normalization factor in Eq.~(\ref{povm2}).  
\item[(iv)] Repetition of the above steps starting with the state
\begin{equation}
    \rho(t=T)=\rho_s^{(\sigma)}(T) 
\otimes |-\rangle\langle -|,
\end{equation}
where the detector is re-initialized.
\end{itemize}
The re-initialization of the detector is crucial for the Lindblad equation for the averaged (over readouts) density matrix, as discussed below. We will also discuss the measurement protocol without this re-initialization.

\subsubsection{Measuring site occupation}

In what follows, we consider the following Hamiltonian for the unitary-evolution step of the protocol:
\begin{equation}\label{Ham1}
H_\text{total}\equiv H_{sd}=M\, \hat{n} \hat{\sigma}_x.
\end{equation} 
Here, $\hat{n}$ is the particle-number operator for the system (which is a projector: $\hat{n}^2=\hat{n}$), $\hat{\sigma}_x$ is the Pauli matrix acting in the detector's space, and $M$ quantifies the strength of the system-detector coupling.
The projective measurements of the detector is 
performed for the Pauli operator 
$\hat{\sigma}_z$. 

The protocol using the system-detector coupling (\ref{Ham1}) can be realized 
with auxiliary particles (or, equivalently, spins) as follows:
\begin{itemize}
\item[(I).] Introduce two ancillary sites, 1 and 2,  with creation (annihilation) operators $\hat{a}_1^\dagger\, (\hat{a}_1)$ and $\hat{a}_2^\dagger\,
(\hat{a}_2)$ respectively.
\item[(II).] Connect these two sites by a tunneling amplitude multiplied by the occupation operator $\hat{n}$ of the main site: 
$$
H_{sd}=M\, \hat{n}\,  \hat{a}^\dagger_{1} \hat{a}_{2} + \text{H.c.}
$$
When the main site is unoccupied, there is no tunneling between the auxiliary sites.
\item[(III).] Initialize the ancillary system by putting a particle in site 1.
\item[(IV).] Evolve the whole system with $H_{sd}$ for time $T$.
\item[(V).] Strongly measure the occupations of the ancillary site 1;
\item[(VI).] Repeat the protocol starting with step (III).
\end{itemize}

In what follows, we continue using a spin as a detector.  The evolution operator for the Hamiltonian
(\ref{Ham1}) takes the form:
\begin{equation}
\label{evolU}
U_t=\exp\left(-i H_{sd} t \right) = \hat{\mathbb{1}} -
\hat{n}\Big(1-\cos(Mt) + i \sin (Mt)  \hat{\sigma}_x \Big).
\end{equation} 
Starting at $t=0$ with the initial pure state 
$$|\Psi(0)\rangle=|\psi_s\rangle \otimes |-\rangle,$$
we get at time $t=T$ (below $\lambda= M T$):
\begin{align}
&|\Psi(T)\rangle = U_T|\Psi(0)\rangle
\notag
\\&\quad =\alpha \cos\lambda \ 
  |1\rangle \otimes  |-\rangle +\beta\ |0\rangle\otimes  |-\rangle
  -i \alpha \sin\lambda\ |1\rangle \otimes  |+\rangle.
  \label{Psi-entire}
\end{align}

A projective measurement of the detector's $\sigma_z$ results in the outcome $\sigma=\pm$ (detector's readout).  
A ``click'' of the detector implies that the outcome ${\sigma=+}$  has been measured (the detector's spin, initially pointing down, flips), whereas ${\sigma=-}$ corresponds to a ``null-measurement'' (no-click outcome). Right after projecting the detector, the system is again described by the wave function:
\begin{align}
&\sigma=-,\ \text{no click:}\notag \\
&|\psi_s(T)\rangle =
 \frac{\alpha \cos\lambda}
 {\sqrt{1-|\alpha|^2\sin^2\lambda}} \ |1\rangle + \frac{\beta}{\sqrt{1-|\alpha|^2 \sin^2\lambda}}\ |0\rangle
 \nonumber
 \\
 &\hspace{1.7cm}= 
 \frac{K_-|\psi_s\rangle}{\sqrt{\langle\psi_s|K_-^\dagger K_-|\psi_s\rangle}},
 \label{eq10}
 \\
 & \sigma=+, \ \text{click:}\notag \\
  &|\psi_s(T)\rangle =
 -i\frac{\alpha  \sin\lambda}{|\alpha  \sin\lambda|}\ |1\rangle
 \nonumber
 \\
 &\hspace{1.7cm}= 
 \frac{K_+|\psi_s\rangle}{\sqrt{\langle\psi_s|K_+^\dagger K_+|\psi_s\rangle}}.
 \label{eq11}
\end{align}
The probabilities $\mathcal{P}_\sigma$ of the measurement outcomes are given by
\begin{align}
 \text{no click:}\
 &\quad 
 \mathcal{P}_- = 1-|\alpha|^2 \sin^2\lambda,
  \label{noclprob}
 \\
 \text{click:}\ &\quad 
 \mathcal{P}_+ = |\alpha|^2 \sin^2\lambda = 1-\mathcal{P}_-.
  \label{clprob}
\end{align}
These probabilities satisfy Eq.~(\ref{povm4}) with
\begin{align}
    K_+&=\langle +|U_T|-\rangle = -i\ \sin\lambda \ \hat{n},
    \label{A+}
    \\
    K_-&=\langle -|U_T|-\rangle = \hat{\mathbb{1}}-(1-\cos\lambda)\hat{n}.
    \label{A-}
\end{align}
In the limit of weak measurements, $MT\ll 1$, which will be used for the derivation of the Lindblad equation, we have
\begin{align}
    K_+&\simeq -i\ MT \ \hat{n},
    \label{A+weak}
    \\
    K_-&\simeq \hat{\mathbb{1}}-\frac{(MT)^2}{2}\hat{n}.
    \label{A-weak}
\end{align}

One sees that a no-click measurement yields a modification of the wave function through the measurement back-action.
The click outcome here drives the system's wave function directly to the state $n=1$ (a particle is measured).
The projective-measurement limit is achieved in this setup by setting $\lambda=\pi/2$, when also the no-click outcome yields a definite occupation $n=0$.
In this case of projective measurements,
the measurement probabilities are given by the Born rule:
\begin{align}
 \text{no click:}\ &\quad 
 \mathcal{P}_- = 1-|\alpha|^2 = 1-\langle \hat{n}\rangle,
 \label{noclprob-strong}
 \\
 \text{click:}\ &\quad 
 \mathcal{P}_+ = |\alpha|^2 = \langle \hat{n}\rangle.
  \label{clprob-strong}
\end{align}
For $MT\ll 1$, we have
\begin{align}
 \text{no click:}\ &\quad 
 \mathcal{P}_- = 1-|\alpha|^2 M^2 T^2,
  \label{noclprob-weak}
 \\
 \text{click:}\ &\quad 
 \mathcal{P}_+ = |\alpha|^2 M^2 T^2,
  \label{clprob-weak}
\end{align}
which can be written through the system's density matrix as
\begin{equation}
    \mathcal{P}_\sigma=\frac{1}{2}(1-\sigma) + \sigma \, \text{tr}(\hat{n} \rho) M^2 T^2.
    \label{weakPtau}
\end{equation}
The measurement probabilities depend on both the system state (through $\rho$) and the measurement strength $M$. Clearly, the probability of no-click measurement is 1 in the absence of coupling to the detector.

\subsection{Lindblad equation for a single site in the conventional measurement setup}\label{sec2}

\subsubsection{Derivation of the Lindblad equation}

Since the system is almost always entangled with the detector (except for discrete times $t_j=j T$ when the projection of the detector is performed), the continuous evolution of the system is described in terms of the density matrix $\rho_s(t)=\text{tr}_d\, \rho(t)$. For a given quantum trajectory
specified by the sequence of detector's readouts, $\sigma_1,\ldots,\sigma_j$, where $\sigma_j=\sigma(t_j)$, the total density matrix evolves
as follows:
\begin{align}
 &t_j<t\leq t_{j+1}:\notag
 \\& 
 \rho(t|\sigma_1,\ldots,\sigma_j)
 \!=\!e^{-i (t-t_j) H_{sd}}  \rho(t_j\!+\!0|\sigma_1,\ldots,\sigma_j) 
 e^{i (t-t_j) H_{sd}} \notag
 \\
 &\quad = e^{-i (t-t_j) H_{sd}}  
 \rho_s(t_j\!+\!0|\sigma_1,\ldots,\sigma_j) \otimes |-\rangle\langle -|
 e^{i (t-t_j) H_{sd}},
 \label{eq21}
 \\
&t=t_{j+1}+0:\notag\\
& \rho(t|\sigma_1,\ldots,\sigma_{j+1}\!=\!+)
 \!=\!\frac{\langle +| \rho(t_{j+1}|\sigma_1,\ldots,\sigma_j)|+ \rangle }{\mathcal{P}_+(t_{j+1})} 
 \!\otimes\!  |-\rangle\langle -|,
 \label{eq21a}
 \\
              & 
 \rho(t|\sigma_1,\ldots,\sigma_{j+1}\!=\!-)
 \!=\!\frac{\langle -| \rho(t_{j+1}|\sigma_1,\ldots,\sigma_j)|-\rangle }{\mathcal{P}_-(t_{j+1})} 
 \!\otimes\!|-\rangle\langle -|,
 \label{eq21b}
\end{align}
where
\begin{align}
    \mathcal{P}_\pm(t_{j+1}) 
= \text{tr}_s 
\left(K^\dagger_\pm(t_{j+1}-t_j)K_\pm(t_{j+1}-t_j)
\rho_s(t_j)
\right).
\label{Psigma-def}
\end{align}

The change of the total density matrix at $t=t_{j+1}$,
resulting from the projection of the detector with the outcome $\sigma_{j+1}=\pm$ and re-initialization of the
detector in state $|-\rangle\langle -|$, is referred to as ``quantum jump''. 

Let us introduce a stochastic variable
\begin{equation}
  \chi(t)=\frac{\sigma_j+1}{2}, \quad t_j<t<t_{j+1},
\end{equation}
such that $\chi=0$ for the no-click measurement and $\chi=1$ for the click at time $t_j$. This variable satisfies $\chi^2=\chi$.
One can now cast the evolution equation for the total density matrix in the form of a stochastic differential equation with sources describing quantum jumps:
\begin{align}
    \frac{d \rho}{d t}&=-i\left[H_{sd},\rho\right]\notag\\
    &+\sum_{j=1} \delta(t-t_j) \Big[(1-\chi_j)\Delta_-(t_j) 
    +\chi_j \Delta_+ (t_j) \Big],
    \label{SME}
\end{align}
where $\left[A,B\right]$ is the commutator, $\chi_j=\chi(t_j)$, and 
\begin{equation}
  \Delta_\pm(t_j) = \rho(t_j+0|\sigma_1,\ldots,\sigma_j=\pm)-\rho(t_j|\sigma_1,\ldots,\sigma_{j-1})  
\end{equation}
describes the quantum jump in the total density matrix at $t=t_j$.

An ensemble average over all possible measurement records $\bar{\chi}$ yields
\begin{equation}
    \bar{\chi}(t) =\mathcal{P}_+(t_j), \quad t_j<t<t_{j+1}.
\label{barchi}
\end{equation}
This average is equivalent to tracing out the detector readouts, which is known as blind measurement, at each step.
The blind measurement can be described by a discrete map, where at the measurement time $t_j$ one averages the density matrix over the two measurement readouts $\sigma=\pm$ with the respective probabilities $ \mathcal{P}_\sigma(t_j)$. 

Consider a fixed sequence of readouts
$\sigma_1,\ldots,\sigma_j$ up to time $t_j$. Then, at time $t_{j+1}-0$, the total density matrix is given by Eq.~(\ref{eq21}).
The blind measurement at time $t_{j+1}$ then yields:
\begin{equation}
\rho_s^{(\text{b})}(t_{j+1}|\sigma_1,\ldots,\sigma_j) = \sum_{\sigma} \mathcal{P}_\sigma(t_{j+1}) \rho_s^{(\sigma)}(t_{j+1}|\sigma_1,\ldots,\sigma_j).
\label{dens-map-prob}
\end{equation}
Here, $\rho_s^{(\sigma)}(t_{j+1}|\sigma_1,\ldots,\sigma_j)$
is defined by Eq.~(\ref{povm2}), and the measurement probabilities are conditioned to the full set of previous readouts through the dependence on the state at time $t_{j+1}-0$ given by Eq.~(\ref{povm4}): $$\mathcal{P}_\sigma(t_{j+1})\equiv\mathcal{P}_\sigma(t_{j+1}|\sigma_1,\ldots,\sigma_j).$$
Using Eqs.~(\ref{povm2}) and (\ref{povm4}),
this can be written as 
\begin{equation}\label{blindDensMat1}
   \rho_s^{(\text{b})}(t_{j+1}|\sigma_1,\ldots,\sigma_j)\!=\!\! \sum_{\sigma=\pm}\!\!  K_\sigma(T) \rho_s(t_j|\sigma_1,\ldots,\sigma_{j}) K^\dagger_\sigma(T),
\end{equation}
where $K_\sigma(T)=K_\sigma(t_{j+1}-t_j)$.
Indeed, for the unitary evolution governed by the Hermitian Hamiltonian $H_{sd}$, the probability of the measurement outcome (\ref{povm4}) exactly cancels the normalization factor in the conditioned density matrix (\ref{povm2}). 
As a result, the prehistory of the readouts remains in Eq.~(\ref{blindDensMat1}) only in the value of the density matrix at time $t_j$, and the averaging of $\rho_s^{(\text{b})}(t_{j+1})$ over this prehistory on the l.h.s of Eq.~(\ref{blindDensMat1}) can be expressed through the averaged density matrix at $t=t_j$.
Thus, one can write a Markovian master equation for
the averaged density matrix  $\bar{\rho}_s$, 
\begin{equation}\label{blindDensMat}
    \bar{\rho}_s(t_{j+1}) = \sum_{\sigma}  K_\sigma(t_{j+1}-t_j) \bar{\rho}_s(t_j) K^\dagger_\sigma(t_{j+1}-t_j).
\end{equation}
This is equivalent to
reinserting the measurement probabilities and normalization factors that depend on $\bar{\rho}_s$ in the equation for the density matrix for blind measurements, Eq.~(\ref{dens-map-prob}):
\begin{equation}\label{blindDensMat0}
    \bar{\rho}_s(t_{j+1}) = \sum_{\sigma} \mathcal{P}_\sigma(t_{j+1}) \bar{\rho}_s^{(\sigma)}(t_{j+1}).
\end{equation}

Representation of the averaged density matrix (\ref{blindDensMat}) is known as a Kraus decomposition of the reduced density matrix corresponding to the conventional quantum-mechanical unitary evolution of the system, with the trace taken over all of the subsystem's (detector's) states that are not distinguished by projection. The effect of measurements here is in the re-initialization of the detector at each measurement step. If no re-initialization is performed, the evolution of the averaged density matrix of the system would be described by the conventional von Neumann equation with $H_{sd}$ and not by the Lindblad equation: the summation over the states of the detector at intermediate steps in this case is equivalent to the resolution of the unity operator. 
At the same time, any particular sequence of the detector's readouts still defines a quantum trajectory of the system, and the full statistics of such trajectories is a highly nontrivial object. This is a clear demonstration of the fact that the Lindblad master equation for the averaged density matrix is in general not sufficient for describing measurement-induced dynamics of the entanglement. Indeed, if the measurements are performed without any re-initialization (as in various models of entanglement phase transitions), the Lindblad equation does not carry any information about the measurements at all.

The resulting equation for the averaged system density matrix $\bar{\rho}_s(t)$ greatly simplifies in the weak-measurement limit $M T\ll 1$, yielding a well-known differential Lindblad equation. Using $K_\sigma(T)$ from Eqs.~(\ref{A+weak}) and (\ref{A-weak}),
we find:
\begin{align}
   & \bar{\rho}_s(t_j+T)=\bar{\rho}_s(t_j)\notag\\
    &\quad +T^2  \left(\mathcal{H}_{+-}\bar{\rho}_s(t_j)\mathcal{H}_{+-}^\dagger
    -\frac{1}{2}\{\mathcal{H}_{+-}^\dagger \mathcal{H}_{+-},\bar{\rho}_s(t_j)\}\right),
    \label{eq31}
\end{align}
where $\left\{A,B\right\}$ is the anticommutator and
\begin{align}
    \mathcal{H}_{\sigma\sigma'}=\langle \sigma|H_{sd}| \sigma'\rangle,
    \quad \mathcal{H}_{-+}=M \hat{n}=\mathcal{H}_{+-}^\dagger.
    \label{eq32}
\end{align}
In the continuous-time limit, we can infer the form of the jump operator in the Lindblad equation for the averaged system density matrix by taking the limit
\begin{equation}
  \frac{d\bar{\rho}_s(t)}{dt} = \lim_{T\to 0}  \frac{\bar{\rho}_s(t\!+\!T)-\bar{\rho}_s(t)}{T}.
  \label{derivative}
\end{equation}
This yields the Lindblad equation [cf. Eq.~(\ref{eq:lindblad}) of the main text]
\begin{equation}
  \frac{d\bar{\rho}_s(t)}{dt}= L \bar{\rho}_s(t) L^\dagger-\frac{1}{2}\{L^\dagger L,\bar{\rho}_s(t)\},
  \label{Lindblad-eq}
\end{equation}
where, for our choice of the system-detector coupling (\ref{Ham1}), the Lindblad jump operator $L$ is given by 
\begin{equation}
    L\equiv M \sqrt{T}\ \hat{n}.
    \label{jumpL}
\end{equation}
In more general cases, several jump operators can be involved in the Lindblad equation, see below.
The continuum limit is realized by 
\begin{equation}
T\to 0, \quad M\to \infty, \quad M \sqrt{T} = \text{const}.
\label{cont-limit}
\end{equation}
while keeping $M \sqrt{T}$ fixed. We see that the Lindblad jump operator in the above conventional setting for continuous measurements of the site density is proportional to the density operator. Using the explicit form of the Lindblad jump operator,  Eq.~(\ref{jumpL}), and its property $L^\dagger L=M^2 T \hat{n}$, we cast Eq.~(\ref{Lindblad-eq}) in the form
\begin{equation}
   \frac{d\bar{\rho}_s}{dt} = M^2 T \left(\hat{n} \bar{\rho}_s \hat{n} - \frac{1}{2}\{\hat{n},\bar{\rho}_s\} \right). 
   \label{LindbEq-usual}
\end{equation}
In the presence of the system's own dynamics governed by $H_s\neq 0$, a standard von Neumann term $-i\left[H_{s},\rho\right]$ is added to the r.h.s. of Eq.~(\ref{LindbEq-usual}). Note that, for the measurement protocol considered here, the system's own dynamics would prevent freezing of the system 
in the $|1\rangle$-state after the first click-readout of the detector.

Above, we have considered the protocol when at each measurement time step of duration $T$ the system is coupled to the re-initialized detector. This corresponds to ``frequent measurements'' with the probability $P=1$ that the site is monitored. If we allow for time steps for which the detector is not coupled to the system, $P<1$, the r.h.s. of the Lindblad equation, Eq.~(\ref{LindbEq-usual}),
is simply multiplied by $P$. Thus, at the level of the Lindblad equation, the measurement frequency can be fully absorbed into the measurement strength $M^2\to P M^2 \equiv \tilde{M}^2$. However, this is generically not so for 
the nonlinear-in-$\rho_s$ quantities. It is also worth emphasizing that the probability of measuring $P$ is not
equivalent to the no-click measurement probability $\mathcal{P}_-$  for the system coupled to the detector
(a famous example is the Elitzur-Vaidman setup \cite{Elitzur1993}).

\subsubsection{Analysis of the solution}

Writing the $2\times 2$ density matrix 
 satisfying $\text{tr} \bar{\rho}=1$ and occupation operator $\hat{n}$ in the basis of $|1\rangle$ and $|0\rangle$ states,
\begin{equation}
   \bar{\rho}= \left( \begin{array}{ll}
   a & b\\
   b^* & 1-a
   \end{array}\right), \qquad 
   \hat{n}= \left( \begin{array}{ll}
   1 & 0\\
   0 & 0
   \end{array}\right), 
   \label{rho-explicit}
\end{equation}
we obtain 
\begin{align}
\hat{n}\,\bar{\rho}\,\hat{n}
    =\left( \begin{array}{ll}
   a & 0\\
   0 & 0
   \end{array}\right)=\hat{n}\, \text{tr}(\hat{n} \bar{\rho}), \qquad  
    \{\hat{n},\bar{\rho}\}=\left( \begin{array}{ll}
   2a & b\\
   b^* & 0
   \end{array}\right),
   \label{nrhon}
\end{align}
and
\begin{align}
\hat{n}\,\bar{\rho}\,\hat{n} - \frac{1}{2} \{\hat{n},\bar{\rho}\}
    =-\frac{1}{2}\left( \begin{array}{ll}
   0 & b \\
   b^* & 0
   \end{array}\right).
   \label{nrhon1}
\end{align}
Equation (\ref{LindbEq-usual}) then reduces to two independent equations for diagonal and non-diagonal elements of the density matrix:
\begin{align}
     \frac{d a}{d t} &= 0,
    \\
    \frac{d b}{d t} &= -\frac{M^2 T}{2} b.
    \label{nonlinear-ab}
\end{align}

We see that the diagonal matrix elements of the averaged density matrix do not change with time: the mean occupation \begin{equation}
    \bar{n} = \text{tr}\,(\hat{n} \bar{\rho}_s )=a
\end{equation}
remains constant.
Multiple weak measurements of the single-site occupation only destroy (on average) the coherence between occupied and unoccupied states. This could have been guessed without any calculation: indeed, we know that measuring the $n=0$ or $n=1$ state with an arbitrary strength, we do not modify the occupation $|\alpha|^2$, see Eqs.~(\ref{eq10})
and (\ref{eq11}) for $\alpha=0$ and $\alpha=1$, respectively. Therefore, the Lindblad equation possesses the two dark states $a=0$ and $a=1$ at $b=0$.
Since the Lindblad equation is linear in the density matrix, it cannot contain the nonlinear product $a(a-1)$ (which would provide the emergence of these two dark states) on the r.h.s. As a result, all states for $b=0$ should belong to the dark space of the Lindbladian.

It is also instructive to consider the averaged purity of the system ($\rho^2=\rho$ for pure states), as an example of the nonlinear-in-$\rho$ characteristics of the system.
In analogy with Eq.~(\ref{dens-map-prob}), we write  $\rho_s^2$ after the blind measurement at time $t_{j+1}$, given the sequence of previous readouts $\sigma_1,\ldots,\sigma_j$:
\begin{align}
\left[\rho_s^2(t_{j+1}|\sigma_1,\ldots,\sigma_j)\right]^{(b)} &= \sum_{\sigma=\pm} \mathcal{P}_\sigma(t_{j+1})
\notag\\
&\times \Big(\rho_s^{(\sigma)}(t_{j+1}|\sigma_1,\ldots,\sigma_j)\Big)^2.
    \label{blindDensMat2}
\end{align}
Here, the measurement probability is again conditioned to all previous readouts, as it is expressed through the exact state at time $t_{j+1}$. 
Importantly, for the blind measurement, the same weight $\mathcal{P}_\sigma$ characterizes the update of all the moments of the density matrix. 
Therefore, the measurement probability in the numerator does not completely cancel the normalization denominator  in Eq.~(\ref{blindDensMat2}), which now contains the 
square of this probability through the squared normalized density matrix $\rho_s^{(\sigma)}$:
\begin{align}
\left[\rho_s^2(t_{j+1})\right]^{(b)} = \sum_{\sigma=\pm}  \frac{K_\sigma(T) \rho_s^2(t_j|\sigma_1,\ldots,\sigma_j) K^\dagger_\sigma(T)}{\mathcal{P}_\sigma(t_{j+1}|\sigma_1,\ldots,\sigma_j)}.
    \label{blindDensMat3}
\end{align}

Importantly, the whole prehistory of readouts is involved in the denominator on the r.h.s. of Eq.~(\ref{blindDensMat3}), which is thus correlated with $\rho_s^2(t_{j})$ in the numerator. Therefore, the averaging of $[\rho_s^2(t_{j+1})]^{(b)}$ on the l.h.s. of Eq.~(\ref{blindDensMat3}) over quantum trajectories does not decouple into the ratio of the averaged density matrix $\rho_s^2(t_{j})$ and averaged measurement probability $\mathcal{P}_\sigma(t_{j+1})$. As a result, in contrast to the averaged density matrix (\ref{blindDensMat}), the discrete map 
for $\rho_s^2$ cannot be, in general, written in terms of a Markovian equation for $\overline{\rho_s^2}$.  
We also note the emergence of the density matrix in the denominator of Eq.~(\ref{blindDensMat2}) through Eq.~(\ref{Psigma-def}), which prevents us from having a linear map for $\rho_s^2$.
On the other hand, starting with a pure state at $t=0$,
we should have the pure state of the system at all discrete times $t_j$, so that $\rho_s^2(t_j)=\rho_s(t_j)$ should hold (also for averages). This follows by induction from Eq.~(\ref{blindDensMat2}) combined with  
Eq.~(\ref{dens-map-prob}). In general, however, in case of continuous monitoring of a large system that is always entangled with detectors, the moments of the system's density matrix are not expressed through the density matrix itself, and their averaging over the prehistory of readouts is highly nontrivial.

\subsubsection{Stochastic Schr\"odinger equation and emergence of non-Hermiticity}

The conventional Lindblad equation (\ref{Lindblad-eq})
contains the two terms that are analogous to the in-going and out-going terms in the Boltzmann equation. These terms 
directly correspond to click and no-click events, respectively. The term with the anticommutator $\{L^\dagger L,\bar{\rho}_s(t)\}$ describes the contribution of null measurements. Indeed, postselecting the sequences of readouts with only $\sigma=-$ outcomes kept (discarding all $\sigma=+$ readouts), we replace Eq.~(\ref{blindDensMat})
by
\begin{equation}\label{postDensMat}
    \rho^\text{null}_s(t_{j+1}) = 
    K_-(T) \rho^\text{null}_s(t_j) K^\dagger_-(T),
\end{equation}
which describes the evolution of such a postselected density matrix $\rho^\text{null}_s$. Note that the norm of the density matrix is reduced in this way, which amounts to a reduction, with each measurement step, of the probability of having a no-click-only sequence of readouts.

Using Eq.~(\ref{A-weak}),
we obtain the map for its evolution,
\begin{equation}\label{postDensMat1}
    \rho^\text{null}_s(t_{j+1}) =  \left(\hat{\mathbb{1}}-\frac{M^2T^2}{2}\hat{n}\right) \rho^\text{null}_s(t_j) \left(\hat{\mathbb{1}}-\frac{M^2T^2}{2}\hat{n}\right),
\end{equation}
which, to leading order in $MT\ll 1$, yields
\begin{equation}\label{postDensMat2}
    \frac{d\rho^\text{null}_s(t)}{dt} =  -M^2T\, \{\hat{n},\rho^\text{null}_s(t)\}.
\end{equation}
This is a Lindblad equation corresponding to an anti-Hermitian Hamiltonian. In general, this Hamiltonian, bilinear in Lindblad jump operators [see Eq.~(\ref{eq31})],
\begin{align}
    H_\text{null}=-\frac{i}{2} L^\dagger L.
    \label{Heff}
\end{align}
describes the loss of the probability in the null-measurement sequence and is the one usually appearing in the context of measurements (see, e.g., Ref.\,\cite{Fuji2020a}). 

Considering pure states at discrete times $t_j$, the stochastic evolution of the density matrix is equivalently represented in the weak-measurement limit by a stochastic Schr{\"o}dinger equation for the system's wave function.
The exact discrete map for the update of the wave function is given by [cf. Eqs.~(\ref{eq21a}), (\ref{eq21b}), and (\ref{SME})]
\begin{align}\label{SDSE}
    |\psi_s(t_{j+1})\rangle
     & = 
  \chi_{j+1}\,\frac{K_+(T)|\psi_s(t_j)\rangle}{\sqrt{\mathcal{P}_+(t_{j+1})}}
  \notag
  \\
  &+ (1-\chi_{j+1})\, \frac{K_-(T)|\psi_s(t_j)\rangle}
   {\sqrt{\mathcal{P}_-(t_{j+1})}}
    \\
     =|\psi_s(t_{j})\rangle 
    &+\chi_{j+1}\!\left[\frac{K_+(T)|\psi_s(t_j)\rangle}{\sqrt{\mathcal{P}_+(t_{j+1})}}-|\psi_s(t_j)\rangle\right]
    \notag\\
    & 
    + (1-\chi_{j+1})\!
    \left[
    \frac{K_-(T)|\psi_s(t_j)\rangle}
    {\sqrt{\mathcal{P}_-(t_{j+1})}}
    -|\psi_s(t_j)\rangle\right],
     \nonumber
\end{align}
with normalization factors that can be written through the norm of the updated wave function: 
$$
\sqrt{\mathcal{P}_\pm(t_{j+1})}=||K_\pm(T)|\psi_s(t_j)\rangle||.
$$
Similar to the derivation of the Lindblad equation, Eqs.~(\ref{eq31})-(\ref{Lindblad-eq}), we consider the continuous-time limit, Eq.~(\ref{cont-limit}). Using 
Eqs.~(\ref{A+weak}) and (\ref{A-weak}) with 
$$|\alpha|^2\to \langle\psi_s(t)|\hat{n}|\psi_s(t)\rangle,$$ we write for $T\to 0$
\begin{align}
 &|\psi_s(t+T)\rangle  \simeq |\psi_s(t)\rangle
 \notag \\
 & + \chi\!\left[\frac{-i MT \hat{n}|\psi_s(t)\rangle}{MT\sqrt{\langle\psi_s(t)|\hat{n}|\psi_s(t)\rangle}}-|\psi_s(t)\rangle\right]   
 \nonumber
 \\
 &+(1-\chi)\! 
    \left[\left
    (1+\frac{M^2 T^2}{2}
    \langle\psi_s(t)|\hat{n}|\psi_s(t)\rangle\right)\right.
    \notag
    \\
    &\qquad\times
    \left.\left(\hat{\mathbb{1}}-\frac{M^2T^2}{2}\hat{n}\right)|\psi_s(t)\rangle
    -|\psi_s(t)\rangle\right]
    \nonumber\\
 &\simeq |\psi_s(t)\rangle+ \chi\left[\frac{-i \hat{n}|\psi_s(t)\rangle}{\sqrt{\langle\psi_s(t)|\hat{n}|\psi_s(t)\rangle}}-|\psi_s(t)\rangle\right]  
 \notag\\
 & +\frac{M^2 T^2}{2}
 \Big
    (\langle\psi_s(t)|\hat{n}|\psi_s(t)\rangle |\psi_s(t)\rangle
    -\hat{n}|\psi_s(t)\rangle\Big).
    \label{SSE-discrete}
\end{align}
Here, in the last term that corresponds to no-click measurements, we approximate $1-\chi \approx 1$. After averaging over trajectories, the term containing $\chi$ would produce a higher power of $T$. Indeed, according to Eqs.~(\ref{barchi}) and (\ref{clprob-weak}), we have in the limit (\ref{cont-limit})
$$ \overline{\chi}=\langle\psi_s(t)|\hat{n}|\psi_s(t)\rangle M^2T^2 \propto T$$
(recall that $\chi^2=\chi$, so that any power of $\chi$ would produce on average $T$).

The no-click term in Eq.~(\ref{SSE-discrete}) can be represented by means of effective state-dependent anti-Hermitian Hamiltonian $H_\text{ncl}$:
\begin{align}
    \frac{M^2 T}{2}
 \Big
    (\langle\psi_s|\hat{n}|\psi_s\rangle |\psi_s\rangle
    -\hat{n}|\psi_s\rangle\Big)=-i H_\text{ncl}|\psi_s\rangle,
    \\
H_\text{ncl}\equiv  \frac{i}{2} \Big( ||L|\psi_s\rangle||^2 - L^\dagger L\Big).
    \label{eff-H}
\end{align}
where we have used the definition of the Lindblad jump operators
for our model, Eq.~(\ref{jumpL}).
The effective non-click Hamiltonian (\ref{eff-H}) differs from Eq.~(\ref{Heff}) by the presence of the first term, which is proportional to the unity operator and depends on the state 
$|\psi_s\rangle$.

Neglecting the click term (postselecting the no-click trajectories, $\chi\to 0$, but restoring the normalization at each step), we obtain the following equation for the normalized no-click-postselected density matrix $\rho^\text{ncl}_s$ instead of Eq.~(\ref{postDensMat2}):
\begin{align}
     \frac{d}{dt}\rho^\text{ncl}_s& \simeq 
    \frac{1}{T}\Big(|\psi_s(t+T)\rangle \langle \psi_s(t+T) | - |\psi_s(t)\rangle \langle \psi_s(t) |  \Big)
      \nonumber
      \\
      &\simeq \frac{M^2 T}{2} 
      \Big
    (|\psi_s\rangle \langle\psi_s|\hat{n}|\psi_s\rangle \langle\psi_s|
    -|\psi_s\rangle\langle\psi_s|\hat{n}
    \notag
    \\
    &\qquad + 
    \langle\psi_s|\hat{n}|\psi_s\rangle |\psi_s\rangle \langle\psi_s|
    -\hat{n}|\psi_s\rangle\langle\psi_s|\Big)
    \nonumber
    \\
    &= \frac{M^2 T}{2} 
      \Big(2\,\text{tr}(\hat{n} \rho^\text{ncl}_s)\,  \rho^\text{ncl}_s - \{\hat{n}, \rho^\text{ncl}_s \}  \Big).
      \label{rho-ncl}
\end{align}
It is easy to check that the normalization $\text{tr}\rho^\text{ncl}_s=1$ is preserved by Eq.~(\ref{rho-ncl}), as it should be by construction:
\begin{align}
     \frac{d}{dt}\text{tr}\rho^\text{ncl}_s
    &= \frac{M^2 T}{2} 
      \Big[2\,\text{tr}(\hat{n} \rho^\text{ncl}_s)\,  \text{tr}\rho^\text{ncl}_s - 2 \,\text{tr}(\hat{n} \rho^\text{ncl}_s)  \Big] =0.
\end{align}
We thus see that postselection accompanied by maintaining normalization gives rise to a nonlinear master equation 
for the postselected density matrix, while not keeping normalization we get a linear equation, Eq.~(\ref{postDensMat2}), corresponding to an anti-Hermitian Hamiltonian, Eq.~(\ref{Heff}). The non-linearity of the Lindblad-type master equation for the averaged density matrix (with normalization maintained) is a generic property of protocols with post-selection.

Now, restoring the click term in Eq.~(\ref{SSE-discrete}) and proceeding along the same lines as for the derivation of Eq.~(\ref{rho-ncl}), we obtain:
\begin{align}
     \frac{d}{dt}\rho_s& \simeq 
    \frac{1}{T}\Big[|\psi_s(t+T)\rangle \langle \psi_s(t+T) | - |\psi_s(t)\rangle \langle \psi_s(t) |  \Big]
      \nonumber
      \\
      &\simeq 
      |\psi_s\rangle
      \frac{\chi}{T}\!\left(\frac{i \langle \psi_s| \hat{n}}{\sqrt{\langle\psi_s|\hat{n}|\psi_s\rangle}}- \langle \psi_s|\right)
      \notag
      \\
      &+
      \frac{\chi}{T}\!\left(\frac{-i \hat{n}|\psi_s\rangle}{\sqrt{\langle\psi_s|\hat{n}|\psi_s\rangle}}-|\psi_s\rangle\right)   
      \langle \psi_s|
      \notag
      \\
      &+\frac{\chi^2}{T}\left(\frac{-i \hat{n}|\psi_s\rangle}{\sqrt{\langle\psi_s|\hat{n}|\psi_s\rangle}}-|\psi_s\rangle\right)\left(\frac{i \langle \psi_s| \hat{n}}{\sqrt{\langle\psi_s|\hat{n}|\psi_s\rangle}}- \langle \psi_s|\right)
      \notag
      \\
      &+
      \frac{M^2 T}{2} 
      \Big
    (|\psi_s\rangle \langle\psi_s|\hat{n}|\psi_s\rangle \langle\psi_s|
    -|\psi_s\rangle\langle\psi_s|\hat{n}
    \notag
    \\
    &\qquad + 
    \langle\psi_s|\hat{n}|\psi_s\rangle |\psi_s\rangle \langle\psi_s|
    -\hat{n}|\psi_s\rangle\langle\psi_s|\Big)
    \nonumber
    \\
    &= \frac{\chi}{T}(
   -2\rho_s)
    +\frac{\chi^2}{T}
    \left[\frac{\hat{n}\rho_s\hat{n}}{\text{tr}(\hat{n} \rho_s)} + \rho_s\right]
    \notag
    \\
    &+
    \frac{M^2 T}{2} 
      \Big(2\,\text{tr}(\hat{n} \rho_s)\,  \rho_s - \{\hat{n}, \rho_s \}  \Big).
      \label{rho-full-Lindblad}
\end{align}
Using $\chi^2=\chi$, we arrive at
\begin{align}
         \frac{d}{dt}\rho_s& \simeq 
         \frac{\chi}{T}\!
      \left[\frac{\hat{n}\rho_s\hat{n}}{\text{tr}(\hat{n} \rho_s)} - \rho_s\right]    
   \notag
    \\
    &+
    \frac{M^2 T}{2} 
      \Big(2\,\text{tr}(\hat{n} \rho_s)\,  \rho_s - \{\hat{n}, \rho_s \}  \Big).
      \label{rho-full-Lindblad-1}      
\end{align}
Averaging $\chi$ on the r.h.s. of Eq.~(\ref{rho-full-Lindblad-1}) over quantum trajectories independently of $\rho_s$, we replace  
\begin{align}
    \overline{\chi}\ \to \ \text{tr}(\hat{n} \bar{\rho}_s)\, M^2 T^2.
\end{align}
In a similar manner, we also decouple the correlations
between the density matrices appearing in the numerator and denominator of the first term in Eq.~(\ref{rho-full-Lindblad-1}), which amounts to the replacement $\rho_s \to \bar{\rho}_s$ everywhere.
The above decoupling of averages is justified in the limit of weak measurements, $MT\ll 1$, which is simultaneously the limit of applicability of the Lindblad equation. 
This decoupling is reminiscent of the decoupling of disorder averages of the distribution functions and the scattering probabilities in the collision integrals of the kinetic equation for weakly disordered systems. It is worth noting that, as the derivation of Eq.~(\ref{blindDensMat}) shows,
the decoupling approximation turns out to be exact, implying that the contriburions of the neglected correlations to the master equation actually cancel each other.

The result of the decoupled average over the quantum trajectories reads:
\begin{align}
         \frac{d}{dt}\bar{\rho}_s& \simeq 
         \frac{\text{tr}(\hat{n} \bar{\rho}_s)\, M^2 T^2}{T}\!
      \left[\frac{\hat{n}\bar{\rho}_s\hat{n}}{\text{tr}(\hat{n} \bar{\rho}_s)} - \bar{\rho}_s\right]    
   \notag
    \\
    &+
    \frac{M^2 T}{2} 
      \Big(2\,\text{tr}(\hat{n} \bar{\rho}_s)\,  \bar{\rho}_s - \{\hat{n}, \bar{\rho}_s \}  \Big)
      \notag
    \\
    &=M^2 T  \left(\hat{n}\bar{\rho}_s\hat{n} - \frac{1}{2} \{\hat{n}, \bar{\rho}_s \}  \right).
      \label{rho-full-Lindblad-2}      
\end{align}
This equation reproduces the conventional (linear) Lindblad equation, Eq.~(\ref{LindbEq-usual}).
The nonlinear (normalization-maintaining) term $\text{tr}(\hat{n} \bar{\rho}_s)\,  \bar{\rho}_s$ from the no-click part is completely cancelled here by the term from the click (``in-going'') part. Importantly, the nonlinear terms would be present in the effective master equation
for any type of postselection accompanied with maintaining the norm of the state, similarly to Eq.~(\ref{rho-ncl}).

It is worth noticing that the $\chi^2$ term was crucially important in Eq.~(\ref{rho-full-Lindblad}). 
Writing
\begin{align}
 \frac{d}{dt}\rho_s&= |\psi_s(t)\rangle \frac{\langle \psi_s(t) |}{dt} + \frac{|\psi_s(t)\rangle}{dt} \langle \psi_s(t) |,
 \label{rho-derivative}
 \end{align}
 and expressing
 \begin{align}
 \frac{d|\psi_s\rangle}{dt}&\simeq \frac{|\psi_s(t+T)\rangle  - |\psi_s(t)\rangle} 
 {T},
\end{align}
through Eq.~(\ref{SSE-discrete}), one would miss the $\chi^2$ contribution to the master equation. This is related to the fact that Eq.~(\ref{rho-derivative}) is only valid for ``well-defined'' differentiable functions, whereas the stochastic measurement process yields the system's wave function only at discrete time steps. The proper way to introduce the derivatives in the effective continuous (``coarse-grained'') limit can be formulated in terms of 
It$\hat{\text{o}}$ stochastic calculus, as described for the case of measurements, e.g., in Ref.~\cite{Wiseman2009}.

\subsection{Non-Hermitian measurement for a single site}\label{sec3}

Following the approach outlined in Sec.~\ref{s2b}, we now imitate the conventional measurement protocol by
introducing a non-Hermitian Hamiltonian acting on a single site. As above, we set the ``own'' 
Hamiltonian of the system to zero. The whole dynamics is then described by the non-unitary evolution
governed by the ``measurement'' Hamiltonian, which, for the time interval $t_j<t<t_{j+1}=t_j+T$,
is defined as
\begin{equation}
H=i M p_j \,\text{sgn}(\langle \hat{n} \rangle - m_j)\ \hat{n},
\label{Hmeas}
\end{equation}
where $M>0$ characterizes the strength of 
``measurement'', 
$p_j=0,1$ is a binary random variable telling whether the site is measured or not for a given time interval
(the probability of measurement is $P$),
and $m_j\in(0,1)$ is a homogeneously distributed random number. In what follow, we will consider the case of frequent measurements, $P=1$, so that $p_j=1$ for all time intervals, as in the previous sections.

In analogy with the conventional measurement protocol of Secs.~\ref{sec1} and \ref{sec2},
we introduce the ``measurement outcomes''
\begin{equation}
\sigma_j=\text{sgn}(\langle \hat{n} \rangle - m_j) =\pm .
\end{equation} 
We will refer to the outcome $\sigma_j=+$ (or $\chi_j=1$) as a click event,
and to $\sigma_j=-$ (equivalently,
$\chi_j=0$) as a no-click event.
The probabilities of the click and no-click measurements are determined by the expectation value of the site occupation:
\begin{align}
\mathcal{P}_+=\langle \hat{n} \rangle, \qquad  \mathcal{P}_-=1-\langle \hat{n} \rangle.
\label{prob-nonherm}
\end{align}
These probabilities correspond to probabilities of having a random number $0<m_j<1$ smaller or larger than 
$\langle \hat{n} \rangle$, respectively.

We stress that, in the limit 
of projective measurements, the probabilities (\ref{noclprob-strong}) and (\ref{clprob-strong}) coincide with (\ref{prob-nonherm}). However, in contrast to conventional measurement probabilities (\ref{noclprob}) and (\ref{clprob}),
the probabilities (\ref{prob-nonherm}) do not depend
on the measurement strength $M$; in particular, there
is a non-zero click probability (at finite $\langle \hat{n} \rangle$) even for $M=0$.  At the same time, in the case of generalized measurements, the ``backaction'' induced by the non-Hermitian ``measurement Hamiltonian'' (\ref{Hmeas}) still depends on $M$, as in the conventional setting.
As a result, the dynamics of the monitored system governed by Hamiltonian (\ref{Hmeas}) does depend on $M$ and is qualitatively similar to the dynamics of the ``truly measured'' system for $MT\lesssim 1$, even in the limit of weak measurements $MT\ll 1$.

Further, in contrast to the conventional setup, where the system is entangled with the detector and is, hence, not described by a pure wave function (except for a discrete set of times when the detector is projected), here we always have a pure system state. This state is maintained normalized to unity, by constantly renormalizing its norm. Although the evolution of the system state governed by Hamiltonian  (\ref{Hmeas}) can be found exactly (see below), it is instructive to consider this evolution for a small time interval $\delta t \ll M^{-1}$,
by expanding the evolution operator 
to second order in $M\delta t$.
For the measurement 
Hamiltonian (\ref{Hmeas}), we obtain
\begin{equation}
    U_t=\exp(-i\, t\, H)=e^{ t M \sigma_j \hat{n}} \simeq 
    1 + t M \sigma_j \hat{n} + \frac{t^2 M^2}{2} \hat{n}.
\end{equation}
Acting with this operator on state $|\psi_s\rangle =\alpha\ |1\rangle +\beta \ |0\rangle$, we find the renormalized state
\begin{align}
  & |\psi_s(\delta t) \rangle  = 
    \frac{U_{\delta t}|\psi_s\rangle}{\sqrt{\langle \psi_s| U_{\delta t} U_{\delta t}|\psi_s\rangle}} = \alpha'\ |1\rangle +\beta' \ |0\rangle \notag
    \\
 &\simeq\!\alpha\!\Big[1+\delta t M \sigma (1-|\alpha|^2) \nonumber \\
 & \!+\!  \frac{t^2 M^2}{2}(1-|\alpha|^2)(1-3|\alpha|^2)\Big]\! 
  |1\rangle 
 \nonumber\\
 &+\!\beta\!\left[ 1-\delta t M \sigma |\alpha|^2-
 \frac{t^2 M^2}{2}|\alpha|^2(2-3|\alpha|^2) \right]\!  |0\rangle .
\end{align}
Clearly, for $|\alpha|=1$ or $\alpha=0$ the ``measurement'' does not affect the state. Note also that if $|\alpha|^2>m$ then also $|\alpha'|^2>m$, and the measurement outcome $\sigma$ is guaranteed to be fixed within the whole measurement step $T$. One can also explicitly check that the normalized state obtained in one iteration of length $2\delta t$ is equal to the state obtained in two consecutive time steps of length $\delta t$. This implies that the renormalization
of the state evolved by the non-unitary evolution operator
can be performed at the end of the whole measurement step of duration $T$.

For the whole measurement step, with an arbitrary value of $MT$, we get
\begin{align}
    U_T|\psi_s\rangle = \alpha e^{M T \sigma}|1\rangle+\beta|0\rangle,
\end{align}
and the normalized state takes the form:
\begin{align}
    |\psi_s(T)\rangle = \frac{\alpha e^{M T \sigma}|1\rangle+\beta|0\rangle}{\sqrt{|\alpha|^2
    \left(e^{2 M T \sigma}-1\right)+1}}.
\end{align}
In the limit $MT\ll 1$, we recover the results above;
in the opposite limit of strong measurement, $MT\gg 1$,
we obtain
\begin{align}
&\text{for}\ \sigma=+,
\notag
 \\
 &|\psi_s(T)\rangle \simeq \frac{\alpha}{|\alpha|}\left(1-\frac{|\beta|^2}{|\alpha|^2} e^{-2 M T}\right)  |1\rangle + \frac{\beta}{|\alpha|}
 e^{- M T} |0\rangle,
 \\
  &\text{for}\ \sigma=-,
  \notag
 \\
 &|\psi_s(T)\rangle \simeq \frac{\alpha}{|\beta|}
 e^{- M T}|1\rangle + \frac{\beta}{|\beta|}
 \left(1-\frac{|\alpha|^2}{|\beta|^2} e^{-2 M T}\right)|0\rangle.
\end{align}
This reproduces the Born rule for strong measurements:
\begin{align}
     \alpha |1\rangle+\beta|0\rangle \ \to \ \frac{\alpha}{|\alpha|} |1\rangle \quad \text{with probability\ $\mathcal{P}_+=|\alpha|^2$},
\end{align}
and
\begin{align}
     \alpha |1\rangle+\beta|0\rangle \ \to \ \frac{\beta}{|\beta|} |0\rangle \quad \text{with probability\ $\mathcal{P}_-=|\beta|^2$}. 
\end{align}

Thus, we see that, in the limit of strong measurement $MT\to \infty$,
the non-Hermitian protocol effectively mimics the conventional measurement implementation based on the state projection (von Neumann's wave-function collapse) and the corresponding Born rule. The non-Hermitian modelling of the projective measurement can be incorporated into the traditional generalized-measurement protocol (described in Sec. \ref{sec1}) by replacing the detector's projection with its stochastic non-Hermitian evolution. Our modeling of the generalized measurement (``incomplete collapse'' of the wave function) as an evolution with a non-Hermitian Hamiltonian is in spirit of the description of the measurement process  in terms of a continuous non-linear stochastic process in Ref.~\cite{Patel2017a}.

The stochastic discrete map for the system's wave function for the non-Hermitian protocol takes the form with $\chi_j=(\sigma_j+1)/2$ after $j$th step:
\begin{align}
    |\psi_s(t_{j+1})\rangle
     & = 
 \chi_j\,\frac{e^{MT \hat{n}}|\psi_s(t_{j})\rangle}{||e^{MT \hat{n}}|\psi_s(t_{j})\rangle||}
 \notag
 \\&
 + (1-\chi_j)\,\frac{e^{-MT \hat{n}}|\psi_s(t_{j})\rangle}{||e^{-MT \hat{n}}|\psi_s(t_{j})\rangle||}.
 \label{SDSENH}
\end{align}

\subsection{Master equation for the non-Hermitian measurement protocol}\label{sec4}

Let us now derive a master equation for the system's density matrix averaged over the ``measurement outcomes'' $\sigma_j$ in the limit of weak measurements, $MT\ll 1$. The evolved non-normalized density matrix for given readout $\sigma$ is given by
\begin{align}
    &\tilde{\rho}^{(\sigma)}(T)=e^{MT \sigma \hat{n}} \rho(0) e^{MT \sigma \hat{n}}
    \notag\\
    &\!\simeq\! \left(\!1+MT \sigma \hat{n} +\frac{M^2 T^2}{2} \hat{n}\!\right)\! \rho(0)\! \left(\!1+MT \sigma \hat{n} +\frac{M^2 T^2}{2} \hat{n}\!\right)
    \nonumber
    \\
    &\!=\!\rho(0)\! +\! MT \sigma\left\{  \hat{n}, \rho(0)\right\}
   \! +\! M^2T^2 \hat{n} \rho(0) \hat{n}
    \!+\! \frac{M^2 T^2}{2}\! \left\{  \hat{n}, \rho(0)\right\}\!.
    \label{rhoT}
\end{align}    
In a general case of an anti-Hermitian Hamiltonian, one obtains 
\begin{align}
    \tilde{\rho}^{(\sigma)}(T)&=\rho(0)-i \left\{ H, \rho(0)\right\} T
    \notag\\
    &- \Big(H \rho(0) H +\frac{1}{2} \left\{ H^2, \rho(0)\right\}\Big) T^2,
    \label{antiH}
\end{align}
and Eq.~(\ref{rhoT}) is an example of such an expansion with $H$ given by Eq.~(\ref{Hmeas}).

The normalized conditioned density matrices are given by
\begin{align}
    \rho^{(+)}(T)=\frac{U_T(\sigma=+) \rho(0) U_T(\sigma=+)}
    {\text{tr}[U_T(\sigma=+) \rho(0) U_T(\sigma=+)] },
    \\
    \rho^{(-)}(T)=\frac{U_T(\sigma=-) \rho(0) U_T(\sigma=-)}
    {\text{tr}[U_T(\sigma=-) \rho(0) U_T(\sigma=-)] }.
    \label{eq46}
\end{align}
The averaged density matrix (the one for blind measurement)
reads [cf. Eq.~(\ref{blindDensMat0})]:
\begin{equation}
    \bar{\rho}(T)=\sum_{\sigma=\pm} \mathcal{P}_\sigma \rho^{(\sigma)}(T)=\sum_{\sigma=\pm} \mathcal{P}_\sigma
    \frac{U_T(\sigma) \rho(0) U_T(\sigma)}
    {\text{tr}[U_T(\sigma) \rho(0) U_T(\sigma)] },
    \label{rho-non-herm}
\end{equation}
where $\mathcal{P}_\sigma$ are measurement probabilities:
\begin{align}
  \mathcal{P}_+&=\langle \hat{n} \rangle = \text{tr}(\hat{n} \rho(0)), 
  \label{eq47}\\
  \mathcal{P}_-&=1-\langle \hat{n} \rangle = 1-\text{tr}(\hat{n} \rho(0)).
  \label{eq48}
\end{align}
Clearly, the averaged density matrix given by Eq.~(\ref{rho-non-herm}) is properly normalized: $\text{tr}\bar{\rho}(T)=1$.

However, the crucial difference of Eq.~(\ref{rho-non-herm})
compared to Eq.~(\ref{blindDensMat}) is that the measurement probabilities are no longer given by the normalization factors
\begin{equation}
   \mathcal{N}_\sigma\equiv \text{tr}[U_T(\sigma) \rho(0) U_T(\sigma)],
   \label{eq49}
\end{equation}
and, hence
\begin{equation}
     \bar{\rho}(T)\neq \sum_{\sigma=\pm} U_T(\sigma) \rho(0) U_T(\sigma).
\end{equation}
In other words, the right-hand side of Eq.~(\ref{rho-non-herm}) is nonlinear in $\rho$,
in contrast to the conventional measurement implementation.
Expanding Eq.~(\ref{rho-non-herm}) to second order in $MT$, by using Eq.~(\ref{rhoT}) and
\begin{equation}
   \mathcal{N}_\sigma\simeq 1+2MT(\sigma+MT)\, \text{tr}(\hat{n} \rho(0)),
   \label{eq51}
\end{equation}
we obtain the following Lindblad-type nonlinear master equation for the averaged density matrix:
\begin{align}
   & \frac{d \bar{\rho}}{d t} = -M\Big[1-2\,\text{tr}(\hat{n} \bar{\rho})\Big]
    \big(\{\hat{n},\bar{\rho}\}-2\bar{\rho}\,
    \text{tr}(\hat{n} \bar{\rho}) \big)\nonumber\\
    & +\! \frac{M^2 T}{2}\!
    \Big[
    \{\hat{n},\bar{\rho}\}\!+\!2\, \hat{n}\bar{\rho}\hat{n}
    \!-\!4\, \text{tr}(\hat{n} \bar{\rho})
    \Big( \bar{\rho}\!+\!
    \{\hat{n},\bar{\rho}\}\!-\! 2 \bar{\rho}\, \text{tr}(\hat{n} \bar{\rho})
    \Big)
    \Big].
    \label{nonlinear}
\end{align}
Equation~\eqref{nonlinear} should be compared with the conventional (linear) Lindblad equation, Eq.~(\ref{Lindblad-eq}), and with its non-linear counterpart for a protocol with post-selection, Eq.~\eqref{rho-ncl}. Strictly speaking, since the probabilities $\mathcal{P}_\pm$ do not cancel the normalization denominators and both depend on the current state of the system, one cannot, in general, replace the density matrix there with the averaged one: the prehistory of readouts matters. Equation (\ref{nonlinear}) was derived by decoupling such correlations for $MT\ll 1$ -- in the spirit of conventional Boltzmann equation, similarly to Eq.~(\ref{rho-full-Lindblad-2}).

For an alternative derivation of Eq.~(\ref{nonlinear}), we use the discrete map for the wave function (\ref{SDSENH}) and follow the steps that led to Eq.~(\ref{rho-full-Lindblad-2}):
\begin{align}
     \frac{d}{dt}\rho& \simeq 
    \frac{1}{T}\Big(|\psi_s(t+T)\rangle \langle \psi_s(t+T) | - |\psi_s(t)\rangle \langle \psi_s(t) |  \Big)
      \nonumber
      \\
      &\simeq 
    \frac{1}{T} \Big(\frac{\chi^2}{\mathcal{N}_+} \Big[1+ MT \hat{n}+\frac{(MT)^2 \hat{n}}{2}\Big]|\psi_s(t)\rangle
    \notag\\
    &\times
    \langle \psi_s(t)|\Big[1+MT \hat{n}+\frac{(MT)^2 \hat{n}}{2}\Big]
    \notag\\
    &+\frac{(1-\chi^2)}{\mathcal{N}_-}
    \Big[1- MT \hat{n}+\frac{(MT)^2 \hat{n}}{2}\Big]|\psi_s(t)\rangle
    \notag\\
    &\times
    \langle \psi_s(t)|\Big[1-MT \hat{n}+\frac{(MT)^2 \hat{n}}{2}\Big]
    \notag\\
    &- |\psi_s(t)\rangle \langle \psi_s(t) |
    \Big).
      \label{rho-full-Lindblad-NH}
\end{align}
Making use of
\begin{align}
    \bar{\chi}=\overline{\chi^2}=\text{tr}(\hat{n} \rho)
\end{align}
and again decoupling the averages of $\chi$ and $\rho$, we reproduce Eq.~(\ref{nonlinear}).

Taking the trace of the left-hand side and right-hand side of Eq.~(\ref{nonlinear}),
one can check explicitly that this equation is trace-preserving with $\text{tr}\bar{\rho}=1$:
\begin{align}
   \frac{d\ \text{tr}\bar{\rho}}{d t} &= -M\Big[1-2\,\text{tr}(\hat{n} \bar{\rho})\Big]
    \big[2\,\text{tr}(\hat{n}\bar{\rho})-2\,\text{tr}\bar{\rho}\,
    \text{tr}(\hat{n} \bar{\rho}) \big]\nonumber\\
    & + \frac{M^2 T}{2}
    \Big[
   2\,\text{tr}(\hat{n}\bar{\rho})
   +2\,\text{tr}(\hat{n}\bar{\rho}) \notag\\
   & -4\, \text{tr}(\hat{n} \bar{\rho})
    \Big(\text{tr}\bar{\rho}+
    2\,\text{tr}(\hat{n}\bar{\rho}) - 2\, \text{tr}\bar{\rho}\, \text{tr}(\hat{n} \bar{\rho})
    \Big)
    \Big]\nonumber
    \\
    &= \frac{M^2 T}{2}\Big[
   2\text{tr}(\hat{n}\bar{\rho})
   +2\text{tr}(\hat{n}\bar{\rho}) 
    -4\, \text{tr}(\hat{n} \bar{\rho})\, 
       \Big] = 0.  
\end{align}
Using the explicit forms of the $2\times 2$ density matrix 
and occupation operator $\hat{n}$, Eq.~(\ref{rho-explicit}),
we get $\text{tr}(\hat{n} \bar{\rho})=a$, $1-2\,\text{tr}(\hat{n} \bar{\rho})=1-2a$, 
\begin{align}
    \{\hat{n},\bar{\rho}\}=\left( \begin{array}{ll}
   2a & b\\
   b^* & 0
   \end{array}\right), \qquad  \{\hat{n},\bar{\rho}\}+2\,\hat{n}\bar{\rho}\hat{n}
    =\left( \begin{array}{ll}
   4a & b\\
   b^* & 0
   \end{array}\right),
\end{align}
\begin{align}
   \{\hat{n},\bar{\rho}\}-2\bar{\rho}\,
    \text{tr}(\hat{n} \bar{\rho})
    =\left( \begin{array}{ll}
   2(1-a)a & -(1-2a)b\\
   (1-2a)b^* & -2(1-a)a
   \end{array}\right),
\end{align}
and
\begin{align}
   \bar{\rho}+\{\hat{n},\bar{\rho}\}-2\bar{\rho}\,
    \text{tr}(\hat{n} \bar{\rho})
    =\left( \begin{array}{ll}
  a(3-2a) & 2(1-a)b\\
   2(1-a)b^* & 1-a(3-2a)
   \end{array}\right).
\end{align}
The resulting nonlinear equations for $a$ and $b,b^*$
\begin{align}
    \frac{d a}{d t} &= -2Ma
    (1-a)(1-2a)(1 - M T),
    \\
    \frac{d b}{d t} &= -M b\Big[(1-2a)^2 + 4 M T 
    \Big(a\,(1-a)-1/2\Big)
    \Big].
    \label{nonlinear-a}
\end{align}
can be solved exactly. The nonlinearity of the equation supports existence of distinct dark states at $b=0$. One sees that, in addition to the natural dark states $a=0,1$ and $b=0$, there is an emergent dark state $a=1/2$ and $b=0$ that corresponds to a maximally mixed state.

It is worth emphasizing that the stochastic anti-Hermitian Hamiltonian used in this approach produces both outcomes of the measurement step, $\sigma=\pm$, in contrast to the effective 
(and non-stochastic) Hamiltonian (\ref{Heff}) associated with postselection 
of a no-click sequence, $\sigma_j=-$ for all $j$. The fact that both are proportional to $\hat{n}$ in the above consideration is a coincidence resulting from $\hat{n}^2=\hat{n}$. In general, these are very different operators.

This consideration can be directly generalized to the case of many sites with inclusion of the system's own Hamiltonian. Note that, when starting from a half-filled pure state, the non-Hermitian measurement protocol does not involve states of other fillings at any step, so that no additional projection to the half-filling manifold is required.

Thus, in the weak-measurement limit, the above non-Hermitian measurement protocol produces a non-linear Lindblad-type master equation for the averaged density matrix. The appearance of the nonlinearity stems from the dependence of the normalization factors $\mathcal{N}_\sigma$ on the system state; this dependence in cancelled by the measurement probabilities chosen for the protocol. This is analogous to the origin
of nonlinearity in the master equation for a stochastic  non-Hermitian Hamiltonian derived in Ref.~\cite{Chen2020a},
where the stochastic variable in the anti-Hermitian part 
of the Hamiltonian was not directly related to the state at all. Nonlinearity of the master equation thus appears to be a general property of the non-Hermitian protocols.

Summarizing the content of this Appendix, the key points are as follows.  A non-Hermitian stochastic Hamiltonian is a useful way to model the generalized measurement process. The anti-Hermitian terms describe an evolution of the system towards an eigenstate of the measurement Hamiltonian, thus implementing an incomplete collapse of the wave function. On the level of the averaged density matrix $\bar{\rho}$ it leads to a Lindblad-type master equation, which has a non-linear form.  A similar nonlinearity of the master equation for $\bar{\rho}$ appears also for a conventional description of measurements, in the case of protocols with a post-selection.  Importantly, the averaged density matrix $\bar{\rho}$ does not in general carry information about the entanglement; one needs knowledge of the averaged moments $\overline{\rho^n}$. Master equations for such moments are generally non-linear \cite{Buchhold2021}.

\bibliography{ref}

\begin{thebibliography}{73}%
\makeatletter
\providecommand \@ifxundefined [1]{%
 \@ifx{#1\undefined}
}%
\providecommand \@ifnum [1]{%
 \ifnum #1\expandafter \@firstoftwo
 \else \expandafter \@secondoftwo
 \fi
}%
\providecommand \@ifx [1]{%
 \ifx #1\expandafter \@firstoftwo
 \else \expandafter \@secondoftwo
 \fi
}%
\providecommand \natexlab [1]{#1}%
\providecommand \enquote  [1]{``#1''}%
\providecommand \bibnamefont  [1]{#1}%
\providecommand \bibfnamefont [1]{#1}%
\providecommand \citenamefont [1]{#1}%
\providecommand \href@noop [0]{\@secondoftwo}%
\providecommand \href [0]{\begingroup \@sanitize@url \@href}%
\providecommand \@href[1]{\@@startlink{#1}\@@href}%
\providecommand \@@href[1]{\endgroup#1\@@endlink}%
\providecommand \@sanitize@url [0]{\catcode `\\12\catcode `\$12\catcode
  `\&12\catcode `\#12\catcode `\^12\catcode `\_12\catcode `\%12\relax}%
\providecommand \@@startlink[1]{}%
\providecommand \@@endlink[0]{}%
\providecommand \url  [0]{\begingroup\@sanitize@url \@url }%
\providecommand \@url [1]{\endgroup\@href {#1}{\urlprefix }}%
\providecommand \urlprefix  [0]{URL }%
\providecommand \Eprint [0]{\href }%
\providecommand \doibase [0]{http://dx.doi.org/}%
\providecommand \selectlanguage [0]{\@gobble}%
\providecommand \bibinfo  [0]{\@secondoftwo}%
\providecommand \bibfield  [0]{\@secondoftwo}%
\providecommand \translation [1]{[#1]}%
\providecommand \BibitemOpen [0]{}%
\providecommand \bibitemStop [0]{}%
\providecommand \bibitemNoStop [0]{.\EOS\space}%
\providecommand \EOS [0]{\spacefactor3000\relax}%
\providecommand \BibitemShut  [1]{\csname bibitem#1\endcsname}%
\let\auto@bib@innerbib\@empty
\bibitem [{\citenamefont {Breuer}\ and\ \citenamefont
  {Petruccione}(2002)}]{Breuer2002}%
  \BibitemOpen
  \bibfield  {author} {\bibinfo {author} {\bibfnamefont {Heinz-Peter}\
  \bibnamefont {Breuer}}\ and\ \bibinfo {author} {\bibfnamefont {Francesco}\
  \bibnamefont {Petruccione}},\ }\href {\doibase
  10.1093/acprof:oso/9780199213900.001.0001} {\emph {\bibinfo {title} {The
  theory of open quantum systems}}}\ (\bibinfo  {publisher} {Oxford University
  Press},\ \bibinfo {address} {Oxford},\ \bibinfo {year} {2002})\BibitemShut
  {NoStop}%
\bibitem [{\citenamefont {Rotter}\ and\ \citenamefont
  {Bird}(2015)}]{Rotter2015a}%
  \BibitemOpen
  \bibfield  {author} {\bibinfo {author} {\bibfnamefont {I.}~\bibnamefont
  {Rotter}}\ and\ \bibinfo {author} {\bibfnamefont {J.~P.}\ \bibnamefont
  {Bird}},\ }\bibfield  {title} {\enquote {\bibinfo {title} {A review of
  progress in the physics of open quantum systems: theory and experiment},}\
  }\href {\doibase 10.1088/0034-4885/78/11/114001} {\bibfield  {journal}
  {\bibinfo  {journal} {Rep. Prog. Phys.}\ }\textbf {\bibinfo {volume} {78}},\
  \bibinfo {pages} {114001} (\bibinfo {year} {2015})}\BibitemShut {NoStop}%
\bibitem [{\citenamefont {Polkovnikov}\ \emph {et~al.}(2011)\citenamefont
  {Polkovnikov}, \citenamefont {Sengupta}, \citenamefont {Silva},\ and\
  \citenamefont {Vengalattore}}]{Polkovnikov2011a}%
  \BibitemOpen
  \bibfield  {author} {\bibinfo {author} {\bibfnamefont {Anatoli}\ \bibnamefont
  {Polkovnikov}}, \bibinfo {author} {\bibfnamefont {Krishnendu}\ \bibnamefont
  {Sengupta}}, \bibinfo {author} {\bibfnamefont {Alessandro}\ \bibnamefont
  {Silva}}, \ and\ \bibinfo {author} {\bibfnamefont {Mukund}\ \bibnamefont
  {Vengalattore}},\ }\bibfield  {title} {\enquote {\bibinfo {title}
  {Colloquium: Nonequilibrium dynamics of closed interacting quantum
  systems},}\ }\href {\doibase 10.1103/RevModPhys.83.863} {\bibfield  {journal}
  {\bibinfo  {journal} {Rev. Mod. Phys.}\ }\textbf {\bibinfo {volume} {83}},\
  \bibinfo {pages} {863} (\bibinfo {year} {2011})}\BibitemShut {NoStop}%
\bibitem [{\citenamefont {Arute~\textit{et al.}}(2019)}]{Arute2019a}%
  \BibitemOpen
  \bibfield  {author} {\bibinfo {author} {\bibfnamefont {Frank}\ \bibnamefont
  {Arute~\textit{et al.}}},\ }\bibfield  {title} {\enquote {\bibinfo {title}
  {Quantum supremacy using a programmable superconducting processor},}\ }\href
  {\doibase 10.1038/s41586-019-1666-5} {\bibfield  {journal} {\bibinfo
  {journal} {Nature}\ }\textbf {\bibinfo {volume} {574}},\ \bibinfo {pages}
  {505} (\bibinfo {year} {2019})}\BibitemShut {NoStop}%
\bibitem [{\citenamefont {Roy}\ \emph {et~al.}(2020)\citenamefont {Roy},
  \citenamefont {Chalker}, \citenamefont {Gornyi},\ and\ \citenamefont
  {Gefen}}]{Roy2020}%
  \BibitemOpen
  \bibfield  {author} {\bibinfo {author} {\bibfnamefont {Sthitadhi}\
  \bibnamefont {Roy}}, \bibinfo {author} {\bibfnamefont {J.~T.}\ \bibnamefont
  {Chalker}}, \bibinfo {author} {\bibfnamefont {I.~V.}\ \bibnamefont {Gornyi}},
  \ and\ \bibinfo {author} {\bibfnamefont {Yuval}\ \bibnamefont {Gefen}},\
  }\bibfield  {title} {\enquote {\bibinfo {title} {Measurement-induced steering
  of quantum systems},}\ }\href {\doibase 10.1103/PhysRevResearch.2.033347}
  {\bibfield  {journal} {\bibinfo  {journal} {Phys. Rev. Research}\ }\textbf
  {\bibinfo {volume} {2}},\ \bibinfo {pages} {033347} (\bibinfo {year}
  {2020})}\BibitemShut {NoStop}%
\bibitem [{\citenamefont {Schlosshauer}(2005)}]{Schlosshauer2005a}%
  \BibitemOpen
  \bibfield  {author} {\bibinfo {author} {\bibfnamefont {Maximilian}\
  \bibnamefont {Schlosshauer}},\ }\bibfield  {title} {\enquote {\bibinfo
  {title} {Decoherence, the measurement problem, and interpretations of quantum
  mechanics},}\ }\href {\doibase 10.1103/RevModPhys.76.1267} {\bibfield
  {journal} {\bibinfo  {journal} {Rev. Mod. Phys.}\ }\textbf {\bibinfo {volume}
  {76}},\ \bibinfo {pages} {1267--1305} (\bibinfo {year} {2005})}\BibitemShut
  {NoStop}%
\bibitem [{\citenamefont {Wiseman}\ and\ \citenamefont
  {Milburn}(2009)}]{Wiseman2009}%
  \BibitemOpen
  \bibfield  {author} {\bibinfo {author} {\bibfnamefont {Howard~M.}\
  \bibnamefont {Wiseman}}\ and\ \bibinfo {author} {\bibfnamefont {Gerard~J.}\
  \bibnamefont {Milburn}},\ }\href@noop {} {\emph {\bibinfo {title} {Quantum
  Measurement and Control}}}\ (\bibinfo  {publisher} {Cambridge University
  Press},\ \bibinfo {address} {Cambridge (UK)},\ \bibinfo {year}
  {2009})\BibitemShut {NoStop}%
\bibitem [{\citenamefont {Calabrese}\ and\ \citenamefont
  {Cardy}(2004)}]{Calabrese2004}%
  \BibitemOpen
  \bibfield  {author} {\bibinfo {author} {\bibfnamefont {Pasquale}\
  \bibnamefont {Calabrese}}\ and\ \bibinfo {author} {\bibfnamefont {John}\
  \bibnamefont {Cardy}},\ }\bibfield  {title} {\enquote {\bibinfo {title}
  {Entanglement entropy and quantum field theory},}\ }\href {\doibase
  10.1088/1742-5468/2004/06/p06002} {\bibfield  {journal} {\bibinfo  {journal}
  {J. Stat. Mech.: Theory Exp.}\ }\textbf {\bibinfo {volume} {2004}},\ \bibinfo
  {pages} {P06002} (\bibinfo {year} {2004})}\BibitemShut {NoStop}%
\bibitem [{\citenamefont {Horodecki}\ \emph {et~al.}(2009)\citenamefont
  {Horodecki}, \citenamefont {Horodecki}, \citenamefont {Horodecki},\ and\
  \citenamefont {Horodecki}}]{Horodecki2009}%
  \BibitemOpen
  \bibfield  {author} {\bibinfo {author} {\bibfnamefont {Ryszard}\ \bibnamefont
  {Horodecki}}, \bibinfo {author} {\bibfnamefont {Pawe\l{}}\ \bibnamefont
  {Horodecki}}, \bibinfo {author} {\bibfnamefont {Micha\l{}}\ \bibnamefont
  {Horodecki}}, \ and\ \bibinfo {author} {\bibfnamefont {Karol}\ \bibnamefont
  {Horodecki}},\ }\bibfield  {title} {\enquote {\bibinfo {title} {Quantum
  entanglement},}\ }\href {\doibase 10.1103/RevModPhys.81.865} {\bibfield
  {journal} {\bibinfo  {journal} {Rev. Mod. Phys.}\ }\textbf {\bibinfo {volume}
  {81}},\ \bibinfo {pages} {865} (\bibinfo {year} {2009})}\BibitemShut
  {NoStop}%
\bibitem [{\citenamefont {Laflorencie}(2016)}]{Laflorencie2016a}%
  \BibitemOpen
  \bibfield  {author} {\bibinfo {author} {\bibfnamefont {Nicolas}\ \bibnamefont
  {Laflorencie}},\ }\bibfield  {title} {\enquote {\bibinfo {title} {Quantum
  entanglement in condensed matter systems},}\ }\href {\doibase
  https://doi.org/10.1016/j.physrep.2016.06.008} {\bibfield  {journal}
  {\bibinfo  {journal} {Phys. Rep.}\ }\textbf {\bibinfo {volume} {646}},\
  \bibinfo {pages} {1 -- 59} (\bibinfo {year} {2016})}\BibitemShut {NoStop}%
\bibitem [{\citenamefont {Li}\ \emph {et~al.}(2018)\citenamefont {Li},
  \citenamefont {Chen},\ and\ \citenamefont {Fisher}}]{Li2018a}%
  \BibitemOpen
  \bibfield  {author} {\bibinfo {author} {\bibfnamefont {Yaodong}\ \bibnamefont
  {Li}}, \bibinfo {author} {\bibfnamefont {Xiao}\ \bibnamefont {Chen}}, \ and\
  \bibinfo {author} {\bibfnamefont {Matthew P.~A.}\ \bibnamefont {Fisher}},\
  }\bibfield  {title} {\enquote {\bibinfo {title} {Quantum {Z}eno effect and
  the many-body entanglement transition},}\ }\href {\doibase
  10.1103/PhysRevB.98.205136} {\bibfield  {journal} {\bibinfo  {journal} {Phys.
  Rev. B}\ }\textbf {\bibinfo {volume} {98}},\ \bibinfo {pages} {205136}
  (\bibinfo {year} {2018})}\BibitemShut {NoStop}%
\bibitem [{\citenamefont {Chan}\ \emph {et~al.}(2019)\citenamefont {Chan},
  \citenamefont {Nandkishore}, \citenamefont {Pretko},\ and\ \citenamefont
  {Smith}}]{Chan2019a}%
  \BibitemOpen
  \bibfield  {author} {\bibinfo {author} {\bibfnamefont {Amos}\ \bibnamefont
  {Chan}}, \bibinfo {author} {\bibfnamefont {Rahul~M.}\ \bibnamefont
  {Nandkishore}}, \bibinfo {author} {\bibfnamefont {Michael}\ \bibnamefont
  {Pretko}}, \ and\ \bibinfo {author} {\bibfnamefont {Graeme}\ \bibnamefont
  {Smith}},\ }\bibfield  {title} {\enquote {\bibinfo {title}
  {Unitary-projective entanglement dynamics},}\ }\href {\doibase
  10.1103/PhysRevB.99.224307} {\bibfield  {journal} {\bibinfo  {journal} {Phys.
  Rev. B}\ }\textbf {\bibinfo {volume} {99}},\ \bibinfo {pages} {224307}
  (\bibinfo {year} {2019})}\BibitemShut {NoStop}%
\bibitem [{\citenamefont {Skinner}\ \emph {et~al.}(2019)\citenamefont
  {Skinner}, \citenamefont {Ruhman},\ and\ \citenamefont
  {Nahum}}]{Skinner2019a}%
  \BibitemOpen
  \bibfield  {author} {\bibinfo {author} {\bibfnamefont {Brian}\ \bibnamefont
  {Skinner}}, \bibinfo {author} {\bibfnamefont {Jonathan}\ \bibnamefont
  {Ruhman}}, \ and\ \bibinfo {author} {\bibfnamefont {Adam}\ \bibnamefont
  {Nahum}},\ }\bibfield  {title} {\enquote {\bibinfo {title}
  {Measurement-induced phase transitions in the dynamics of entanglement},}\
  }\href {\doibase 10.1103/PhysRevX.9.031009} {\bibfield  {journal} {\bibinfo
  {journal} {Phys. Rev. X}\ }\textbf {\bibinfo {volume} {9}},\ \bibinfo {pages}
  {031009} (\bibinfo {year} {2019})}\BibitemShut {NoStop}%
\bibitem [{\citenamefont {Li}\ \emph {et~al.}(2019)\citenamefont {Li},
  \citenamefont {Chen},\ and\ \citenamefont {Fisher}}]{Li2019a}%
  \BibitemOpen
  \bibfield  {author} {\bibinfo {author} {\bibfnamefont {Yaodong}\ \bibnamefont
  {Li}}, \bibinfo {author} {\bibfnamefont {Xiao}\ \bibnamefont {Chen}}, \ and\
  \bibinfo {author} {\bibfnamefont {Matthew P.~A.}\ \bibnamefont {Fisher}},\
  }\bibfield  {title} {\enquote {\bibinfo {title} {Measurement-driven
  entanglement transition in hybrid quantum circuits},}\ }\href {\doibase
  10.1103/PhysRevB.100.134306} {\bibfield  {journal} {\bibinfo  {journal}
  {Phys. Rev. B}\ }\textbf {\bibinfo {volume} {100}},\ \bibinfo {pages}
  {134306} (\bibinfo {year} {2019})}\BibitemShut {NoStop}%
\bibitem [{\citenamefont {Szyniszewski}\ \emph {et~al.}(2019)\citenamefont
  {Szyniszewski}, \citenamefont {Romito},\ and\ \citenamefont
  {Schomerus}}]{Szyniszewski2019a}%
  \BibitemOpen
  \bibfield  {author} {\bibinfo {author} {\bibfnamefont {M.}~\bibnamefont
  {Szyniszewski}}, \bibinfo {author} {\bibfnamefont {A.}~\bibnamefont
  {Romito}}, \ and\ \bibinfo {author} {\bibfnamefont {H.}~\bibnamefont
  {Schomerus}},\ }\bibfield  {title} {\enquote {\bibinfo {title} {Entanglement
  transition from variable-strength weak measurements},}\ }\href {\doibase
  10.1103/PhysRevB.100.064204} {\bibfield  {journal} {\bibinfo  {journal}
  {Phys. Rev. B}\ }\textbf {\bibinfo {volume} {100}},\ \bibinfo {pages}
  {064204} (\bibinfo {year} {2019})}\BibitemShut {NoStop}%
\bibitem [{\citenamefont {Szyniszewski}\ \emph {et~al.}(2020)\citenamefont
  {Szyniszewski}, \citenamefont {Romito},\ and\ \citenamefont
  {Schomerus}}]{Szyniszewski2020a}%
  \BibitemOpen
  \bibfield  {author} {\bibinfo {author} {\bibfnamefont {M.}~\bibnamefont
  {Szyniszewski}}, \bibinfo {author} {\bibfnamefont {A.}~\bibnamefont
  {Romito}}, \ and\ \bibinfo {author} {\bibfnamefont {H.}~\bibnamefont
  {Schomerus}},\ }\bibfield  {title} {\enquote {\bibinfo {title} {Universality
  of entanglement transitions from stroboscopic to continuous measurements},}\
  }\href {\doibase 10.1103/PhysRevLett.125.210602} {\bibfield  {journal}
  {\bibinfo  {journal} {Phys. Rev. Lett.}\ }\textbf {\bibinfo {volume} {125}},\
  \bibinfo {pages} {210602} (\bibinfo {year} {2020})}\BibitemShut {NoStop}%
\bibitem [{\citenamefont {Bao}\ \emph {et~al.}(2020)\citenamefont {Bao},
  \citenamefont {Choi},\ and\ \citenamefont {Altman}}]{Bao2020a}%
  \BibitemOpen
  \bibfield  {author} {\bibinfo {author} {\bibfnamefont {Yimu}\ \bibnamefont
  {Bao}}, \bibinfo {author} {\bibfnamefont {Soonwon}\ \bibnamefont {Choi}}, \
  and\ \bibinfo {author} {\bibfnamefont {Ehud}\ \bibnamefont {Altman}},\
  }\bibfield  {title} {\enquote {\bibinfo {title} {Theory of the phase
  transition in random unitary circuits with measurements},}\ }\href {\doibase
  10.1103/PhysRevB.101.104301} {\bibfield  {journal} {\bibinfo  {journal}
  {Phys. Rev. B}\ }\textbf {\bibinfo {volume} {101}},\ \bibinfo {pages}
  {104301} (\bibinfo {year} {2020})}\BibitemShut {NoStop}%
\bibitem [{\citenamefont {Gullans}\ and\ \citenamefont
  {Huse}(2020{\natexlab{a}})}]{Gullans2020a}%
  \BibitemOpen
  \bibfield  {author} {\bibinfo {author} {\bibfnamefont {Michael~J.}\
  \bibnamefont {Gullans}}\ and\ \bibinfo {author} {\bibfnamefont {David~A.}\
  \bibnamefont {Huse}},\ }\bibfield  {title} {\enquote {\bibinfo {title}
  {Dynamical purification phase transition induced by quantum measurements},}\
  }\href {\doibase 10.1103/PhysRevX.10.041020} {\bibfield  {journal} {\bibinfo
  {journal} {Phys. Rev. X}\ }\textbf {\bibinfo {volume} {10}},\ \bibinfo
  {pages} {041020} (\bibinfo {year} {2020}{\natexlab{a}})}\BibitemShut
  {NoStop}%
\bibitem [{\citenamefont {Jian}\ \emph
  {et~al.}(2020{\natexlab{a}})\citenamefont {Jian}, \citenamefont {You},
  \citenamefont {Vasseur},\ and\ \citenamefont {Ludwig}}]{Jian2020a}%
  \BibitemOpen
  \bibfield  {author} {\bibinfo {author} {\bibfnamefont {Chao-Ming}\
  \bibnamefont {Jian}}, \bibinfo {author} {\bibfnamefont {Yi-Zhuang}\
  \bibnamefont {You}}, \bibinfo {author} {\bibfnamefont {Romain}\ \bibnamefont
  {Vasseur}}, \ and\ \bibinfo {author} {\bibfnamefont {Andreas W.~W.}\
  \bibnamefont {Ludwig}},\ }\bibfield  {title} {\enquote {\bibinfo {title}
  {Measurement-induced criticality in random quantum circuits},}\ }\href
  {\doibase 10.1103/PhysRevB.101.104302} {\bibfield  {journal} {\bibinfo
  {journal} {Phys. Rev. B}\ }\textbf {\bibinfo {volume} {101}},\ \bibinfo
  {pages} {104302} (\bibinfo {year} {2020}{\natexlab{a}})}\BibitemShut
  {NoStop}%
\bibitem [{\citenamefont {Jian}\ \emph
  {et~al.}(2020{\natexlab{b}})\citenamefont {Jian}, \citenamefont {Bauer},
  \citenamefont {Keselman},\ and\ \citenamefont {Ludwig}}]{Jian2020b}%
  \BibitemOpen
  \bibfield  {author} {\bibinfo {author} {\bibfnamefont {Chao-Ming}\
  \bibnamefont {Jian}}, \bibinfo {author} {\bibfnamefont {Bela}\ \bibnamefont
  {Bauer}}, \bibinfo {author} {\bibfnamefont {Anna}\ \bibnamefont {Keselman}},
  \ and\ \bibinfo {author} {\bibfnamefont {Andreas W.~W.}\ \bibnamefont
  {Ludwig}},\ }\href@noop {} {\enquote {\bibinfo {title} {Criticality and
  entanglement in non-unitary quantum circuits and tensor networks of
  non-interacting fermions},}\ } (\bibinfo {year} {2020}{\natexlab{b}}),\
  \Eprint {http://arxiv.org/abs/2012.04666} {arXiv:2012.04666
  [cond-mat.stat-mech]} \BibitemShut {NoStop}%
\bibitem [{\citenamefont {Choi}\ \emph {et~al.}(2020)\citenamefont {Choi},
  \citenamefont {Bao}, \citenamefont {Qi},\ and\ \citenamefont
  {Altman}}]{Choi2020a}%
  \BibitemOpen
  \bibfield  {author} {\bibinfo {author} {\bibfnamefont {Soonwon}\ \bibnamefont
  {Choi}}, \bibinfo {author} {\bibfnamefont {Yimu}\ \bibnamefont {Bao}},
  \bibinfo {author} {\bibfnamefont {Xiao-Liang}\ \bibnamefont {Qi}}, \ and\
  \bibinfo {author} {\bibfnamefont {Ehud}\ \bibnamefont {Altman}},\ }\bibfield
  {title} {\enquote {\bibinfo {title} {Quantum error correction in scrambling
  dynamics and measurement-induced phase transition},}\ }\href {\doibase
  10.1103/PhysRevLett.125.030505} {\bibfield  {journal} {\bibinfo  {journal}
  {Phys. Rev. Lett.}\ }\textbf {\bibinfo {volume} {125}},\ \bibinfo {pages}
  {030505} (\bibinfo {year} {2020})}\BibitemShut {NoStop}%
\bibitem [{\citenamefont {Fan}\ \emph {et~al.}(2021)\citenamefont {Fan},
  \citenamefont {Vijay}, \citenamefont {Vishwanath},\ and\ \citenamefont
  {You}}]{Fan2020a}%
  \BibitemOpen
  \bibfield  {author} {\bibinfo {author} {\bibfnamefont {Ruihua}\ \bibnamefont
  {Fan}}, \bibinfo {author} {\bibfnamefont {Sagar}\ \bibnamefont {Vijay}},
  \bibinfo {author} {\bibfnamefont {Ashvin}\ \bibnamefont {Vishwanath}}, \ and\
  \bibinfo {author} {\bibfnamefont {Yi-Zhuang}\ \bibnamefont {You}},\
  }\bibfield  {title} {\enquote {\bibinfo {title} {Self-organized error
  correction in random unitary circuits with measurement},}\ }\href {\doibase
  10.1103/PhysRevB.103.174309} {\bibfield  {journal} {\bibinfo  {journal}
  {Phys. Rev. B}\ }\textbf {\bibinfo {volume} {103}},\ \bibinfo {pages}
  {174309} (\bibinfo {year} {2021})}\BibitemShut {NoStop}%
\bibitem [{\citenamefont {Chen}\ \emph {et~al.}(2020)\citenamefont {Chen},
  \citenamefont {Li}, \citenamefont {Fisher},\ and\ \citenamefont
  {Lucas}}]{Chen2020a}%
  \BibitemOpen
  \bibfield  {author} {\bibinfo {author} {\bibfnamefont {Xiao}\ \bibnamefont
  {Chen}}, \bibinfo {author} {\bibfnamefont {Yaodong}\ \bibnamefont {Li}},
  \bibinfo {author} {\bibfnamefont {Matthew P.~A.}\ \bibnamefont {Fisher}}, \
  and\ \bibinfo {author} {\bibfnamefont {Andrew}\ \bibnamefont {Lucas}},\
  }\bibfield  {title} {\enquote {\bibinfo {title} {Emergent conformal symmetry
  in nonunitary random dynamics of free fermions},}\ }\href {\doibase
  10.1103/PhysRevResearch.2.033017} {\bibfield  {journal} {\bibinfo  {journal}
  {Phys. Rev. Research}\ }\textbf {\bibinfo {volume} {2}},\ \bibinfo {pages}
  {033017} (\bibinfo {year} {2020})}\BibitemShut {NoStop}%
\bibitem [{\citenamefont {Sang}\ and\ \citenamefont {Hsieh}(2021)}]{Sang2020a}%
  \BibitemOpen
  \bibfield  {author} {\bibinfo {author} {\bibfnamefont {Shengqi}\ \bibnamefont
  {Sang}}\ and\ \bibinfo {author} {\bibfnamefont {Timothy~H.}\ \bibnamefont
  {Hsieh}},\ }\bibfield  {title} {\enquote {\bibinfo {title}
  {Measurement-protected quantum phases},}\ }\href {\doibase
  10.1103/PhysRevResearch.3.023200} {\bibfield  {journal} {\bibinfo  {journal}
  {Phys. Rev. Research}\ }\textbf {\bibinfo {volume} {3}},\ \bibinfo {pages}
  {023200} (\bibinfo {year} {2021})}\BibitemShut {NoStop}%
\bibitem [{\citenamefont {Zabalo}\ \emph {et~al.}(2020)\citenamefont {Zabalo},
  \citenamefont {Gullans}, \citenamefont {Wilson}, \citenamefont
  {Gopalakrishnan}, \citenamefont {Huse},\ and\ \citenamefont
  {Pixley}}]{Zabalo2020a}%
  \BibitemOpen
  \bibfield  {author} {\bibinfo {author} {\bibfnamefont {Aidan}\ \bibnamefont
  {Zabalo}}, \bibinfo {author} {\bibfnamefont {Michael~J.}\ \bibnamefont
  {Gullans}}, \bibinfo {author} {\bibfnamefont {Justin~H.}\ \bibnamefont
  {Wilson}}, \bibinfo {author} {\bibfnamefont {Sarang}\ \bibnamefont
  {Gopalakrishnan}}, \bibinfo {author} {\bibfnamefont {David~A.}\ \bibnamefont
  {Huse}}, \ and\ \bibinfo {author} {\bibfnamefont {J.~H.}\ \bibnamefont
  {Pixley}},\ }\bibfield  {title} {\enquote {\bibinfo {title} {Critical
  properties of the measurement-induced transition in random quantum
  circuits},}\ }\href {\doibase 10.1103/PhysRevB.101.060301} {\bibfield
  {journal} {\bibinfo  {journal} {Phys. Rev. B}\ }\textbf {\bibinfo {volume}
  {101}},\ \bibinfo {pages} {060301(R)} (\bibinfo {year} {2020})}\BibitemShut
  {NoStop}%
\bibitem [{\citenamefont {Ippoliti}\ \emph {et~al.}(2021)\citenamefont
  {Ippoliti}, \citenamefont {Gullans}, \citenamefont {Gopalakrishnan},
  \citenamefont {Huse},\ and\ \citenamefont {Khemani}}]{Ippoliti2021a}%
  \BibitemOpen
  \bibfield  {author} {\bibinfo {author} {\bibfnamefont {Matteo}\ \bibnamefont
  {Ippoliti}}, \bibinfo {author} {\bibfnamefont {Michael~J.}\ \bibnamefont
  {Gullans}}, \bibinfo {author} {\bibfnamefont {Sarang}\ \bibnamefont
  {Gopalakrishnan}}, \bibinfo {author} {\bibfnamefont {David~A.}\ \bibnamefont
  {Huse}}, \ and\ \bibinfo {author} {\bibfnamefont {Vedika}\ \bibnamefont
  {Khemani}},\ }\bibfield  {title} {\enquote {\bibinfo {title} {Entanglement
  phase transitions in measurement-only dynamics},}\ }\href {\doibase
  10.1103/PhysRevX.11.011030} {\bibfield  {journal} {\bibinfo  {journal} {Phys.
  Rev. X}\ }\textbf {\bibinfo {volume} {11}},\ \bibinfo {pages} {011030}
  (\bibinfo {year} {2021})}\BibitemShut {NoStop}%
\bibitem [{\citenamefont {Lavasani}\ \emph {et~al.}(2021)\citenamefont
  {Lavasani}, \citenamefont {Alavirad},\ and\ \citenamefont
  {Barkeshli}}]{Lavasani2021a}%
  \BibitemOpen
  \bibfield  {author} {\bibinfo {author} {\bibfnamefont {A.}~\bibnamefont
  {Lavasani}}, \bibinfo {author} {\bibfnamefont {Y.}~\bibnamefont {Alavirad}},
  \ and\ \bibinfo {author} {\bibfnamefont {M.}~\bibnamefont {Barkeshli}},\
  }\bibfield  {title} {\enquote {\bibinfo {title} {Measurement-induced
  topological entanglement transitions in symmetric random quantum circuits},}\
  }\href {\doibase 10.1038/s41567-020-01112-z} {\bibfield  {journal} {\bibinfo
  {journal} {Nat. Phys.}\ }\textbf {\bibinfo {volume} {17}},\ \bibinfo {pages}
  {342–347} (\bibinfo {year} {2021})}\BibitemShut {NoStop}%
\bibitem [{\citenamefont {Ippoliti}\ and\ \citenamefont
  {Khemani}(2021)}]{Ippoliti2021b}%
  \BibitemOpen
  \bibfield  {author} {\bibinfo {author} {\bibfnamefont {Matteo}\ \bibnamefont
  {Ippoliti}}\ and\ \bibinfo {author} {\bibfnamefont {Vedika}\ \bibnamefont
  {Khemani}},\ }\bibfield  {title} {\enquote {\bibinfo {title}
  {Postselection-free entanglement dynamics via spacetime duality},}\ }\href
  {\doibase 10.1103/PhysRevLett.126.060501} {\bibfield  {journal} {\bibinfo
  {journal} {Phys. Rev. Lett.}\ }\textbf {\bibinfo {volume} {126}},\ \bibinfo
  {pages} {060501} (\bibinfo {year} {2021})}\BibitemShut {NoStop}%
\bibitem [{\citenamefont {Lunt}\ and\ \citenamefont {Pal}(2020)}]{Lunt2020a}%
  \BibitemOpen
  \bibfield  {author} {\bibinfo {author} {\bibfnamefont {Oliver}\ \bibnamefont
  {Lunt}}\ and\ \bibinfo {author} {\bibfnamefont {Arijeet}\ \bibnamefont
  {Pal}},\ }\bibfield  {title} {\enquote {\bibinfo {title} {Measurement-induced
  entanglement transitions in many-body localized systems},}\ }\href {\doibase
  10.1103/PhysRevResearch.2.043072} {\bibfield  {journal} {\bibinfo  {journal}
  {Phys. Rev. Research}\ }\textbf {\bibinfo {volume} {2}},\ \bibinfo {pages}
  {043072} (\bibinfo {year} {2020})}\BibitemShut {NoStop}%
\bibitem [{\citenamefont {Lang}\ and\ \citenamefont
  {B\"uchler}(2020)}]{Lang2020a}%
  \BibitemOpen
  \bibfield  {author} {\bibinfo {author} {\bibfnamefont {Nicolai}\ \bibnamefont
  {Lang}}\ and\ \bibinfo {author} {\bibfnamefont {Hans~Peter}\ \bibnamefont
  {B\"uchler}},\ }\bibfield  {title} {\enquote {\bibinfo {title} {Entanglement
  transition in the projective transverse field {I}sing model},}\ }\href
  {\doibase 10.1103/PhysRevB.102.094204} {\bibfield  {journal} {\bibinfo
  {journal} {Phys. Rev. B}\ }\textbf {\bibinfo {volume} {102}},\ \bibinfo
  {pages} {094204} (\bibinfo {year} {2020})}\BibitemShut {NoStop}%
\bibitem [{\citenamefont {Rossini}\ and\ \citenamefont
  {Vicari}(2020)}]{Rossini2020a}%
  \BibitemOpen
  \bibfield  {author} {\bibinfo {author} {\bibfnamefont {Davide}\ \bibnamefont
  {Rossini}}\ and\ \bibinfo {author} {\bibfnamefont {Ettore}\ \bibnamefont
  {Vicari}},\ }\bibfield  {title} {\enquote {\bibinfo {title}
  {Measurement-induced dynamics of many-body systems at quantum criticality},}\
  }\href {\doibase 10.1103/PhysRevB.102.035119} {\bibfield  {journal} {\bibinfo
   {journal} {Phys. Rev. B}\ }\textbf {\bibinfo {volume} {102}},\ \bibinfo
  {pages} {035119} (\bibinfo {year} {2020})}\BibitemShut {NoStop}%
\bibitem [{\citenamefont {Biella}\ and\ \citenamefont
  {Schir{\'{o}}}(2021)}]{Biella2021a}%
  \BibitemOpen
  \bibfield  {author} {\bibinfo {author} {\bibfnamefont {Alberto}\ \bibnamefont
  {Biella}}\ and\ \bibinfo {author} {\bibfnamefont {Marco}\ \bibnamefont
  {Schir{\'{o}}}},\ }\bibfield  {title} {\enquote {\bibinfo {title}
  {Many-{B}ody {Q}uantum {Z}eno {E}ffect and {M}easurement-{I}nduced
  {S}ubradiance {T}ransition},}\ }\href {\doibase 10.22331/q-2021-08-19-528}
  {\bibfield  {journal} {\bibinfo  {journal} {{Quantum}}\ }\textbf {\bibinfo
  {volume} {5}},\ \bibinfo {pages} {528} (\bibinfo {year} {2021})}\BibitemShut
  {NoStop}%
\bibitem [{\citenamefont {Turkeshi}\ \emph {et~al.}(2021)\citenamefont
  {Turkeshi}, \citenamefont {Biella}, \citenamefont {Fazio}, \citenamefont
  {Dalmonte},\ and\ \citenamefont {Schir\'o}}]{Turkeshi2021}%
  \BibitemOpen
  \bibfield  {author} {\bibinfo {author} {\bibfnamefont {Xhek}\ \bibnamefont
  {Turkeshi}}, \bibinfo {author} {\bibfnamefont {Alberto}\ \bibnamefont
  {Biella}}, \bibinfo {author} {\bibfnamefont {Rosario}\ \bibnamefont {Fazio}},
  \bibinfo {author} {\bibfnamefont {Marcello}\ \bibnamefont {Dalmonte}}, \ and\
  \bibinfo {author} {\bibfnamefont {Marco}\ \bibnamefont {Schir\'o}},\
  }\bibfield  {title} {\enquote {\bibinfo {title} {Measurement-induced
  entanglement transitions in the quantum ising chain: From infinite to zero
  clicks},}\ }\href {\doibase 10.1103/PhysRevB.103.224210} {\bibfield
  {journal} {\bibinfo  {journal} {Phys. Rev. B}\ }\textbf {\bibinfo {volume}
  {103}},\ \bibinfo {pages} {224210} (\bibinfo {year} {2021})}\BibitemShut
  {NoStop}%
\bibitem [{\citenamefont {Cao}\ \emph {et~al.}(2019)\citenamefont {Cao},
  \citenamefont {Tilloy},\ and\ \citenamefont {Luca}}]{Cao2019a}%
  \BibitemOpen
  \bibfield  {author} {\bibinfo {author} {\bibfnamefont {Xiangyu}\ \bibnamefont
  {Cao}}, \bibinfo {author} {\bibfnamefont {Antoine}\ \bibnamefont {Tilloy}}, \
  and\ \bibinfo {author} {\bibfnamefont {Andrea~De}\ \bibnamefont {Luca}},\
  }\bibfield  {title} {\enquote {\bibinfo {title} {{Entanglement in a fermion
  chain under continuous monitoring}},}\ }\href {\doibase
  10.21468/SciPostPhys.7.2.024} {\bibfield  {journal} {\bibinfo  {journal}
  {SciPost Phys.}\ }\textbf {\bibinfo {volume} {7}},\ \bibinfo {pages} {24}
  (\bibinfo {year} {2019})}\BibitemShut {NoStop}%
\bibitem [{\citenamefont {Alberton}\ \emph {et~al.}(2021)\citenamefont
  {Alberton}, \citenamefont {Buchhold},\ and\ \citenamefont
  {Diehl}}]{Alberton2020a}%
  \BibitemOpen
  \bibfield  {author} {\bibinfo {author} {\bibfnamefont {O.}~\bibnamefont
  {Alberton}}, \bibinfo {author} {\bibfnamefont {M.}~\bibnamefont {Buchhold}},
  \ and\ \bibinfo {author} {\bibfnamefont {S.}~\bibnamefont {Diehl}},\
  }\bibfield  {title} {\enquote {\bibinfo {title} {Entanglement transition in a
  monitored free-fermion chain: From extended criticality to area law},}\
  }\href {\doibase 10.1103/PhysRevLett.126.170602} {\bibfield  {journal}
  {\bibinfo  {journal} {Phys. Rev. Lett.}\ }\textbf {\bibinfo {volume} {126}},\
  \bibinfo {pages} {170602} (\bibinfo {year} {2021})}\BibitemShut {NoStop}%
\bibitem [{\citenamefont {Buchhold}\ \emph {et~al.}(2021)\citenamefont
  {Buchhold}, \citenamefont {Minoguchi}, \citenamefont {Altland},\ and\
  \citenamefont {Diehl}}]{Buchhold2021}%
  \BibitemOpen
  \bibfield  {author} {\bibinfo {author} {\bibfnamefont {M.}~\bibnamefont
  {Buchhold}}, \bibinfo {author} {\bibfnamefont {Y.}~\bibnamefont {Minoguchi}},
  \bibinfo {author} {\bibfnamefont {A.}~\bibnamefont {Altland}}, \ and\
  \bibinfo {author} {\bibfnamefont {S.}~\bibnamefont {Diehl}},\ }\bibfield
  {title} {\enquote {\bibinfo {title} {Effective theory for the
  measurement-induced phase transition of {D}irac fermions},}\ }\href {\doibase
  10.1103/PhysRevX.11.041004} {\bibfield  {journal} {\bibinfo  {journal} {Phys.
  Rev. X}\ }\textbf {\bibinfo {volume} {11}},\ \bibinfo {pages} {041004}
  (\bibinfo {year} {2021})}\BibitemShut {NoStop}%
\bibitem [{\citenamefont {Fuji}\ and\ \citenamefont
  {Ashida}(2020)}]{Fuji2020a}%
  \BibitemOpen
  \bibfield  {author} {\bibinfo {author} {\bibfnamefont {Yohei}\ \bibnamefont
  {Fuji}}\ and\ \bibinfo {author} {\bibfnamefont {Yuto}\ \bibnamefont
  {Ashida}},\ }\bibfield  {title} {\enquote {\bibinfo {title}
  {Measurement-induced quantum criticality under continuous monitoring},}\
  }\href {\doibase 10.1103/PhysRevB.102.054302} {\bibfield  {journal} {\bibinfo
   {journal} {Phys. Rev. B}\ }\textbf {\bibinfo {volume} {102}},\ \bibinfo
  {pages} {054302} (\bibinfo {year} {2020})}\BibitemShut {NoStop}%
\bibitem [{\citenamefont {Minato}\ \emph {et~al.}(2022)\citenamefont {Minato},
  \citenamefont {Sugimoto}, \citenamefont {Kuwahara},\ and\ \citenamefont
  {Saito}}]{Minato2021a}%
  \BibitemOpen
  \bibfield  {author} {\bibinfo {author} {\bibfnamefont {Takaaki}\ \bibnamefont
  {Minato}}, \bibinfo {author} {\bibfnamefont {Koudai}\ \bibnamefont
  {Sugimoto}}, \bibinfo {author} {\bibfnamefont {Tomotaka}\ \bibnamefont
  {Kuwahara}}, \ and\ \bibinfo {author} {\bibfnamefont {Keiji}\ \bibnamefont
  {Saito}},\ }\bibfield  {title} {\enquote {\bibinfo {title} {Fate of
  measurement-induced phase transition in long-range interactions},}\ }\href
  {\doibase 10.1103/PhysRevLett.128.010603} {\bibfield  {journal} {\bibinfo
  {journal} {Phys. Rev. Lett.}\ }\textbf {\bibinfo {volume} {128}},\ \bibinfo
  {pages} {010603} (\bibinfo {year} {2022})}\BibitemShut {NoStop}%
\bibitem [{\citenamefont {Goto}\ and\ \citenamefont
  {Danshita}(2020)}]{Goto2020a}%
  \BibitemOpen
  \bibfield  {author} {\bibinfo {author} {\bibfnamefont {Shimpei}\ \bibnamefont
  {Goto}}\ and\ \bibinfo {author} {\bibfnamefont {Ippei}\ \bibnamefont
  {Danshita}},\ }\bibfield  {title} {\enquote {\bibinfo {title}
  {Measurement-induced transitions of the entanglement scaling law in ultracold
  gases with controllable dissipation},}\ }\href {\doibase
  10.1103/PhysRevA.102.033316} {\bibfield  {journal} {\bibinfo  {journal}
  {Phys. Rev. A}\ }\textbf {\bibinfo {volume} {102}},\ \bibinfo {pages}
  {033316} (\bibinfo {year} {2020})}\BibitemShut {NoStop}%
\bibitem [{\citenamefont {Snizhko}\ \emph
  {et~al.}(2021{\natexlab{a}})\citenamefont {Snizhko}, \citenamefont {Kumar},
  \citenamefont {Rao},\ and\ \citenamefont {Gefen}}]{Snizhko2020a}%
  \BibitemOpen
  \bibfield  {author} {\bibinfo {author} {\bibfnamefont {Kyrylo}\ \bibnamefont
  {Snizhko}}, \bibinfo {author} {\bibfnamefont {Parveen}\ \bibnamefont
  {Kumar}}, \bibinfo {author} {\bibfnamefont {Nihal}\ \bibnamefont {Rao}}, \
  and\ \bibinfo {author} {\bibfnamefont {Yuval}\ \bibnamefont {Gefen}},\
  }\bibfield  {title} {\enquote {\bibinfo {title} {Weak-measurement-induced
  asymmetric dephasing: Manifestation of intrinsic measurement chirality},}\
  }\href {\doibase 10.1103/PhysRevLett.127.170401} {\bibfield  {journal}
  {\bibinfo  {journal} {Phys. Rev. Lett.}\ }\textbf {\bibinfo {volume} {127}},\
  \bibinfo {pages} {170401} (\bibinfo {year} {2021}{\natexlab{a}})}\BibitemShut
  {NoStop}%
\bibitem [{\citenamefont {Snizhko}\ \emph
  {et~al.}(2021{\natexlab{b}})\citenamefont {Snizhko}, \citenamefont {Rao},
  \citenamefont {Kumar},\ and\ \citenamefont {Gefen}}]{Snizhko2020b}%
  \BibitemOpen
  \bibfield  {author} {\bibinfo {author} {\bibfnamefont {Kyrylo}\ \bibnamefont
  {Snizhko}}, \bibinfo {author} {\bibfnamefont {Nihal}\ \bibnamefont {Rao}},
  \bibinfo {author} {\bibfnamefont {Parveen}\ \bibnamefont {Kumar}}, \ and\
  \bibinfo {author} {\bibfnamefont {Yuval}\ \bibnamefont {Gefen}},\ }\bibfield
  {title} {\enquote {\bibinfo {title} {Weak-measurement-induced phases and
  dephasing: Broken symmetry of the geometric phase},}\ }\href {\doibase
  10.1103/PhysRevResearch.3.043045} {\bibfield  {journal} {\bibinfo  {journal}
  {Phys. Rev. Research}\ }\textbf {\bibinfo {volume} {3}},\ \bibinfo {pages}
  {043045} (\bibinfo {year} {2021}{\natexlab{b}})}\BibitemShut {NoStop}%
\bibitem [{\citenamefont {Kumar}\ \emph {et~al.}(2020)\citenamefont {Kumar},
  \citenamefont {Snizhko},\ and\ \citenamefont {Gefen}}]{Kumar2020b}%
  \BibitemOpen
  \bibfield  {author} {\bibinfo {author} {\bibfnamefont {Parveen}\ \bibnamefont
  {Kumar}}, \bibinfo {author} {\bibfnamefont {Kyrylo}\ \bibnamefont {Snizhko}},
  \ and\ \bibinfo {author} {\bibfnamefont {Yuval}\ \bibnamefont {Gefen}},\
  }\bibfield  {title} {\enquote {\bibinfo {title} {Engineering two-qubit mixed
  states with weak measurements},}\ }\href {\doibase
  10.1103/PhysRevResearch.2.042014} {\bibfield  {journal} {\bibinfo  {journal}
  {Phys. Rev. Research}\ }\textbf {\bibinfo {volume} {2}},\ \bibinfo {pages}
  {042014 (R)} (\bibinfo {year} {2020})}\BibitemShut {NoStop}%
\bibitem [{\citenamefont {Gebhart}\ \emph {et~al.}(2020)\citenamefont
  {Gebhart}, \citenamefont {Snizhko}, \citenamefont {Wellens}, \citenamefont
  {Buchleitner}, \citenamefont {Romito},\ and\ \citenamefont
  {Gefen}}]{Gebhart2020a}%
  \BibitemOpen
  \bibfield  {author} {\bibinfo {author} {\bibfnamefont {Valentin}\
  \bibnamefont {Gebhart}}, \bibinfo {author} {\bibfnamefont {Kyrylo}\
  \bibnamefont {Snizhko}}, \bibinfo {author} {\bibfnamefont {Thomas}\
  \bibnamefont {Wellens}}, \bibinfo {author} {\bibfnamefont {Andreas}\
  \bibnamefont {Buchleitner}}, \bibinfo {author} {\bibfnamefont {Alessandro}\
  \bibnamefont {Romito}}, \ and\ \bibinfo {author} {\bibfnamefont {Yuval}\
  \bibnamefont {Gefen}},\ }\bibfield  {title} {\enquote {\bibinfo {title}
  {Topological transition in measurement-induced geometric phases},}\ }\href
  {\doibase 10.1073/pnas.1911620117} {\bibfield  {journal} {\bibinfo  {journal}
  {Proc. Natl. Acad. Sci. U.S.A.}\ }\textbf {\bibinfo {volume} {117}},\
  \bibinfo {pages} {5706--5713} (\bibinfo {year} {2020})}\BibitemShut {NoStop}%
\bibitem [{\citenamefont {Xu}\ \emph {et~al.}(2020)\citenamefont {Xu},
  \citenamefont {Liu}, \citenamefont {Datta}, \citenamefont {Knee},
  \citenamefont {Lundeen}, \citenamefont {Lu},\ and\ \citenamefont
  {Zhang}}]{Xu2020a}%
  \BibitemOpen
  \bibfield  {author} {\bibinfo {author} {\bibfnamefont {Liang}\ \bibnamefont
  {Xu}}, \bibinfo {author} {\bibfnamefont {Zexuan}\ \bibnamefont {Liu}},
  \bibinfo {author} {\bibfnamefont {Animesh}\ \bibnamefont {Datta}}, \bibinfo
  {author} {\bibfnamefont {George~C.}\ \bibnamefont {Knee}}, \bibinfo {author}
  {\bibfnamefont {Jeff~S.}\ \bibnamefont {Lundeen}}, \bibinfo {author}
  {\bibfnamefont {Yan-qing}\ \bibnamefont {Lu}}, \ and\ \bibinfo {author}
  {\bibfnamefont {Lijian}\ \bibnamefont {Zhang}},\ }\bibfield  {title}
  {\enquote {\bibinfo {title} {Approaching quantum-limited metrology with
  imperfect detectors by using weak-value amplification},}\ }\href {\doibase
  10.1103/PhysRevLett.125.080501} {\bibfield  {journal} {\bibinfo  {journal}
  {Phys. Rev. Lett.}\ }\textbf {\bibinfo {volume} {125}},\ \bibinfo {pages}
  {080501} (\bibinfo {year} {2020})}\BibitemShut {NoStop}%
\bibitem [{\citenamefont {Ivanov}\ \emph {et~al.}(2020)\citenamefont {Ivanov},
  \citenamefont {Ivanova}, \citenamefont {Caballero-Benitez},\ and\
  \citenamefont {Mekhov}}]{Ivanov2020a}%
  \BibitemOpen
  \bibfield  {author} {\bibinfo {author} {\bibfnamefont {D.~A.}\ \bibnamefont
  {Ivanov}}, \bibinfo {author} {\bibfnamefont {T.~Yu.}\ \bibnamefont
  {Ivanova}}, \bibinfo {author} {\bibfnamefont {S.~F.}\ \bibnamefont
  {Caballero-Benitez}}, \ and\ \bibinfo {author} {\bibfnamefont {I.~B.}\
  \bibnamefont {Mekhov}},\ }\bibfield  {title} {\enquote {\bibinfo {title}
  {Feedback-induced quantum phase transitions using weak measurements},}\
  }\href {\doibase 10.1103/PhysRevLett.124.010603} {\bibfield  {journal}
  {\bibinfo  {journal} {Phys. Rev. Lett.}\ }\textbf {\bibinfo {volume} {124}},\
  \bibinfo {pages} {010603} (\bibinfo {year} {2020})}\BibitemShut {NoStop}%
\bibitem [{\citenamefont {Manousakis}\ \emph {et~al.}(2020)\citenamefont
  {Manousakis}, \citenamefont {Wille}, \citenamefont {Altland}, \citenamefont
  {Egger}, \citenamefont {Flensberg},\ and\ \citenamefont
  {Hassler}}]{Manousakis2020a}%
  \BibitemOpen
  \bibfield  {author} {\bibinfo {author} {\bibfnamefont {J.}~\bibnamefont
  {Manousakis}}, \bibinfo {author} {\bibfnamefont {C.}~\bibnamefont {Wille}},
  \bibinfo {author} {\bibfnamefont {A.}~\bibnamefont {Altland}}, \bibinfo
  {author} {\bibfnamefont {R.}~\bibnamefont {Egger}}, \bibinfo {author}
  {\bibfnamefont {K.}~\bibnamefont {Flensberg}}, \ and\ \bibinfo {author}
  {\bibfnamefont {F.}~\bibnamefont {Hassler}},\ }\bibfield  {title} {\enquote
  {\bibinfo {title} {Weak measurement protocols for {M}ajorana bound state
  identification},}\ }\href {\doibase 10.1103/PhysRevLett.124.096801}
  {\bibfield  {journal} {\bibinfo  {journal} {Phys. Rev. Lett.}\ }\textbf
  {\bibinfo {volume} {124}},\ \bibinfo {pages} {096801} (\bibinfo {year}
  {2020})}\BibitemShut {NoStop}%
\bibitem [{\citenamefont {Mu\~noz Arias}\ \emph {et~al.}(2020)\citenamefont
  {Mu\~noz Arias}, \citenamefont {Poggi}, \citenamefont {Jessen},\ and\
  \citenamefont {Deutsch}}]{Munoz2020a}%
  \BibitemOpen
  \bibfield  {author} {\bibinfo {author} {\bibfnamefont {Manuel~H.}\
  \bibnamefont {Mu\~noz Arias}}, \bibinfo {author} {\bibfnamefont {Pablo~M.}\
  \bibnamefont {Poggi}}, \bibinfo {author} {\bibfnamefont {Poul~S.}\
  \bibnamefont {Jessen}}, \ and\ \bibinfo {author} {\bibfnamefont {Ivan~H.}\
  \bibnamefont {Deutsch}},\ }\bibfield  {title} {\enquote {\bibinfo {title}
  {Simulating nonlinear dynamics of collective spins via quantum measurement
  and feedback},}\ }\href {\doibase 10.1103/PhysRevLett.124.110503} {\bibfield
  {journal} {\bibinfo  {journal} {Phys. Rev. Lett.}\ }\textbf {\bibinfo
  {volume} {124}},\ \bibinfo {pages} {110503} (\bibinfo {year}
  {2020})}\BibitemShut {NoStop}%
\bibitem [{\citenamefont {Monroe}\ \emph {et~al.}(2021)\citenamefont {Monroe},
  \citenamefont {Yunger~Halpern}, \citenamefont {Lee},\ and\ \citenamefont
  {Murch}}]{Monroe2021a}%
  \BibitemOpen
  \bibfield  {author} {\bibinfo {author} {\bibfnamefont {Jonathan~T.}\
  \bibnamefont {Monroe}}, \bibinfo {author} {\bibfnamefont {Nicole}\
  \bibnamefont {Yunger~Halpern}}, \bibinfo {author} {\bibfnamefont {Taeho}\
  \bibnamefont {Lee}}, \ and\ \bibinfo {author} {\bibfnamefont {Kater~W.}\
  \bibnamefont {Murch}},\ }\bibfield  {title} {\enquote {\bibinfo {title} {Weak
  measurement of a superconducting qubit reconciles incompatible operators},}\
  }\href {\doibase 10.1103/PhysRevLett.126.100403} {\bibfield  {journal}
  {\bibinfo  {journal} {Phys. Rev. Lett.}\ }\textbf {\bibinfo {volume} {126}},\
  \bibinfo {pages} {100403} (\bibinfo {year} {2021})}\BibitemShut {NoStop}%
\bibitem [{\citenamefont {Wang}\ \emph {et~al.}(2021)\citenamefont {Wang},
  \citenamefont {Snizhko}, \citenamefont {Romito}, \citenamefont {Gefen},\ and\
  \citenamefont {Murch}}]{Wang2021a}%
  \BibitemOpen
  \bibfield  {author} {\bibinfo {author} {\bibfnamefont {Yunzhao}\ \bibnamefont
  {Wang}}, \bibinfo {author} {\bibfnamefont {Kyrylo}\ \bibnamefont {Snizhko}},
  \bibinfo {author} {\bibfnamefont {Alessandro}\ \bibnamefont {Romito}},
  \bibinfo {author} {\bibfnamefont {Yuval}\ \bibnamefont {Gefen}}, \ and\
  \bibinfo {author} {\bibfnamefont {Kater}\ \bibnamefont {Murch}},\ }\href@noop
  {} {\enquote {\bibinfo {title} {Observing a topological transition in
  weak-measurement-induced geometric phases},}\ } (\bibinfo {year} {2021}),\
  \Eprint {http://arxiv.org/abs/2102.05660} {arXiv:2102.05660 [quant-ph]}
  \BibitemShut {NoStop}%
\bibitem [{\citenamefont {Van~Regemortel}\ \emph {et~al.}(2021)\citenamefont
  {Van~Regemortel}, \citenamefont {Cian}, \citenamefont {Seif}, \citenamefont
  {Dehghani},\ and\ \citenamefont {Hafezi}}]{VanRegemortel2021a}%
  \BibitemOpen
  \bibfield  {author} {\bibinfo {author} {\bibfnamefont {Mathias}\ \bibnamefont
  {Van~Regemortel}}, \bibinfo {author} {\bibfnamefont {Ze-Pei}\ \bibnamefont
  {Cian}}, \bibinfo {author} {\bibfnamefont {Alireza}\ \bibnamefont {Seif}},
  \bibinfo {author} {\bibfnamefont {Hossein}\ \bibnamefont {Dehghani}}, \ and\
  \bibinfo {author} {\bibfnamefont {Mohammad}\ \bibnamefont {Hafezi}},\
  }\bibfield  {title} {\enquote {\bibinfo {title} {Entanglement entropy scaling
  transition under competing monitoring protocols},}\ }\href {\doibase
  10.1103/PhysRevLett.126.123604} {\bibfield  {journal} {\bibinfo  {journal}
  {Phys. Rev. Lett.}\ }\textbf {\bibinfo {volume} {126}},\ \bibinfo {pages}
  {123604} (\bibinfo {year} {2021})}\BibitemShut {NoStop}%
\bibitem [{\citenamefont {Gullans}\ and\ \citenamefont
  {Huse}(2020{\natexlab{b}})}]{Gullans2020b}%
  \BibitemOpen
  \bibfield  {author} {\bibinfo {author} {\bibfnamefont {Michael~J.}\
  \bibnamefont {Gullans}}\ and\ \bibinfo {author} {\bibfnamefont {David~A.}\
  \bibnamefont {Huse}},\ }\bibfield  {title} {\enquote {\bibinfo {title}
  {Scalable probes of measurement-induced criticality},}\ }\href {\doibase
  10.1103/PhysRevLett.125.070606} {\bibfield  {journal} {\bibinfo  {journal}
  {Phys. Rev. Lett.}\ }\textbf {\bibinfo {volume} {125}},\ \bibinfo {pages}
  {070606} (\bibinfo {year} {2020}{\natexlab{b}})}\BibitemShut {NoStop}%
\bibitem [{\citenamefont {Sang}\ \emph {et~al.}(2021)\citenamefont {Sang},
  \citenamefont {Li}, \citenamefont {Zhou}, \citenamefont {Chen}, \citenamefont
  {Hsieh},\ and\ \citenamefont {Fisher}}]{Sang2020b}%
  \BibitemOpen
  \bibfield  {author} {\bibinfo {author} {\bibfnamefont {Shengqi}\ \bibnamefont
  {Sang}}, \bibinfo {author} {\bibfnamefont {Yaodong}\ \bibnamefont {Li}},
  \bibinfo {author} {\bibfnamefont {Tianci}\ \bibnamefont {Zhou}}, \bibinfo
  {author} {\bibfnamefont {Xiao}\ \bibnamefont {Chen}}, \bibinfo {author}
  {\bibfnamefont {Timothy~H.}\ \bibnamefont {Hsieh}}, \ and\ \bibinfo {author}
  {\bibfnamefont {Matthew~P.A.}\ \bibnamefont {Fisher}},\ }\bibfield  {title}
  {\enquote {\bibinfo {title} {Entanglement negativity at measurement-induced
  criticality},}\ }\href {\doibase 10.1103/PRXQuantum.2.030313} {\bibfield
  {journal} {\bibinfo  {journal} {PRX Quantum}\ }\textbf {\bibinfo {volume}
  {2}},\ \bibinfo {pages} {030313} (\bibinfo {year} {2021})}\BibitemShut
  {NoStop}%
\bibitem [{\citenamefont {Li}\ and\ \citenamefont {Fisher}(2021)}]{Li2021a}%
  \BibitemOpen
  \bibfield  {author} {\bibinfo {author} {\bibfnamefont {Yaodong}\ \bibnamefont
  {Li}}\ and\ \bibinfo {author} {\bibfnamefont {Matthew P.~A.}\ \bibnamefont
  {Fisher}},\ }\bibfield  {title} {\enquote {\bibinfo {title} {Statistical
  mechanics of quantum error correcting codes},}\ }\href {\doibase
  10.1103/PhysRevB.103.104306} {\bibfield  {journal} {\bibinfo  {journal}
  {Phys. Rev. B}\ }\textbf {\bibinfo {volume} {103}},\ \bibinfo {pages}
  {104306} (\bibinfo {year} {2021})}\BibitemShut {NoStop}%
\bibitem [{\citenamefont {Bao}\ \emph {et~al.}(2021)\citenamefont {Bao},
  \citenamefont {Choi},\ and\ \citenamefont {Altman}}]{Bao2021a}%
  \BibitemOpen
  \bibfield  {author} {\bibinfo {author} {\bibfnamefont {Yimu}\ \bibnamefont
  {Bao}}, \bibinfo {author} {\bibfnamefont {Soonwon}\ \bibnamefont {Choi}}, \
  and\ \bibinfo {author} {\bibfnamefont {Ehud}\ \bibnamefont {Altman}},\
  }\bibfield  {title} {\enquote {\bibinfo {title} {Symmetry enriched phases of
  quantum circuits},}\ }\href {\doibase
  https://doi.org/10.1016/j.aop.2021.168618} {\bibfield  {journal} {\bibinfo
  {journal} {Annals of Physics}\ }\textbf {\bibinfo {volume} {435}},\ \bibinfo
  {pages} {168618} (\bibinfo {year} {2021})},\ \bibinfo {note} {special issue
  on Philip W. Anderson}\BibitemShut {NoStop}%
\bibitem [{\citenamefont {Jian}\ \emph {et~al.}(2021)\citenamefont {Jian},
  \citenamefont {Liu}, \citenamefont {Chen}, \citenamefont {Swingle},\ and\
  \citenamefont {Zhang}}]{Jian2021a}%
  \BibitemOpen
  \bibfield  {author} {\bibinfo {author} {\bibfnamefont {Shao-Kai}\
  \bibnamefont {Jian}}, \bibinfo {author} {\bibfnamefont {Chunxiao}\
  \bibnamefont {Liu}}, \bibinfo {author} {\bibfnamefont {Xiao}\ \bibnamefont
  {Chen}}, \bibinfo {author} {\bibfnamefont {Brian}\ \bibnamefont {Swingle}}, \
  and\ \bibinfo {author} {\bibfnamefont {Pengfei}\ \bibnamefont {Zhang}},\
  }\bibfield  {title} {\enquote {\bibinfo {title} {Measurement-induced phase
  transition in the monitored {S}achdev-{Y}e-{K}itaev model},}\ }\href
  {\doibase 10.1103/PhysRevLett.127.140601} {\bibfield  {journal} {\bibinfo
  {journal} {Phys. Rev. Lett.}\ }\textbf {\bibinfo {volume} {127}},\ \bibinfo
  {pages} {140601} (\bibinfo {year} {2021})}\BibitemShut {NoStop}%
\bibitem [{\citenamefont {Calabrese}\ and\ \citenamefont
  {Cardy}(2009)}]{Calabrese2009}%
  \BibitemOpen
  \bibfield  {author} {\bibinfo {author} {\bibfnamefont {Pasquale}\
  \bibnamefont {Calabrese}}\ and\ \bibinfo {author} {\bibfnamefont {John}\
  \bibnamefont {Cardy}},\ }\bibfield  {title} {\enquote {\bibinfo {title}
  {Entanglement entropy and conformal field theory},}\ }\href {\doibase
  10.1088/1751-8113/42/50/504005} {\bibfield  {journal} {\bibinfo  {journal}
  {J. Phys. A}\ }\textbf {\bibinfo {volume} {42}},\ \bibinfo {pages} {504005}
  (\bibinfo {year} {2009})}\BibitemShut {NoStop}%
\bibitem [{\citenamefont {Gopalakrishnan}\ and\ \citenamefont
  {Gullans}(2021)}]{Gopalakrishnan2020a}%
  \BibitemOpen
  \bibfield  {author} {\bibinfo {author} {\bibfnamefont {Sarang}\ \bibnamefont
  {Gopalakrishnan}}\ and\ \bibinfo {author} {\bibfnamefont {Michael~J.}\
  \bibnamefont {Gullans}},\ }\bibfield  {title} {\enquote {\bibinfo {title}
  {Entanglement and purification transitions in non-hermitian quantum
  mechanics},}\ }\href {\doibase 10.1103/PhysRevLett.126.170503} {\bibfield
  {journal} {\bibinfo  {journal} {Phys. Rev. Lett.}\ }\textbf {\bibinfo
  {volume} {126}},\ \bibinfo {pages} {170503} (\bibinfo {year}
  {2021})}\BibitemShut {NoStop}%
\bibitem [{\citenamefont {Schollw\"ock}(2011)}]{Schollwock2011a}%
  \BibitemOpen
  \bibfield  {author} {\bibinfo {author} {\bibfnamefont {U.}~\bibnamefont
  {Schollw\"ock}},\ }\bibfield  {title} {\enquote {\bibinfo {title} {The
  density-matrix renormalization group in the age of matrix product states},}\
  }\href {\doibase 10.1016/j.aop.2010.09.012} {\bibfield  {journal} {\bibinfo
  {journal} {Ann. Phys. (N. Y.)}\ }\textbf {\bibinfo {volume} {326}},\ \bibinfo
  {pages} {96 -- 192} (\bibinfo {year} {2011})}\BibitemShut {NoStop}%
\bibitem [{\citenamefont {Paeckel}\ \emph {et~al.}(2019)\citenamefont
  {Paeckel}, \citenamefont {Köhler}, \citenamefont {Swoboda}, \citenamefont
  {Manmana}, \citenamefont {Schollwöck},\ and\ \citenamefont
  {Hubig}}]{Paeckel2019a}%
  \BibitemOpen
  \bibfield  {author} {\bibinfo {author} {\bibfnamefont {Sebastian}\
  \bibnamefont {Paeckel}}, \bibinfo {author} {\bibfnamefont {Thomas}\
  \bibnamefont {Köhler}}, \bibinfo {author} {\bibfnamefont {Andreas}\
  \bibnamefont {Swoboda}}, \bibinfo {author} {\bibfnamefont {Salvatore~R.}\
  \bibnamefont {Manmana}}, \bibinfo {author} {\bibfnamefont {Ulrich}\
  \bibnamefont {Schollwöck}}, \ and\ \bibinfo {author} {\bibfnamefont
  {Claudius}\ \bibnamefont {Hubig}},\ }\bibfield  {title} {\enquote {\bibinfo
  {title} {Time-evolution methods for matrix-product states},}\ }\href
  {\doibase https://doi.org/10.1016/j.aop.2019.167998} {\bibfield  {journal}
  {\bibinfo  {journal} {Ann. Phys. (N. Y.)}\ }\textbf {\bibinfo {volume}
  {411}},\ \bibinfo {pages} {167998} (\bibinfo {year} {2019})}\BibitemShut
  {NoStop}%
\bibitem [{\citenamefont {Tang}\ and\ \citenamefont {Zhu}(2020)}]{Tang2020a}%
  \BibitemOpen
  \bibfield  {author} {\bibinfo {author} {\bibfnamefont {Qicheng}\ \bibnamefont
  {Tang}}\ and\ \bibinfo {author} {\bibfnamefont {W.}~\bibnamefont {Zhu}},\
  }\bibfield  {title} {\enquote {\bibinfo {title} {Measurement-induced phase
  transition: A case study in the nonintegrable model by density-matrix
  renormalization group calculations},}\ }\href {\doibase
  10.1103/PhysRevResearch.2.013022} {\bibfield  {journal} {\bibinfo  {journal}
  {Phys. Rev. Research}\ }\textbf {\bibinfo {volume} {2}},\ \bibinfo {pages}
  {013022} (\bibinfo {year} {2020})}\BibitemShut {NoStop}%
\bibitem [{\citenamefont {Haegeman}\ \emph {et~al.}(2016)\citenamefont
  {Haegeman}, \citenamefont {Lubich}, \citenamefont {Oseledets}, \citenamefont
  {Vandereycken},\ and\ \citenamefont {Verstraete}}]{Haegeman2016a}%
  \BibitemOpen
  \bibfield  {author} {\bibinfo {author} {\bibfnamefont {Jutho}\ \bibnamefont
  {Haegeman}}, \bibinfo {author} {\bibfnamefont {Christian}\ \bibnamefont
  {Lubich}}, \bibinfo {author} {\bibfnamefont {Ivan}\ \bibnamefont
  {Oseledets}}, \bibinfo {author} {\bibfnamefont {Bart}\ \bibnamefont
  {Vandereycken}}, \ and\ \bibinfo {author} {\bibfnamefont {Frank}\
  \bibnamefont {Verstraete}},\ }\bibfield  {title} {\enquote {\bibinfo {title}
  {Unifying time evolution and optimization with matrix product states},}\
  }\href {\doibase 10.1103/PhysRevB.94.165116} {\bibfield  {journal} {\bibinfo
  {journal} {Phys. Rev. B}\ }\textbf {\bibinfo {volume} {94}},\ \bibinfo
  {pages} {165116} (\bibinfo {year} {2016})}\BibitemShut {NoStop}%
\bibitem [{\citenamefont {Doggen}\ \emph {et~al.}(2020)\citenamefont {Doggen},
  \citenamefont {Gornyi}, \citenamefont {Mirlin},\ and\ \citenamefont
  {Polyakov}}]{Doggen2020a}%
  \BibitemOpen
  \bibfield  {author} {\bibinfo {author} {\bibfnamefont {Elmer V.~H.}\
  \bibnamefont {Doggen}}, \bibinfo {author} {\bibfnamefont {Igor~V.}\
  \bibnamefont {Gornyi}}, \bibinfo {author} {\bibfnamefont {Alexander~D.}\
  \bibnamefont {Mirlin}}, \ and\ \bibinfo {author} {\bibfnamefont {Dmitry~G.}\
  \bibnamefont {Polyakov}},\ }\bibfield  {title} {\enquote {\bibinfo {title}
  {Slow many-body delocalization beyond one dimension},}\ }\href {\doibase
  10.1103/PhysRevLett.125.155701} {\bibfield  {journal} {\bibinfo  {journal}
  {Phys. Rev. Lett.}\ }\textbf {\bibinfo {volume} {125}},\ \bibinfo {pages}
  {155701} (\bibinfo {year} {2020})}\BibitemShut {NoStop}%
\bibitem [{\citenamefont {Daley}(2014)}]{Daley2014a}%
  \BibitemOpen
  \bibfield  {author} {\bibinfo {author} {\bibfnamefont {Andrew~J.}\
  \bibnamefont {Daley}},\ }\bibfield  {title} {\enquote {\bibinfo {title}
  {Quantum trajectories and open many-body quantum systems},}\ }\href {\doibase
  10.1080/00018732.2014.933502} {\bibfield  {journal} {\bibinfo  {journal}
  {Adv. Phys.}\ }\textbf {\bibinfo {volume} {63}},\ \bibinfo {pages} {77--149}
  (\bibinfo {year} {2014})}\BibitemShut {NoStop}%
\bibitem [{\citenamefont {Dalibard}\ \emph {et~al.}(1992)\citenamefont
  {Dalibard}, \citenamefont {Castin},\ and\ \citenamefont
  {M\o{}lmer}}]{Dalibard1992a}%
  \BibitemOpen
  \bibfield  {author} {\bibinfo {author} {\bibfnamefont {Jean}\ \bibnamefont
  {Dalibard}}, \bibinfo {author} {\bibfnamefont {Yvan}\ \bibnamefont {Castin}},
  \ and\ \bibinfo {author} {\bibfnamefont {Klaus}\ \bibnamefont {M\o{}lmer}},\
  }\bibfield  {title} {\enquote {\bibinfo {title} {Wave-function approach to
  dissipative processes in quantum optics},}\ }\href {\doibase
  10.1103/PhysRevLett.68.580} {\bibfield  {journal} {\bibinfo  {journal} {Phys.
  Rev. Lett.}\ }\textbf {\bibinfo {volume} {68}},\ \bibinfo {pages} {580--583}
  (\bibinfo {year} {1992})}\BibitemShut {NoStop}%
\bibitem [{\citenamefont {Patel}\ and\ \citenamefont
  {Kumar}(2017)}]{Patel2017a}%
  \BibitemOpen
  \bibfield  {author} {\bibinfo {author} {\bibfnamefont {Apoorva}\ \bibnamefont
  {Patel}}\ and\ \bibinfo {author} {\bibfnamefont {Parveen}\ \bibnamefont
  {Kumar}},\ }\bibfield  {title} {\enquote {\bibinfo {title} {Weak
  measurements, quantum-state collapse, and the born rule},}\ }\href {\doibase
  10.1103/PhysRevA.96.022108} {\bibfield  {journal} {\bibinfo  {journal} {Phys.
  Rev. A}\ }\textbf {\bibinfo {volume} {96}},\ \bibinfo {pages} {022108}
  (\bibinfo {year} {2017})}\BibitemShut {NoStop}%
\bibitem [{\citenamefont {De~Chiara}\ \emph {et~al.}(2006)\citenamefont
  {De~Chiara}, \citenamefont {Montangero}, \citenamefont {Calabrese},\ and\
  \citenamefont {Fazio}}]{DeChiara2006a}%
  \BibitemOpen
  \bibfield  {author} {\bibinfo {author} {\bibfnamefont {Gabriele}\
  \bibnamefont {De~Chiara}}, \bibinfo {author} {\bibfnamefont {Simone}\
  \bibnamefont {Montangero}}, \bibinfo {author} {\bibfnamefont {Pasquale}\
  \bibnamefont {Calabrese}}, \ and\ \bibinfo {author} {\bibfnamefont {Rosario}\
  \bibnamefont {Fazio}},\ }\bibfield  {title} {\enquote {\bibinfo {title}
  {Entanglement entropy dynamics of {H}eisenberg chains},}\ }\href {\doibase
  10.1088/1742-5468/2006/03/P03001} {\bibfield  {journal} {\bibinfo  {journal}
  {J. Stat. Mech.: Theory Exp.}\ }\textbf {\bibinfo {volume} {2006}},\ \bibinfo
  {pages} {P03001} (\bibinfo {year} {2006})}\BibitemShut {NoStop}%
\bibitem [{\citenamefont {Doggen}\ \emph {et~al.}(2021)\citenamefont {Doggen},
  \citenamefont {Gornyi},\ and\ \citenamefont {Polyakov}}]{Doggen2021a}%
  \BibitemOpen
  \bibfield  {author} {\bibinfo {author} {\bibfnamefont {Elmer V.~H.}\
  \bibnamefont {Doggen}}, \bibinfo {author} {\bibfnamefont {Igor~V.}\
  \bibnamefont {Gornyi}}, \ and\ \bibinfo {author} {\bibfnamefont {Dmitry~G.}\
  \bibnamefont {Polyakov}},\ }\bibfield  {title} {\enquote {\bibinfo {title}
  {Stark many-body localization: Evidence for {H}ilbert-space shattering},}\
  }\href {\doibase 10.1103/PhysRevB.103.L100202} {\bibfield  {journal}
  {\bibinfo  {journal} {Phys. Rev. B}\ }\textbf {\bibinfo {volume} {103}},\
  \bibinfo {pages} {L100202} (\bibinfo {year} {2021})}\BibitemShut {NoStop}%
\bibitem [{\citenamefont {Snizhko}\ \emph {et~al.}(2020)\citenamefont
  {Snizhko}, \citenamefont {Kumar},\ and\ \citenamefont {Romito}}]{Kumar2020}%
  \BibitemOpen
  \bibfield  {author} {\bibinfo {author} {\bibfnamefont {Kyrylo}\ \bibnamefont
  {Snizhko}}, \bibinfo {author} {\bibfnamefont {Parveen}\ \bibnamefont
  {Kumar}}, \ and\ \bibinfo {author} {\bibfnamefont {Alessandro}\ \bibnamefont
  {Romito}},\ }\bibfield  {title} {\enquote {\bibinfo {title} {Quantum {Z}eno
  effect appears in stages},}\ }\href {\doibase
  10.1103/PhysRevResearch.2.033512} {\bibfield  {journal} {\bibinfo  {journal}
  {Phys. Rev. Research}\ }\textbf {\bibinfo {volume} {2}},\ \bibinfo {pages}
  {033512} (\bibinfo {year} {2020})}\BibitemShut {NoStop}%
\bibitem [{\citenamefont {Bakr}\ \emph {et~al.}(2009)\citenamefont {Bakr},
  \citenamefont {Gillen}, \citenamefont {Peng}, \citenamefont {Fölling},\ and\
  \citenamefont {Greiner}}]{Bakr2009a}%
  \BibitemOpen
  \bibfield  {author} {\bibinfo {author} {\bibfnamefont {Waseem~S.}\
  \bibnamefont {Bakr}}, \bibinfo {author} {\bibfnamefont {Jonathon~I.}\
  \bibnamefont {Gillen}}, \bibinfo {author} {\bibfnamefont {Amy}\ \bibnamefont
  {Peng}}, \bibinfo {author} {\bibfnamefont {Simon}\ \bibnamefont {Fölling}},
  \ and\ \bibinfo {author} {\bibfnamefont {Markus}\ \bibnamefont {Greiner}},\
  }\bibfield  {title} {\enquote {\bibinfo {title} {A quantum gas microscope for
  detecting single atoms in a {H}ubbard-regime optical lattice},}\ }\href
  {\doibase 10.1038/nature08482} {\bibfield  {journal} {\bibinfo  {journal}
  {Nature}\ }\textbf {\bibinfo {volume} {462}},\ \bibinfo {pages} {74--77}
  (\bibinfo {year} {2009})}\BibitemShut {NoStop}%
\bibitem [{\citenamefont {Kamakari}\ \emph {et~al.}(2022)\citenamefont
  {Kamakari}, \citenamefont {Sun}, \citenamefont {Motta},\ and\ \citenamefont
  {Minnich}}]{Kamakari2021a}%
  \BibitemOpen
  \bibfield  {author} {\bibinfo {author} {\bibfnamefont {Hirsh}\ \bibnamefont
  {Kamakari}}, \bibinfo {author} {\bibfnamefont {Shi-Ning}\ \bibnamefont
  {Sun}}, \bibinfo {author} {\bibfnamefont {Mario}\ \bibnamefont {Motta}}, \
  and\ \bibinfo {author} {\bibfnamefont {Austin~J.}\ \bibnamefont {Minnich}},\
  }\bibfield  {title} {\enquote {\bibinfo {title} {Digital quantum simulation
  of open quantum systems using quantum imaginary--time evolution},}\ }\href
  {\doibase 10.1103/PRXQuantum.3.010320} {\bibfield  {journal} {\bibinfo
  {journal} {PRX Quantum}\ }\textbf {\bibinfo {volume} {3}},\ \bibinfo {pages}
  {010320} (\bibinfo {year} {2022})}\BibitemShut {NoStop}%
\bibitem [{\citenamefont {Hauschild}\ and\ \citenamefont
  {Pollmann}(2018)}]{tenpy}%
  \BibitemOpen
  \bibfield  {author} {\bibinfo {author} {\bibfnamefont {Johannes}\
  \bibnamefont {Hauschild}}\ and\ \bibinfo {author} {\bibfnamefont {Frank}\
  \bibnamefont {Pollmann}},\ }\bibfield  {title} {\enquote {\bibinfo {title}
  {{Efficient numerical simulations with Tensor Networks: Tensor Network Python
  (TeNPy)}},}\ }\href {\doibase 10.21468/SciPostPhysLectNotes.5} {\bibfield
  {journal} {\bibinfo  {journal} {SciPost Phys. Lect. Notes}\ ,\ \bibinfo
  {pages} {5}} (\bibinfo {year} {2018})},\ \bibinfo {note} {code available from
  \url{https://github.com/tenpy/tenpy}}\BibitemShut {NoStop}%
\bibitem [{\citenamefont {Jacobs}\ and\ \citenamefont
  {Steck}(2006)}]{Jacobs2006a}%
  \BibitemOpen
  \bibfield  {author} {\bibinfo {author} {\bibfnamefont {Kurt}\ \bibnamefont
  {Jacobs}}\ and\ \bibinfo {author} {\bibfnamefont {Daniel~A.}\ \bibnamefont
  {Steck}},\ }\bibfield  {title} {\enquote {\bibinfo {title} {A straightforward
  introduction to continuous quantum measurement},}\ }\href {\doibase
  10.1080/00107510601101934} {\bibfield  {journal} {\bibinfo  {journal}
  {Contemp. Phys.}\ }\textbf {\bibinfo {volume} {47}},\ \bibinfo {pages}
  {279--303} (\bibinfo {year} {2006})}\BibitemShut {NoStop}%
\bibitem [{\citenamefont {Elitzur}\ and\ \citenamefont
  {Vaidman}(1993)}]{Elitzur1993}%
  \BibitemOpen
  \bibfield  {author} {\bibinfo {author} {\bibfnamefont {A.C.}\ \bibnamefont
  {Elitzur}}\ and\ \bibinfo {author} {\bibfnamefont {L.}~\bibnamefont
  {Vaidman}},\ }\bibfield  {title} {\enquote {\bibinfo {title} {Quantum
  mechanical interaction-free measurements},}\ }\href {\doibase
  10.1007/BF00736012} {\bibfield  {journal} {\bibinfo  {journal} {Found.
  Phys.}\ }\textbf {\bibinfo {volume} {23}},\ \bibinfo {pages} {987} (\bibinfo
  {year} {1993})}\BibitemShut {NoStop}%
\end{thebibliography}%

\end{document}